\documentclass{aa}  

\usepackage{graphicx}
\usepackage{lscape}
\usepackage{color}
\usepackage{longtable}
\usepackage{morefloats}
\usepackage{marvosym}
\usepackage{txfonts}
\usepackage{color,hyperref}
\definecolor{darkblue}{rgb}{0.0,0.0,0.3}
\hypersetup{colorlinks,breaklinks,
            linkcolor=blue,urlcolor=blue,
            anchorcolor=blue,citecolor=blue}

\newcommand{\kms}{km s$^{-1}$}
\usepackage{floatrow}
\usepackage{mathtools}
\usepackage{amssymb}
 
\begin{document}

\title{\textit{Herschel}/HIFI spectral line survey of the Orion Bar}
\subtitle{Temperature and density differentiation near the PDR surface}
\authorrunning{Z. Nagy et al.}

\author{Z. Nagy\inst{1,2};
        Y. Choi\inst{3,4,5};
        V. Ossenkopf-Okada\inst{1};
        F. F. S. van der Tak\inst{3,4};
        E. A. Bergin\inst{6};
        M. Gerin\inst{7};
        C. Joblin\inst{8,9};
        M. R{\"o}llig\inst{1};
        R. Simon\inst{1};
        \and           
        J. Stutzki\inst{1}
        }
          
\institute{
I. Physikalisches Institut, Universit\"at zu K\"oln, Z\"ulpicher Str. 77, 50937 K\"oln, Germany
\and
Department of Physics and Astronomy, University of Toledo, 2801 West Bancroft Street, Toledo, OH 43606, USA \\
\email{zsofia.nagy.astro@gmail.com} 
\and
Kapteyn Astronomical Institute, University of Groningen, PO box 800, 9700 AV Groningen, The Netherlands 
\and
SRON Netherlands Institute for Space Research, Landleven 12, 9747 AD Groningen, The Netherlands
\and
School of Space Research, Kyung Hee University, 1732, Deogyeong-daero, Giheung-gu, Yongin-si, Gyeonggi-do 17104, Republic of Korea
\and
University of Michigan, Ann Arbor, MI 48197, USA
\and
LERMA, UMR 8112 du CNRS, Observatoire de Paris, \'Ecole Normale Sup\'erieure, France
\and
Universit\'e de Toulouse, UPS-OMP, IRAP, Toulouse, France
\and
CNRS, IRAP, 9 Av. Colonel Roche, BP 44346, 31028 Toulouse Cedex 4, France
}


 
\abstract
{Photon dominated regions (PDRs) are interfaces between the mainly ionized and mainly molecular material around young massive stars. Analysis of the physical and chemical structure of such regions traces the impact of far-ultraviolet radiation of young massive stars on their environment.}
{We present results on the physical and chemical structure of the prototypical high UV-illumination edge-on Orion Bar PDR from an unbiased spectral line survey with a wide spectral coverage which includes lines of many important gas coolants such as [C{\sc{ii}}], [C{\sc{i}}], and CO and other key molecules such as H$_2$CO, H$_2$O, HCN, HCO$^+$, and SO.}
{A spectral scan from 480-1250 GHz and 1410-1910 GHz at 1.1 MHz resolution was obtained by the HIFI instrument on board the \textit{Herschel} Space Observatory. 
We obtained physical parameters for the observed molecules. For molecules with multiple transitions we used rotational diagrams to obtain excitation temperatures and column densities. For species with a single detected transition we used an optically thin LTE approximation. In the case of species with available collisional rates, we also performed a non-LTE analysis to obtain kinetic temperatures, H$_2$ volume densities, and column densities.
}
{About 120 lines corresponding to 29 molecules (including isotopologues) have been detected in the \textit{Herschel}/HIFI line survey, including 11 transitions of CO, 7 transitions of $^{13}$CO, 6 transitions of C$^{18}$O, 10 transitions of H$_2$CO, and 6 transitions of H$_2$O.
The rotational temperatures are in the range between $\sim$22 and $\sim$146~K and the column densities are in the range between 1.8$\times$10$^{12}$ cm$^{-2}$ and 4.5$\times$10$^{17}$ cm$^{-2}$.
For species with at least three detected transitions and available collisional excitation rates we derived a best fit kinetic temperature and H$_2$ volume density. Most species trace kinetic temperatures in the range between 100 and 150~K and H$_2$ volume densities in the range between 10$^5$ and 10$^6$ cm$^{-3}$. The species with temperatures and / or densities outside  this range include the H$_2$CO transitions tracing a very high temperature (315~K) and density ($1.4\times10^6$ cm$^{-3}$) component and SO corresponding to the lowest temperature (56~K) measured as a part of this line survey.
}
{
The observed lines/species reveal a range of physical conditions (gas density /temperature) involving structures at high density / high pressure, making the traditional clump/interclump picture of the Orion Bar obsolete.
}


\maketitle
%

\section{Introduction}

Massive stars have a strong impact on their environment and play an important role in influencing the physical and chemical structure of the interstellar medium. One of the most important feedback effects is radiation feedback by the far-ultraviolet (FUV) radiation of massive stars, which dissociates molecules in their surroundings and creates different chemical layers as a function of depth into the molecular clouds. These regions between the fully ionized and fully molecular clouds are photon dominated regions (PDRs), which can be studied well at sub-mm and far-infrared wavelengths. 
See \citet{hollenbachtielens1997} for a review.
 
\begin{figure}[h]
\centering
\includegraphics[width=\textwidth, trim=0.0cm 0.0cm 0.0cm -0.8cm,clip=true]{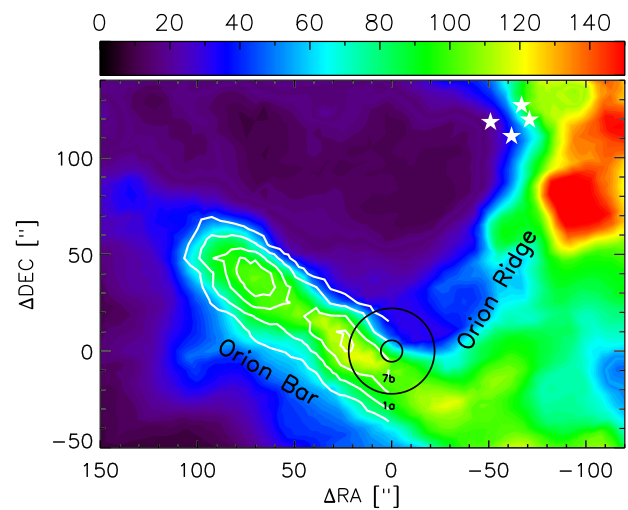}
\caption{$^{13}$CO 3-2 integrated intensity (in K \kms units) of the Orion Bar observed with the JCMT. The circles at the position of the CO$^+$ peak ($\alpha_\mathrm{J2000}=\rm{05^h35^m20.6^s}$, $\delta_\mathrm{J2000}=-05^\circ 25'14''$) indicate the beam sizes corresponding to the smallest (band 7b, $\sim$11$''$) and largest (band 1a, $\sim$44$''$) beam sizes of the data presented in this line survey. The white symbols show the position of the Trapezium stars.
The white contours show the distribution of the integrated intensity (in K \kms units) of HCN 4-3 measured with the JCMT \citep{vanderwiel2009}. The contours are 25\%, 40\%, 60\%, and 80\% of the peak HCN intensity of $\sim$24 K \kms.}
\label{orionbar_map}
\end{figure}

In this paper we study the physical and chemical structure of the well-known Orion Bar PDR based on data from an unbiased spectral line survey. 
The Orion Bar is part of the Orion Molecular cloud (OMC) 1, located at a distance of $\sim$420 pc (\citealt{menten2007}, \citealt{hirota2007}). Parts of the OMC 1 region are irradiated by the Trapezium cluster which consists of four massive stars including the O6 star $\Theta^1$ Ori C. The H{\sc{ii}} region around the Trapezium is surrounded by the Orion Bar on its south-eastern side and the Orion Ridge on its western side (Fig. \ref{orionbar_map}). In addition to the PDR interfaces around the Trapezium, there is ongoing star formation in two parts of the OMC 1 region: the Orion BN/KL and Orion S regions. While Orion BN/KL (e.g. \citealp{tercero2010}) is part of the OMC 1, Orion S is located within the ionized nebula in front of the OMC 1 cloud \citep{odell2009}.
Thanks to its nearly edge-on orientation, the Orion Bar is an ideal source to study PDR physics and chemistry. With its high FUV field ($1-4 \times 10^4 \chi_0$), in \citet{draine1978} units and average kinetic temperature of 85 K \citep{hogerheijde1995}, it is a prototypical high UV-illumination, warm PDR. 
Part of the molecular line emission measured toward the Orion Bar corresponds to an `interclump medium' with densities between a few 10$^4$ and 2$\times$10$^5$ cm$^{-3}$ \citep{simon1997}. It has been suggested that other molecular lines originate in clumps with densities in the range between 1.5$\times$10$^6$ and 6$\times$10$^6$ cm$^{-3}$ \citep{lisschilke2003}. Evidence for densities higher than the average gas density of 10$^5$ cm$^{-3}$ \citep{hogerheijde1995} has also been found based on OH transitions, tracing a warm ($T_{\rm{kin}}\sim$160-220 K) and dense ($n_{\rm{H}}\sim$10$^6$-10$^7$ cm$^{-3}$) gas component with a small filling factor \citep{goicoechea2011}.
The clumpiness of the Orion Bar was found to be consistent with the intensity of [C{\sc{ii}}] and different $^{12}$/$^{13}$CO and HCO$^+$ transitions using a three-dimensional PDR model representing the Orion Bar PDR by a clumpy edge-on cavity wall \citep{andreelabsch2014}.

Previous spectral line surveys of the Orion Bar were carried out at mm- and sub-mm wavelengths including the 279--308 GHz \citep{leurini2006} and 330-360 GHz \citep{vanderwiel2009} ranges, and in different bands of the IRAM 30m telescope in the 80--359 GHz frequency range \citep{cuadrado2015}. 
In addition to the line surveys performed with ground-based telescopes, a line survey in the 447--1545 GHz range was carried out at low spectral resolution using \textit{Herschel}/SPIRE \citep{habart2010}. 

In this paper, we present a \textit{Herschel}/HIFI line survey of spectrally resolved data covering the 480--1250 GHz and 1410--1910 GHz frequency ranges. Previous papers based on this line survey include the first detection of HF in emission toward a Galactic source \citep{vandertak2012}. This line survey also resulted in the analysis of the properties of reactive ions such as CH$^+$ and SH$^+$ \citep{nagy2013a}, and OH$^+$ \citep{vandertak2013}. Two H$_2^{18}$O transitions detected as part of this line survey allowed the ortho-to-para ratio of water to be measured toward the Orion Bar \citep{choi2014}. 
We also  analysed six high-$N$ transitions ($N$=6-5,...,10-9) of C$_2$H as a part of this line survey and found them to trace warm and dense gas including a component with densities of $n$(H$_2$)$\sim$5$\times$10$^6$ cm$^{-3}$ and temperatures of $T_{\rm{kin}}\sim$400 K \citep{nagy2015b}.
The already studied molecules detected in this line survey suggest a complex structure with disparate gas components traced by different species.

In this paper we give a general description of the line survey and aim to provide a complete picture of the physical and chemical structure of the Orion Bar based on our \textit{Herschel}/HIFI line survey.

\section{Observations and data reduction}
\label{sect:obs}

The CO$^+$ peak ($\alpha_\mathrm{J2000}=\rm{05^h35^m20.6^s}$, $\delta_\mathrm{J2000}=-05^\circ 25'14''$) of the Orion Bar \citep{stoerzer1995} was observed using the Heterodyne Instrument for the Far-Infrared (HIFI, \citealp{degraauw2010}) on board the \textit{Herschel} Space Observatory \citet{pilbratt2010} as part of the HEXOS\footnote{Herschel observations of EXtra-Ordinary Sources} guaranteed-time key program \citep{bergin2010}. The observations were carried out as spectral scans, and cover the full frequency coverage of HIFI (480--1250 GHz and 1410--1910 GHz).
The total integration times of the observations are 2.4 h (band 1a), 2.2 h (band 1b), 3.4 h (band 2a), 2.6 h (band 2b), 1.6 h (band 3a), 2.8 h (band 3b), 4.3 h (band 4a), 2.6 h (band 4b), 5.7 h (band 5a), 4.5 h (band 5b), 7.5 h (band 6a), 5.6 h (band 6b), 4.4 h (band 7a), and 5.9 h (band 7b).
The average beam sizes of the scans are in the range between $\sim$39$''$ (band 1) and 11$''$ (band 7) corresponding to 0.08 pc and 0.02 pc, respectively.
The Wide-Band Spectrometer (WBS) backend was used which covers the 4 GHz bandwidth in four 1140 MHz sub-bands at 1.1 MHz resolution, equivalent to 0.66 \kms at 500 GHz and 0.18 \kms at 1850 GHz.

\begin{figure*}[ht]
\centering
\includegraphics[width=\textwidth, trim=0.0cm 0.4cm 0.0cm 0.2cm,clip=true]
{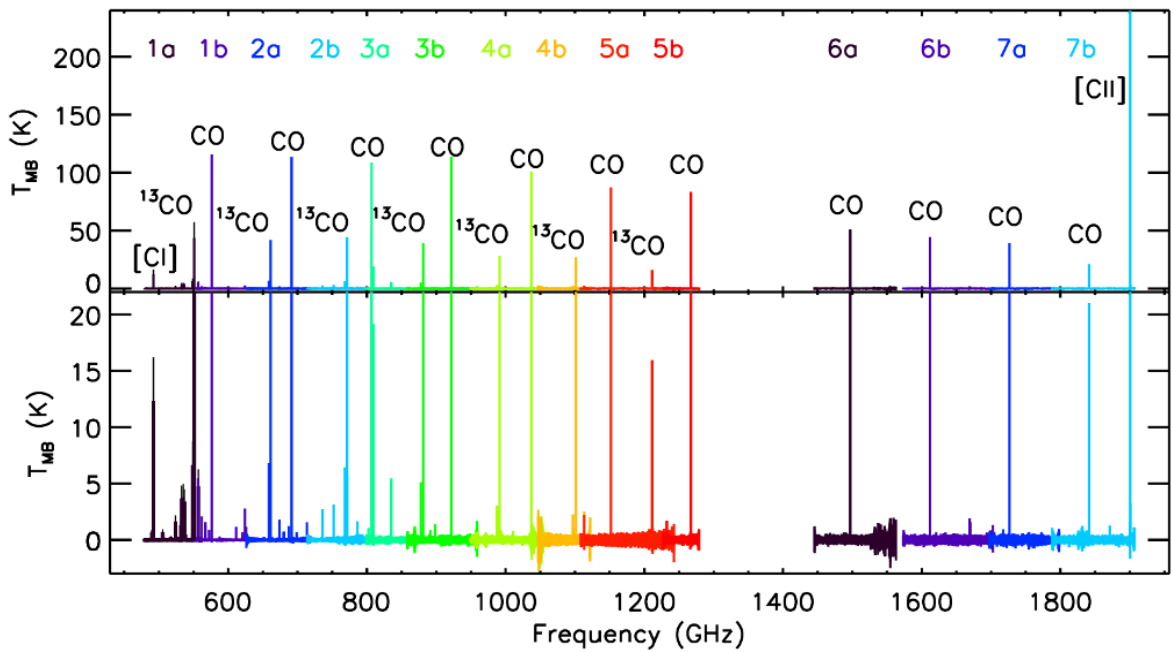}
\caption{Full baseline-subtracted spectrum toward the CO$^+$ peak of the Orion Bar. Bands 5, 6, and 7 were smoothed from the original velocity resolutions of $\sim$0.12 \kms, $\sim$0.09 \kms, and $\sim$0.08 \kms\ to the $\sim$1 \kms, $\sim$1.5 \kms, and $\sim$1.4 \kms\ channels, respectively.}
\label{full_linesurvey}
\end{figure*}

The data were reduced using the \textit{Herschel} interactive processing environment (HIPE, \citealp{ott2010}) pipeline versions 9.0 and 10.0 and are calibrated to $T_{\rm{A}}$ scale. 
First, the double sideband (DSB) scans were deconvolved using the \textit{doDeconvolution} task in HIPE with the strongest lines ($T_{\rm{A}}$ $>$ 10 K) removed, reducing the probability of ghost lines in the resultant single sideband (SSB) spectrum.
Second, we performed another deconvolution with the strongest lines present to recover the data at strong line frequencies. Finally, the strong lines were incorporated into the weak line SSB spectrum. 
Before the deconvolution the baseline subtraction was done using the fitBaseline task of HIPE. Spurs and other bad data such as spectra with very high rms noise levels (compared to the typical rms noise levels of the scans) were flagged using the flagTool in HIPE.
The main error of the deconvolution step is the uncertainty from the sideband gain divided by two, i.e. 5\% for most frequencies and about a factor of two higher at a few frequencies (around 630, 870, 1260 GHz) based on \citet{higgins2014}.
The velocity calibration of HIFI data is accurate to $\sim$0.5 km~s$^{-1}$ or better.
Based on fitting second-order polynomials, the continuum at the observed frequencies is negligible as it is similar to the measured rms noise levels. The rms noise levels at the frequency of the detected lines is shown in Table \ref{line_ident}. The average rms noise levels of the spectral scans are 0.02-0.04 K for band 1, 0.04-0.09 K for band 2, 0.09-0.15 K for band 3, 0.08-0.27 K for band 4, 0.30-0.42 K for band 5, 0.35-0.46 K for band 6, and 0.30-0.48 K for band 7.
High rms noise levels (compared to the average values) were measured in the three frequency ranges that were affected by problems in the HIFI gain stability, namely for LO frequencies around 863 GHz and 1046 GHz, and between 1540 and 1560 GHz. Apart from these frequencies, the baseline quality is good, as shown in Appendix \ref{sect:obs_data}.
The resulting spectrum toward the Orion Bar CO$^+$ peak is shown in Fig. \ref{full_linesurvey}.

Finally, we derived correction factors for the integrated intensities measured with different beam sizes, which is important when comparing the line intensities of such data, as is done in sections \ref{sect_rotdiagram} and \ref{sect_radex}. 
The largest and smallest average beam sizes for the HIFI bands ($\sim$39$''$ and 11$''$) corresponding to the line survey data are shown in Fig. \ref{orionbar_map} on a $^{13}$CO 3-2 map observed with the JCMT\footnote{The James Clerk Maxwell Telescope is operated by The Joint Astronomy Centre on behalf of the Science and Technology Facilities Council of the United Kingdom, the Netherlands Organisation for Scientific Research, and the National Research Council of Canada.} (JCMT archive\footnote{\href{http://www1.cadc-ccda.hia-iha.nrc-cnrc.gc.ca/jcmt}{http://www1.cadc-ccda.hia-iha.nrc-cnrc.gc.ca/jcmt}}).
To derive correction factors for the integrated intensities measured with different beam sizes corresponding to the different bands, we need a model or proxy for the spatial extent of the emission (a simple scaling by the beam size is only possible for point sources). To this end, we used JCMT images at a spatial resolution of $\sim$15$''$, as we did in \citet{nagy2013a}. In \citet{nagy2013a} we used an HCN $J$=4-3 map. In addition to HCN 4-3, we also used $^{13}$CO 3-2 (Fig. \ref{orionbar_map}) in this paper. 
The beam size of 15$''$ is slightly above but comparable to the smallest beam size of the scans corresponding to band 7 (11$''$).
For this method we assume that the spatial distribution of molecular species observed in this line survey is similar to those of HCN or $^{13}$CO. The two species represent limiting cases: HCN predominantly traces gas near the PDR surface, while $^{13}$CO represents the bulk of the emission from the molecular material. We use the $^{13}$CO integrated intensity distribution as a reference to derive correction factors, and use the HCN integrated intensity distribution to estimate error bars of the integrated intensities converted to the different beam sizes.
We convolved the $^{13}$CO 3-2 and HCN 4-3 images to spatial resolutions equivalent to the values corresponding to the HIFI bands, i.e. 39$''$ for band 1, 30$''$ for band 2, 25$''$ for band 3, 21$''$ for band 4, and 19$''$ for band 5. To trace the spatial distribution of the gas at the resolutions of band 6 (15$''$) and 7 (11$''$) we used the original (15$''$ beam) JCMT images. 
We compared the integrated intensities corresponding to the different spatial resolutions at a position close to the CO$^+$ peak to derive correction factors between the different beam sizes.
Based on the derived correction factors, we converted the observed integrated intensities to a $\sim$39$''$ beam size, equivalent to the beam size of band 1 \citep{roelfsema2012} when using LTE and non-LTE methods to estimate physical parameters in Sections \ref{sect_rotdiagram} and \ref{sect_radex}.
The derived correction factors based on HCN between the integrated intensities of the different bands and band 1 are 0.84 for band 2, 0.77 for band 3, 0.70 for band 4, 0.67 for band 5, and 0.59 for bands 6 and 7.
The derived correction factors based on $^{13}$CO between the integrated intensities of the different bands and band 1 are 0.94 for band 2, 0.91 for band 3, 0.88 for band 4, 0.87 for band 5, and 0.84 for bands 6 and 7.
In this paper we used the conversion factors based on $^{13}$CO, and used the values based on HCN to give an estimate of the errors of the integrated intensities resulting from applying conversion factors to correct for the changing beam size over the observed frequency range. 
The errors of the integrated intensities which have been corrected using the conversion factors between each band and band 1 derived from $^{13}$CO: 
15\% for band 1, 22\% for band 2, 26\% for band 3, 30\% for band 4, 31\% for band 5, and 36\% for bands 6 and 7.  The error for band 1 includes the calibration error and the error from the Gaussian fitting used to obtain the line parameters. The errors for the other bands also include a contribution from the scaling factors used to convert intensities to the same beam size.
Table \ref{line_ident}. contains the original intensity values, which have not been converted to the spatial resolution of band 1, unlike the values used in Sects. \ref{sect_rotdiagram} and \ref{sect_radex}.
As the HIFI beam efficiencies were re-calculated compared to those listed in \citet{roelfsema2012}, we used the updated beam efficiencies\footnote{The updated beam efficiencies are listed in the technical note ''Measured beam efficiencies on Mars (revision v1.1, 1 October 2014)''
(\href{http://herschel.esac.esa.int/twiki/bin/view/Public/HifiCalibrationWeb}{http://herschel.esac.esa.int/twiki/bin/view/Public/HifiCalibrationWeb})
}
in this paper. In earlier works based on this line survey the \citet{roelfsema2012} beam efficiency was used (\citealp{vandertak2012}, \citealp{nagy2013a}, \citealp{vandertak2013}, \citealp{choi2014}, \citealp{nagy2015b}).
The reduced data are available for download from the Herschel Science Archive at the user provided products.\footnote{
\href{http://www.cosmos.esa.int/web/herschel/user-provided-data-products}{http://www.cosmos.esa.int/web/herschel/user-provided-data-products}
}

\section{Detected species}

The line identification is based on the Cologne Database for Molecular Spectroscopy (CDMS, \citealp{muller2005})\footnote{\href{http://www.astro.uni-koeln.de/cdms/catalog}{http://www.astro.uni-koeln.de/cdms/catalog}} and the Jet Propulsion Laboratory (JPL, \citealp{pickett1998})\footnote{\href{http://spec.jpl.nasa.gov}{http://spec.jpl.nasa.gov}} molecular databases. 
We have identified about 120 lines corresponding to 29 molecules. There are two unidentified lines with a line detection threshold of about 3$\sigma$ RMS noise level.
Several molecules show multiple transitions detected with different beam sizes. To correct for the effect of the changing beam size on the integrated intensities obtained with different beams we apply the method explained above (Sect. \ref{sect:obs}). 
The effect of the changing beam size is most important for CO and its isotopologues, for which transitions were detected across a wide frequency range. For many species, the detected transitions fall in the frequency range of band 1, and for those cases the changing beam size does not need to be taken into account. These species include SO, H$_2$CO, SH$^+$, and three of the four detected CS transitions.
The parameters of the lines obtained by a Gaussian fit are shown in Table \ref{line_ident}.

We detected 11 transitions of CO (from $J$=5-4 up to $J$=16-15), 7 transitions of $^{13}$CO, 6 transitions of C$^{18}$O, and 5 transitions of C$^{17}$O. The CO line emission observed as a part of this line survey will be analysed in detail using PDR models by Joblin et al. (in prep.). 
Six transitions of H$_2$O have  been detected toward the Orion Bar CO$^+$ peak (Choi et al., in prep.), and two transitions of H$_2^{18}$O, which were used by \citet{choi2014} to measure the ortho-to-para ratio of water, have also been detected toward the Orion Bar.
Ten transitions of formaldehyde (H$_2$CO) have  been detected in the line survey, nine of them being ortho and one of them a para transition.
Several transitions of sulphur-bearing species have been detected including four transitions of H$_2$S and three transitions of SO.
Five doublets of C$_2$H have been detected in the line survey from $N$=6-5 to $N$=10-9, and have been found to trace warm and dense gas \citep{nagy2015b}, most of it at temperatures of $T_{\rm{kin}} \sim 100-150$ K and densities of $n$(H$_2$)$\sim$10$^5-10^6$ cm$^{-3}$, and the highest-$N$ transitions tracing densities of $n$(H$_2$)$\sim$5$\times$10$^6$ cm$^{-3}$ and temperatures of $T_{\rm{kin}} \sim$400 K.
Two molecules which are expected to originate in the outer PDR layers, HCO$^+$ and CH, have also been detected with six and five lines, respectively. 
The H$_2$Cl$^+$ line detected in this line survey was previously reported by \citet{melnick2012} and was analysed in more detail by \citet{neufeld2012}. Two other species of the chlorine chemistry have also been detected, namely the H$^{35}$Cl and the H$^{37}$Cl isotopologues of hydrogen chloride.
The rotational ground state line of ammonia, o-NH$_3$ (J, K) = (1, 0) $\rightarrow$ (0, 0) at 572.5 GHz detected with Odin \citep{larsson2003} is confirmed by our observations with a higher spatial resolution compared to the large Odin beam (about 2 arcminutes) . The observed NH$_3$ line parameters are consistent with the $V_{\rm{LSR}}$=10.5 \kms\ and full width at half maximum (FWHM) of 3.3 \kms\ observed by the Odin telescope, which are also consistent with the CO $J$=4--3 line parameters \citep{wilson2001}. 
Other $N$-bearing molecules detected in the HIFI line survey include NO, CN, HCN, and HNC.
The first detection of HF emission in a Galactic source has been reported by \citet{vandertak2012} based on data from the line survey summarized in this paper, who found that HF toward the Orion Bar is most likely excited by collisions with electrons.
The CH$^+$, SH$^+$, and CF$^+$ ions from this line survey toward the Orion Bar have been analysed in \citet{nagy2013a}. Emission of OH$^+$ and upper limits for H$_2$O$^+$ and H$_3$O$^+$ based on data from this line survey were presented in \citet{vandertak2013}. HIFI data of C$^+$ and $^{13}$C$^+$ toward the Orion Bar have been presented by \citet{ossenkopf2013}.
Both fine structure transitions of atomic carbon and four transitions of CS are also covered by this line survey.
The strongest hyperfine transition ($F$=11/2-9/2 at 589872.2 MHz) of the CO$^+$ $N$=5-4 line is tentatively detected in the line survey. This transition has also been detected with \textit{Herschel}/HIFI toward another position of the OMC-1 cloud, the Orion S molecular core \citep{nagy2013b}, which is located in the H{\sc{ii}} region cavity around the Trapezium cluster \citep{odell2009}.

Most previous line surveys of the Orion Bar were carried out at lower frequencies, such as the line survey presented in \citet{hogerheijde1995} using different mm / sub-mm instruments  between 90 and 492 GHz. This line survey was carried out toward five positions perpendicular to the Orion Bar, one of which near the ionization front close to the CO$^+$ peak ($\Delta$RA=-6.0$''$, $\Delta$Dec=6.8$''$).
The line survey of \citet{leurini2006} was carried out between 279 and 361.5 GHz using APEX toward the most massive clump identified using H$^{13}$CN by \citet{lisschilke2003} at offsets of $\Delta$RA=70.0$''$, $\Delta$Dec=40.0$''$ from the CO$^+$ peak.
The line survey of \citet{cuadrado2015} with the IRAM 30m was carried out in several frequency ranges between 80 and 359 GHz toward a position near the CO$^+$ peak ($\Delta$RA=3.0$''$, $\Delta$Dec=-3.0$''$), and mostly includes transitions of hydrocarbons. The JCMT line survey of \citet{vanderwiel2009} includes transitions in the range between 330 and 350 GHz and two more transitions around 230 GHz for a 2$\times$2 arcminute region. The CO$^+$ peak is at the edge of the region covered by the JCMT line survey, which is centred on $\Delta$RA=58.0$''$, $\Delta$Dec=14.0$''$ relative to the CO$^+$ peak. 
The line survey of \citet{habart2010} used the SPIRE instrument of \textit{Herschel} in a frequency range similar to our line survey, between 447 and 1545 GHz, but with a very low spectral resolution ($\geq$0.04 cm$^{-1}$ or 1.2 GHz), so that they identified only 43 lines. Their observed position is $\Delta$RA=33.3$''$, $\Delta$Dec=16.3$''$ relative to the CO$^+$ peak position.
The spectral line survey analysed in this paper represents the dataset with the largest frequency coverage at the highest frequencies and spectral resolution (1.1 MHz).
While the line survey presented in this paper contains the largest number of molecular lines detected toward the Orion Bar, a few molecules which were detected in earlier line surveys at lower frequencies,  such as CH$_3$OH, are not detected in
this line survey \citep{leurini2006}. CH$_3$OH was detected toward the nearby Orion S region, which was demonstrated to have a strong UV-irradiation and similar abundances to the Orion Bar (\citet{tahani2016}; see Sect \ref{sect:discussion}).

In this paper we present the data corresponding to the \textit{Herschel}/HIFI line survey with conclusions on the physical and chemical structure of the Orion Bar. We do not go into details about the molecules which have already been analysed in earlier works based on this line survey.
                     
\section{Results} 

\subsection{Line properties}   
\label{sect_line_properties}

\begin{figure*}[ht]
\centering
\includegraphics[width=18 cm, trim=0cm 0cm 0cm 4cm,clip=true]{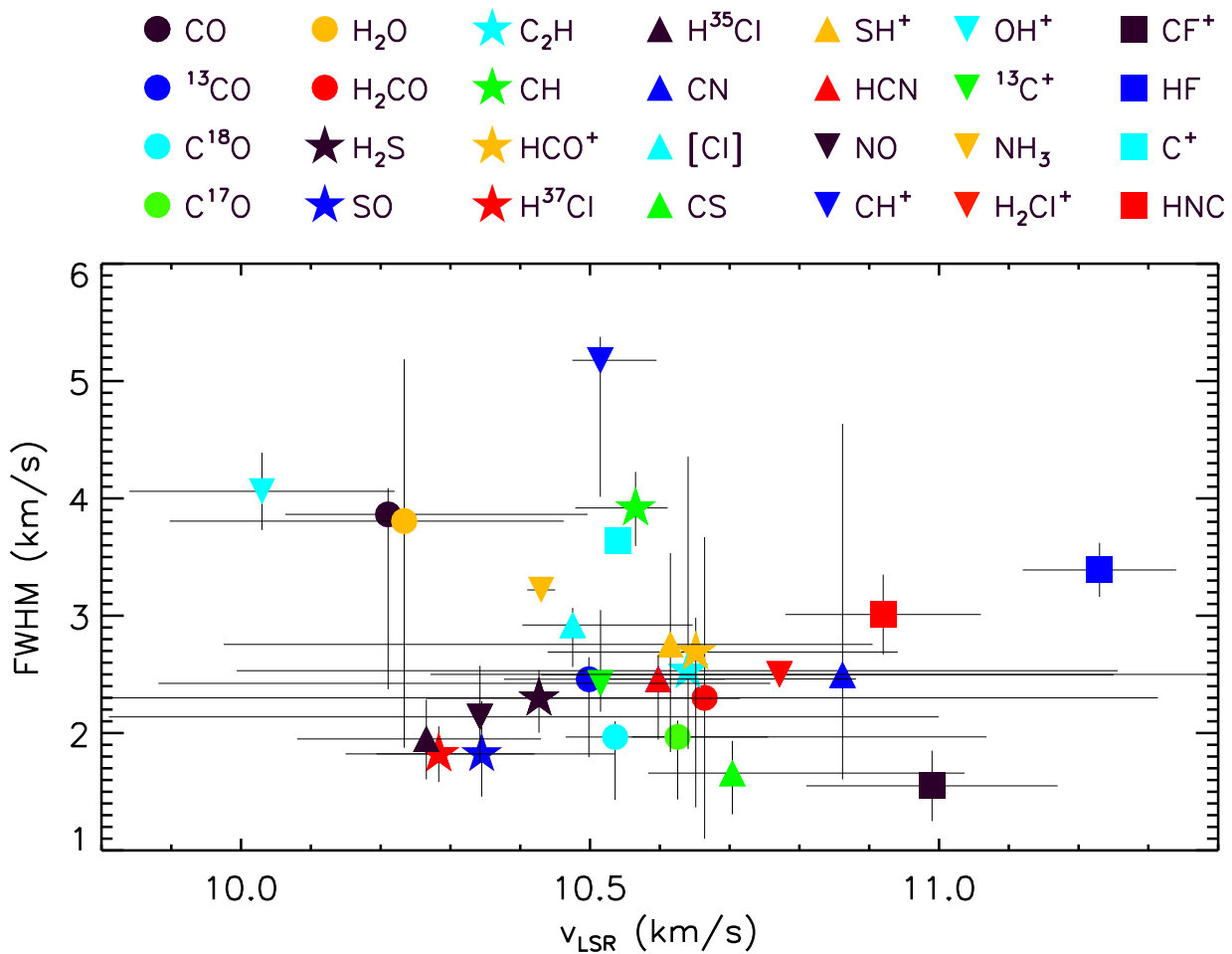}
\caption{Average line width versus average velocity (weighted by the signal-to-noise ratio
of the lines) of all of the detected species in the HEXOS Orion Bar line survey. The error bars represent the ranges of line widths and LSR velocities for the species with multiple detected transitions including an error which is based on the errors of the parameters for the individual transitions from the Gaussian fitting. 
For the species with only one detected transition, the error bars are from the Gaussian fitting of the line profiles.}
\label{vlsr_fwhm}
\end{figure*}

Molecules with similar kinematical properties may originate in the same gas component, i.e. species tracing diffuse or dense, cold or hot gas may have different kinematical properties (peak velocities and line widths).
Owing to the high spectral resolution, the fitted line properties (line width, $V_{\rm{LSR}}$) can be used to obtain information on the kinematics / structure of the source. 
Even though the spectral line survey presented here corresponds to one position, the kinematical properties of the observed species may provide information on the structure of the PDR.
In Fig. \ref{vlsr_fwhm} the fitted peak LSR velocities of the observed lines are plotted against their observed line width (FWHM). 
In the case of molecules with multiple transitions, the average $V_{\rm{LSR}}$ and FWHM were weighted by the signal-to-noise ratio of the lines.
The error bars of these species represent the range in the observed line parameters including an error which is based on the errors of the parameters for the individual transitions from the Gaussian fitting. The data points without (upper or lower) error bars are due to the weighted average of the observed width or velocity being closer to the maximum or the minimum value of the width or velocity.
For the species with only one detected transition the error bars are from the Gaussian fitting of the line profiles.
For OH$^+$ we only used its transition at 1033118.6 MHz as the other observed transition (at 971805.3 MHz) is an unresolved blend of multiple hyperfine components.
For most detected molecules there is a decreasing trend for the FWHM values with increasing velocities between 9.5 and 11 \kms, as can be seen in Fig. \ref{vlsr_fwhm}. 

Molecules which do not follow this trend include CH$^+$. This was expected considering that the line-broadening of this ion is strongly related to its chemistry rather than the physical properties of its emitting region, as discussed in \citet{nagy2013a}. 
Other species with FWHM values above the typical 2-3 \kms\ for the Orion Bar include OH$^+$, HF, and H$_2$O. The width of the lines is expected to be affected by a combination of thermal line broadening, opacity, and chemical pumping such as for CH$^+$.

In Appendix \ref{appendix:lineparam} we show that the spread of the velocity widths is partially due to a low signal-to-noise ratio for the weak lines and partially due to opacity broadening for CO. The variation in the line position is mainly produced by the change in the pickup of signal from the Orion Ridge for different beam widths.

In summary, the $V_{\rm{LSR}}$-line width relation is likely affected by the combination of the changing beam size, opacity broadening, and the species originating in different temperature and density components. Lines stemming from dense molecular material have a typical velocity of 10.6 \kms\ and a line width between 1.5 and 3 \kms.  The higher line velocities shown by  HF and CF$^+$  may stem  from some diffuse gas; to some degree this may also apply to CN. OH$^{+}$, CO (and $^{13}$CO), and the HCl isotopologues show significant velocity contributions from the Orion Ridge. Broader line profiles are also observed for optically thick lines.

\subsection{Physical parameters: Single excitation temperature}
\label{sect_rotdiagram}

In this section, we estimate column densities for the observed molecules in LTE approximation. 
As not every molecule detected in this line survey has collisional excitation rates for a non-LTE analysis, we use this as a common method to obtain a column density estimate for each observed molecule.   

For some of the species detected in this line survey, multiple transitions also provide an estimate on the excitation temperature  by applying the rotational diagram method. 
We have created rotational diagrams for species with at least three observed transitions, covering a sufficient range in upper level energy. In the case of SO the transitions cover an upper level energy range between $\sim$166 and 174~K, which is too low to result in an accurate estimate of the SO rotational temperature and column density.
This method assumes optically thin lines, uniform beam filling, and a single excitation temperature to describe the observed transitions. For most observed transitions the optically thin assumption is reasonable. The single excitation temperature is a simplification as we expect the gas covered by the \textit{Herschel} beam to originate in a range of temperatures. In some cases, a non-uniform beam filling may also affect the excitation of the observed molecules and the different contributions from the Orion Ridge.
In the rotational diagram method the measured integrated main-beam temperatures of lines ($\int T_{\mathrm{MB}} \mathrm{d}V$ K km s$^{-1}$) can be converted to the column densities of the molecules in the upper level ($N_\mathrm{u}$) using

\begin{equation}
        \label{rotdiagram}
                \frac{N_\mathrm{u}}{g_\mathrm{u}}=\frac{N_{\mathrm{tot}}}{Q(T_{\mathrm{rot}})} \exp\left({-\frac{E_\mathrm{u}}{k T_{\mathrm{rot}}}}\right)
\end{equation}
and
\begin{equation}
\label{coldens_lte}
N_{\rm{tot}}=\frac{8\pi k \nu^2}{h c^3} \frac{Q(T_{\rm{rot}})}{g_u A_{\rm{ul}}} e^{E_u/k T_{\rm{rot}}} \int{T_{\rm{mb}} d{\rm{v}}}
,\end{equation}
with $g_\mathrm{u}$ the statistical weight of level u, $N_{\mathrm{tot}}$ the total column density, $Q(T_{\mathrm{rot}})$ the partition function for $T_{\mathrm{rot}}$, $E_\mathrm{u}$ the upper level energy, $\nu$ the frequency, and $A_{\rm{ul}}$ the spontaneous decay rate.
A linear fit to $\ln(N_\mathrm{u}/g_\mathrm{u})$ - $E_\mathrm{u}$ gives $T_\mathrm{rot}$ as the inverse of the slope, and $N_\mathrm{tot}$  can be derived. The rotational temperature is expected to be equal to the kinetic temperature if all levels are thermalized.
We use the rotational diagram method for molecules which have more than three detected transitions. For the rest of the detected species, we derive column densities in the LTE approximation (see below) using the integrated intensity of their strongest detected transition.
Equation \ref{coldens_lte} represents the Rayleigh-Jeans limit without background, which is very accurate for frequencies below 1~THz, where most observed lines in this line survey are.
Figure \ref{rot_diagram} shows the rotational diagrams of molecules observed toward the Orion Bar CO$^+$ peak with at least three observed transitions. The derived excitation temperatures and column densities are shown in Table \ref{coldens}.

\begin{figure*}[ht]
\centering
\includegraphics[width=14.0cm,trim=0.0cm 0cm 0cm 5.5cm,clip=true]{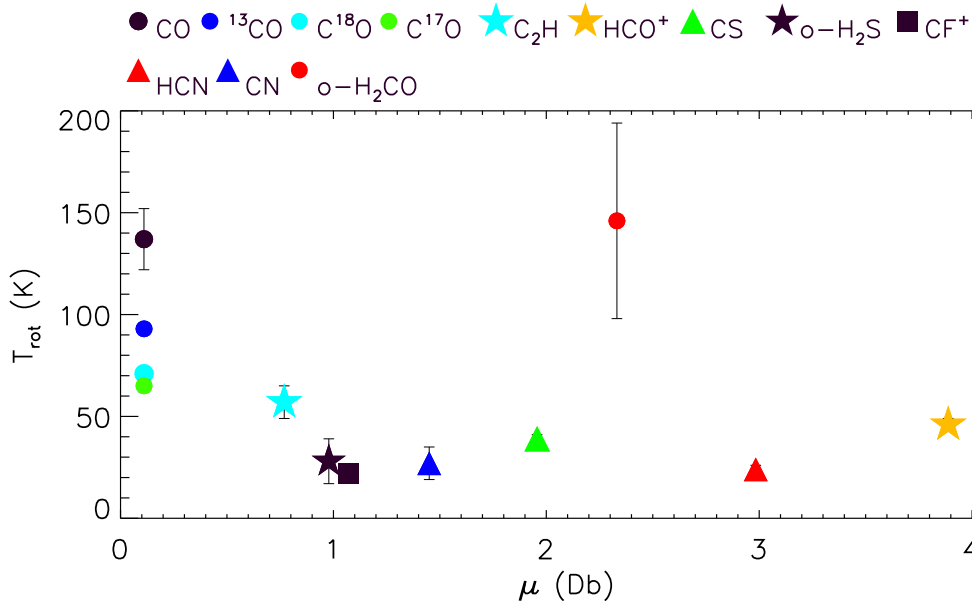}    
\caption{Derived rotational temperatures as a function of the dipole moment.}
\label{rot_temp_summary}
\end{figure*}

Figure \ref{rot_temp_summary} shows a summary of the derived excitation temperatures as a function of the dipole moment. The highest values are consistent with or slightly above the 85$\pm$30 K average kinetic temperature derived by \citet{hogerheijde1995}. The value found for CO is close to the $\sim$150~K kinetic temperature measured near the ionization front of the Orion Bar by \citet{batrlawilson2003} and \citet{goicoechea2011}.
The lower values most likely correspond to non-LTE conditions. In the non-LTE case, the radiative decay dominates the collisional excitation, resulting in excitation temperatures below the kinetic temperature. As seen in Fig. \ref{rot_temp_summary}, the rotational temperatures found for HCO$^+$, HCN, and CS are below the gas temperature in the regions where these species originate, as indicated by their high dipole moments.  CF$^+$, o-H$_2$S, and C$_2$H have lower dipole moments than  HCO$^+$, HCN, and CS, and are therefore  expected to be closer to thermalized.

For molecules with no information available on their excitation (less than three observed transitions), we derived a column density in the LTE approximation for a range of excitation temperatures using Eqn. \ref{coldens_lte}.
Table \ref{coldens} includes the column densities derived in LTE approximation, assuming excitation temperatures in the range between $\sim$20 and 150 K. 

\subsection{Physical parameters: Non-LTE approximation}
\label{sect_radex}

For some of the detected molecules, collisional excitation rates are available, and therefore, non-LTE calculations with the RADEX \citep{vandertak2007} code can be performed. We adopt a background radiation field which was used for the non-LTE analysis of OH$^+$ in \citet{vandertak2013} and which is based on the results of \citet{arab2012}, who describe the continuum toward the Orion Bar with a modified blackbody distribution with a dust temperature of 50 K and a dust emissivity index of $\beta$=1.6. 
For molecules with at least three detected transitions we aim to derive a best fit kinetic temperature and H$_2$ volume density. 
In the case of species with at least three transitions, we compared the observed line intensities to the values calculated with RADEX. In the case of species with only two transitions such as [C{\sc{i}}] we compared the observed line ratio to those predicted by the model. When fitting the line intensities we assumed uniform beam filling.
Most species observed toward the Orion Bar are expected to originate in both ``clumps'' with a lower filling factor and in the ``interclump'' medium with a filling factor close to one. Probing the contribution of the components with different filling factors is beyond the scope of this work, as many of the species modelled with RADEX have a low number (3-4) of transitions, which makes it difficult to constrain a beam filling factor.\footnote{For the species with the highest number of transitions (CO and its isotopologues) the changing beam size introduces an uncertainty from the varying Orion ridge pickup that would make a two-component fit questionable.}
We ran a 30$\times$30 grid of RADEX models with kinetic temperatures in the range between 50~K and 300~K and H$_2$ volume densities between 10$^4$ and 10$^6$ cm$^{-3}$ to cover conditions that are expected in the clump and interclump gas. In a few cases we extended the grid to densities up to 10$^7$ cm$^{-3}$.
Based on previous observations most temperatures are expected to be in the range between 50 K and 300 K. As the CO$^+$ peak is near the ionization front of the Orion Bar, some of the molecules originate in regions with temperatures of a few 100 K, such as those measured for OH (160-220 K, \citealp{goicoechea2011}). However, some of the detected species,  such as H$_2$O which may trace temperatures as low as 60 K, may trace larger depths into the PDR \citep{choi2014}. Most species detected toward the Orion Bar are expected to trace the average density of 10$^5$ cm$^{-3}$, but some of the species may trace densities of a few times 10$^4$ cm$^{-3}$ \citep{simon1997} or densities around 10$^6$-10$^7$ cm$^{-3}$ (e.g. \citealp{goicoechea2011}).
The logarithmic temperature and density steps in the grid are 0.026 dex and 0.067 dex, respectively.
We also adopt a column density based on the LTE calculations, the best fit value for the molecules with rotational diagram results, and the average value for the rest. We apply a $\chi^2$ minimization to the results of the grid. When the minimum $\chi^2$ is larger than a few $\times$ 1, we change the column density and repeat the method. The error bars of the best fit H$_2$ volume densities and kinetic temperatures given in Table \ref{coldens} are determined as the range of densities and temperatures within $\chi^2 \leq {\rm{min}}(\chi^2)+1$.
Aiming to derive best fit kinetic temperatures and H$_2$ volume densities traced by the observed molecules and transitions is a simplification as a range of physical conditions  is expected  to be sampled in the HIFI beam. Therefore, only beam-averaged densities and temperatures can be derived.
In Sect. \ref{appendix_radex} we discuss each molecule in Table \ref{coldens}. The applied inelastic collision rates are listed in Table \ref{coll_data}. All our tracers are expected to show their maximum abundance at visual extinctions well above $A_{\rm{V}}=0.1$ where most of the hydrogen is in molecular form, which means that collisions with atomic hydrogen and electrons can be neglected.

\begin{table}[ht]
\begin{minipage}[!h]{\linewidth}\centering
\caption{Inelastic collision data used for the RADEX modelling.}
\label{coll_data}
\renewcommand{\footnoterule}{}
\renewcommand{\thefootnote}{\alph{footnote}}
\resizebox{\textwidth}{!}{
\begin{tabular}{lllll}
\hline
\footnotetext[1]{Scaled using rates with He.}
\footnotetext[2]{Scaled from o--H$_2$O--p--H$_2$, assuming thermal o/p for H$_2$.}
Molecule& $E_{\rm{max}}$& $T$ (K)& Collision& Reference\\
        &    (cm$^{-1}$)&        & partner&            \\
\hline
CF$^+$&     188&  10--300&   H$_2$\footnotemark[1]&     \citet{ajilihammami2013}\\
CH&        1001&  10--300&   H$_2$\footnotemark[1]&     \citet{marinakis2015}\\
o--H$_2$CO& 207&  10--300&   H$_2$&                     \citet{wiesenfeldfaure2013}\\
o--NH$_3$&  420&  15--300&   p--H$_2$&                  \citet{danby1988}\\
NO&         375&   5--300&   H$_2$\footnotemark[1]&     \citet{lique2009}\\
SO&         676&  60--300&   H$_2$&                     \citet{lique2006a}\\
HCO$^+$&   1381&  10--400&   H$_2$&                     \citet{flower1999}\\
CS&         759&  10--300&   H$_2$\footnotemark[1]&     \citet{lique2006b}\\
C{\sc{i}} &   44&  10--1200&  p--H$_2$&                  \citet{schroeder1991}\\
o--H$_2$S& 1400&    5-1500&  p--H$_2$\footnotemark[2]&  \citet{dubernet2009}\\
CN&         793&    5--300&  H$_2$\footnotemark[1]&     \citet{lique2010}\\
HCN&        960&    5--500&  H$_2$\footnotemark[1]&     \citet{dumouchel2010}\\
C$^{17}$O& 3058&   2--3000&  p--H$_2$&                  \citet{yang2010}\\
C$^{18}$O& 2988&   2--3000&  p--H$_2$&                  \citet{yang2010}\\
$^{13}$CO& 2999&   2--3000&  p--H$_2$&                  \citet{yang2010}\\
CO&        3137&   2--3000&  p--H$_2$&                  \citet{yang2010}\\
\hline
\end{tabular}
}
\end{minipage}
\end{table}

Figure \ref{radex_fits} lists the results from RADEX described above, showing the comparison of the observed and modelled line intensities. Table \ref{coldens} lists the parameters used for the non-LTE analysis and the results. Previous results based on this line survey are also shown in Table \ref{coldens} for C$_2$H, HF, H$_2$O, CH$^+$, SH$^+$, and OH$^+$.

\begin{figure*}[ht]
\centering
\includegraphics[width=14cm, angle=0]{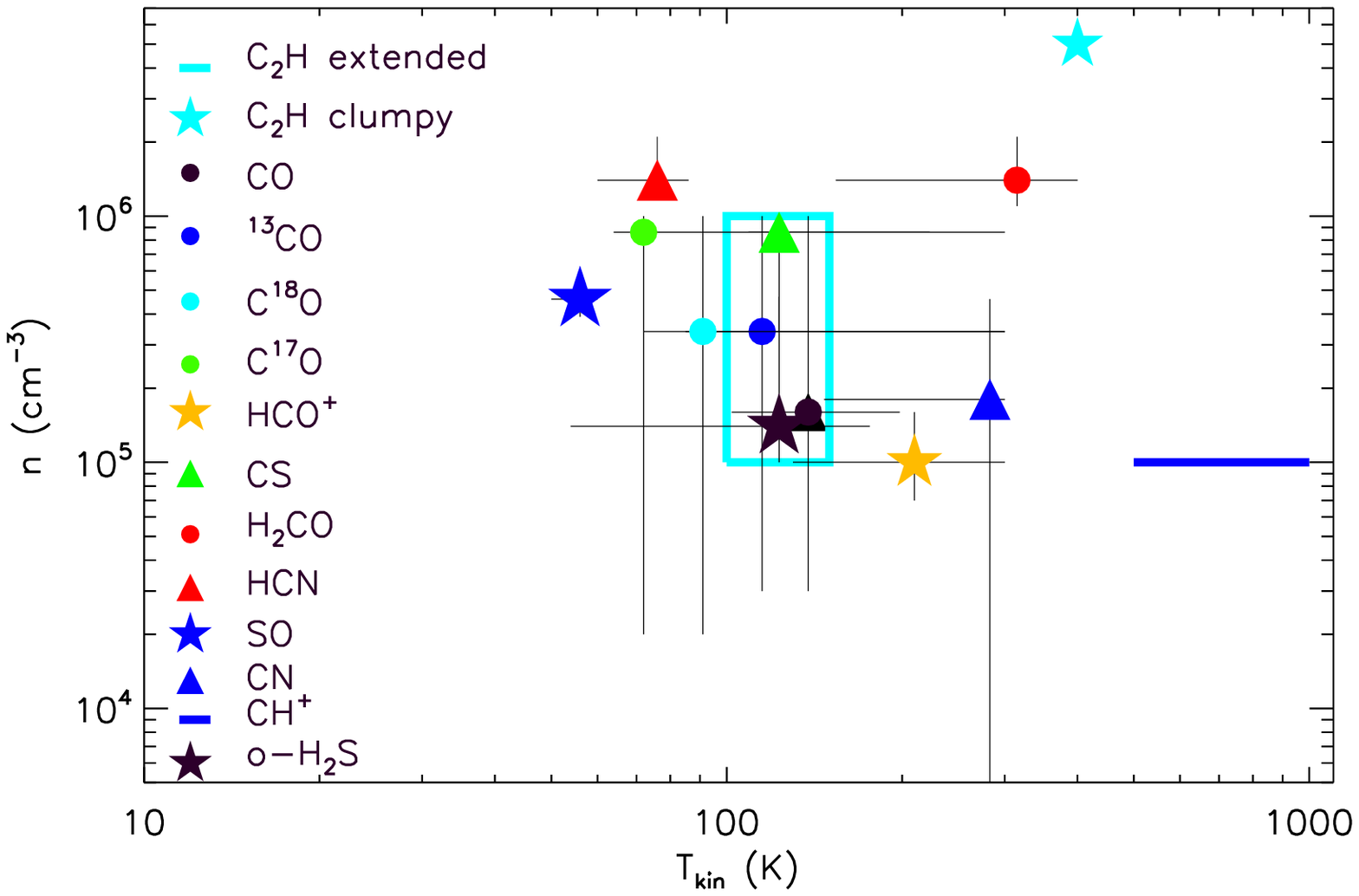}
\caption{Summary of the kinetic temperatures and H$_2$ volume densities derived from the RADEX models for the molecules with at least three detected transitions. The results for CH$^+$ from \citet{nagy2013a} and the results for C$_2$H from \citet{nagy2015b} are also included. The results for C$_2$H include two components, one with temperatures of 100-150~K and densities of 10$^5$-10$^6$ cm$^{-3}$ representing most C$_2$H column density, and a small fraction with a temperature of 400~K and density of 5$\times$10$^6$ cm$^{-3}$ required to fit the highest $N$ transitions of C$_2$H.
}
\label{temp_dens}
\end{figure*}

The summary of the RADEX results is shown in Fig. \ref{temp_dens}, including the earlier results based on this line survey for CH$^+$ \citep{nagy2013a} and C$_2$H \citep{nagy2015b}.
As seen in Fig. \ref{temp_dens}, most species trace temperatures in the range between 100-150~K and densities in the range between 10$^5$ and 10$^6$ cm$^{-3}$. The best fit kinetic temperature close to 150~K for CO (138~K) is similar to those found by \citet{goicoechea2011} and \citet{batrlawilson2003} near the surface of the Orion Bar. The observed CO$^+$ peak position is indeed close to the PDR surface; therefore, a kinetic temperature above the average value of 85$\pm$30~K \citep{hogerheijde1995} was expected.

\begin{figure*}[ht]
\centering
\includegraphics[width=6.5cm, angle=-90, trim=0cm 0cm 0cm 0cm,clip=true]{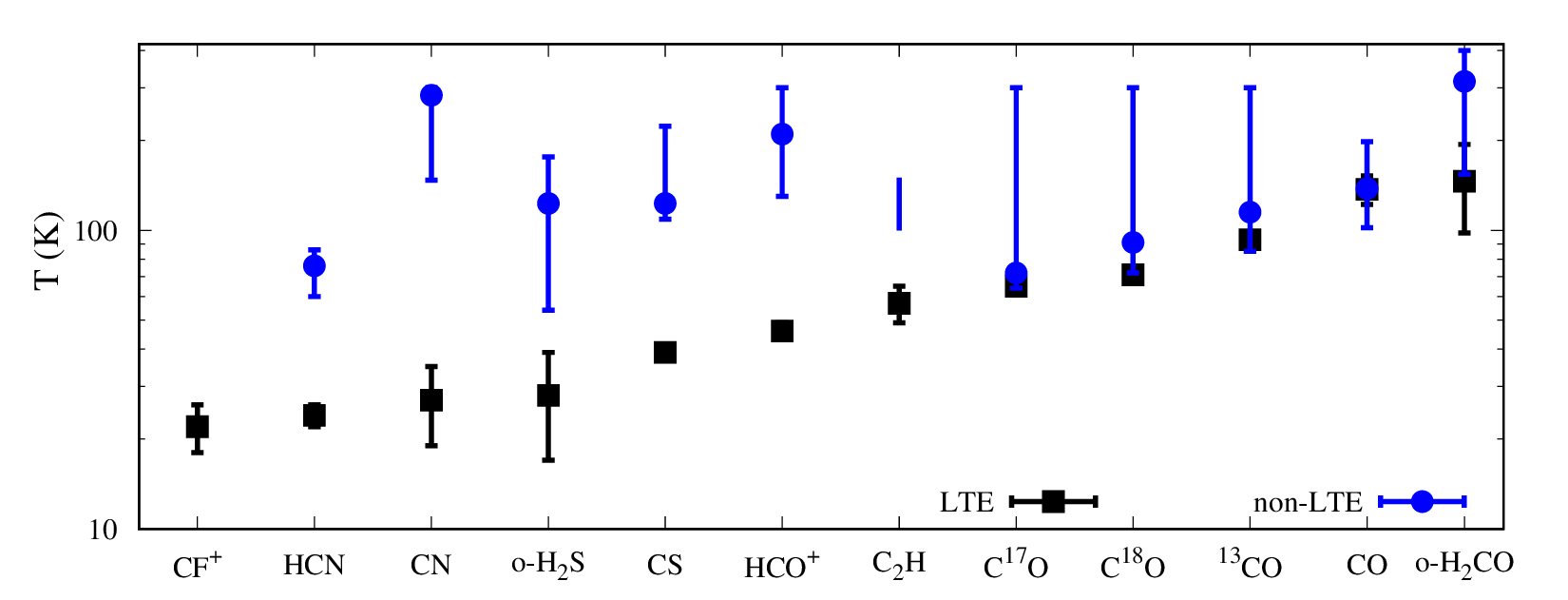}    
\caption{Summary of the rotational temperatures derived in Sect. \ref{sect_rotdiagram} and the kinetic temperatures derived in \ref{sect_radex}. For C$_2$H we show the range of kinetic temperatures which corresponds to the highest  C$_2$H column density for the observed transitions \citep{nagy2015b}.
}
\label{rot_temp_kin_temp}
\end{figure*}

A comparison between the rotational temperatures from the LTE and the kinetic temperatures from the non-LTE analysis is shown in Fig. \ref{rot_temp_kin_temp}.
The warm surface of the PDR is mainly traced in the optically thick CO; the inner shielded regions produce most of the observed C$^{17}$O and C$^{18}$O emission. As expected for partially subthermal excitation, the kinetic temperatures are always higher than the excitation temperatures. For species with a small dipole moment, the gas densities are sufficient for a good coupling so that we find small differences between kinetic and rotation temperature. For species with higher dipole moment, the two temperatures deviate much more. HCO$^+$ with the highest dipole moment shows the greatest difference; CS with its high dipole moment is a similar case.

We found a range of temperatures and densities, indicating a morphology of the PDR that is more complex than the clump/interclump model.

\begin{table*}[ht]
\begin{minipage}[!h]{\linewidth}\centering
\caption{Results derived from the single excitation temperature and non-LTE assumptions, when applicable. The gas pressures are based on the parameters used for or derived from the non-LTE estimates. A single value is shown for the excitation temperature when inferred from the rotational diagram, and the assumed range of parameters is shown when the number of transitions and/or the energy coverage was not high enough to perform a rotational diagram fit.}
\label{coldens}
\renewcommand{\footnoterule}{}
\renewcommand{\thefootnote}{\alph{footnote}}
\renewcommand{\arraystretch}{1.5}
\resizebox{\textwidth}{!}{%
\begin{tabular}{p{1.2cm}lrrllrrl}
\hline
\footnotetext[1]{The non-LTE column density of CO is based on the C$^{17}$O column density and isotopic ratios of C$^{18}$O/C$^{17}$O of 3.2 and CO/C$^{18}$O of 560 \citep{wilsonrood1994}.}
\footnotetext[2]{The difference between the LTE values quoted here and in \citet{nagy2015b} is due to the combination of the updated beam efficiencies and conversion factors between the different beam sizes applied in this paper.}
\footnotetext[3]{Based on the 3-2, 2-1, and 1-0 transitions from \citet{neufeld2006} and the 5-4 transition detected as a part of this HIFI line survey. The difference in the rotation temperature compared to that quoted in \citet{nagy2013a} is due to the re-reduction of the CF$^+$ 5-4 data with a more recent HIPE version.}
\footnotetext[4]{The LTE estimate for CH$^+$ is based on the $J$=1-0 transition, while the non-LTE estimate \citep{nagy2013a} is based on six transitions including four transitions observed with PACS.}
\footnotetext[5]{The LTE estimate for OH$^+$ is from \citet{vandertak2013}.}
Molecule& Number of& & LTE& \multicolumn{3}{c}{non-LTE}& Gas pressure& Reference for the\\
& transitions& $T_{\rm{ex}}$ (K)& $N$ (cm$^{-2}$)& $n$(H$_2$) (cm$^{-3}$)& $T_{\rm{kin}}$ (K)& $N$ (cm$^{-2}$)& (K cm$^{-3}$)& non-LTE estimate\\
\hline

CO\footnotemark[1]&        11& $137^{+16}_{-14}$&  $4.5^{+2.5}_{-1.1}\times10^{17}$& 
$1.6^{+8.4}_{-1.3}\times10^5$& 138$^{+60}_{-36}$& 3.2$\times$10$^{18}$& 2.2$\times$10$^7$& This paper\\

$^{13}$CO&  7& 93$\pm$3& (4.7$\pm$0.2)$\times$10$^{16}$&
$3.4^{+6.6}_{-3.1}\times10^5$& 115$^{+185}_{-30}$& 6.0$\times$10$^{16}$& 2.9$\times$10$^7$& This paper\\

C$^{18}$O&  6&  71$\pm$2& (6.2$\pm$0.2)$\times$10$^{15}$& 
$3.4^{+6.6}_{-3.2}\times10^5$& 91$^{+209}_{-19}$& 6.0$\times$10$^{15}$& 3.1$\times$10$^7$& This paper\\

C$^{17}$O&  5&  65$\pm$3& (1.8$\pm$0.1)$\times10^{15}$& 
$8.6^{+1.4}_{-8.4}\times10^5$& $72^{+228}_{-8}$& 1.8$\times$10$^{15}$& 6.2$\times$10$^7$& This paper\\

o-H$_2$CO&  9&  146$\pm$48& $(1.5\pm0.5)\times10^{12}$&     
$1.4^{+0.7}_{-0.3}\times10^6$& $315^{+85}_{-161}$& $4.0\times10^{12}$& 4.4$\times$10$^8$& This paper\\

HCO$^+$&    6&  46$\pm$3&  $(5.4\pm0.3)\times10^{12}$&           
$(1.0^{+0.6}_{-0.3})\times10^5$& $210^{+90}_{-80}$& $1.0\times10^{14}$& 2.1$\times$10$^7$& This paper\\

CS&         4&  39$\pm$2&  $(1.0\pm0.1)\times10^{13}$&     
$8.6^{+1.4}_{-5.2}\times10^5$& $123^{+100}_{-14}$& 2.0$\times$10$^{13}$& 1.1$\times$10$^8$& This paper\\ 

o-H$_2$S&   3&  27$\pm$10&  $(1.5\pm0.6)\times10^{14}$&     
$1.4^{+3.3}_{-0.4}\times10^5$& $123^{+53}_{-69}$& $5.0\times10^{14}$& 1.7$\times$10$^7$& This paper\\

HCN&        3&  24$\pm$2& (1.1$\pm$0.1)$\times$10$^{13}$&     
$1.4^{+0.7}_{-0.2}\times10^6$& $76^{+10}_{-16}$& $6\times10^{13}$& 8.5$\times$10$^7$& This paper\\

C$_2$H&     5&  57$\pm$8& $(8.2\pm1.2)\times10^{13}$\footnotemark[2]&    
$10^5-10^6$& 100-150& $8\times10^{13}$& (1-15)$\times$10$^7$& \citet{nagy2015b}\\ 

      &      &          &                          &
$5\times10^6$& 400& $2\times10^{13}$& 2$\times$10$^9$&  \\    

CF$^+$&     1&  22$\pm$4& $(1.8\pm0.3)\times10^{12}$& 
$3.4^{+0.6}_{-2.2}\times10^5$& $50^{+250}_{-0}$& $1.8\times10^{12}$& 1.7$\times$10$^7$& This paper\footnotemark[3]\\  

o-H$_2$O&   3&  53$\pm$18& $(2.6\pm0.9)\times10^{12}$&
10$^6$&  60& 1.7$\times$10$^{12}$& 6.0$\times$10$^7$& \citet{choi2014}\\

p-H$_2$O&   3&  139$\pm$92& $(2.0\pm1.3)\times10^{13}$&  
10$^6$&  60& 1.8$\times$10$^{13}$& 6.0$\times$10$^7$& \citet{choi2014}\\

CN&         7& 27$\pm$8& ($4.7\pm1.3)\times10^{13}$&     
$1.8^{+2.8}_{-2.6}\times10^5$& $283^{+17}_{-136}$& $2.0\times10^{14}$& 5.2$\times$10$^7$& This paper\\

HNC&        1& 18.75--150& (0.2-1.6)$\times$10$^{12}$&
10$^5$& 100& $5\times10^{12}$& 10$^7$& This paper\\

SO&         3& 18.75--150& (0.01-1.0)$\times$10$^{15}$& 
$4.6^{+0.8}_{-0.7}\times10^5$& $56^{+4}_{-6}$& $5\times10^{14}$& 2.5$\times$10$^7$& This paper\\

NO&         4& 18.75--150& (9.4-50.6)$\times$10$^{14}$&     
$2.9^{+7.1}_{-0.8}\times10^5$& $300^{+0}_{-154}$& $2\times10^{14}$& 8.8$\times$10$^7$& This paper\\

NH$_3$&     1& 18.75--150& (0.5-2.7)$\times$10$^{13}$&     
$2\times10^5$& 145& $1.1\times10^{14}$& 2.9$\times$10$^7$& This paper\\

[C{\sc{i}}]&  2& 18.75--150& (5.1-27.7)$\times$10$^{17}$&     
$8.6_{-7.6}^{+91.0}\times10^4$& $86_{-33}^{+214}$& $1\times10^{18}$& 7.3$\times$10$^6$& This paper\\ 

CH&         5& 18.75--150& (1.4-2.1)$\times$10$^{14}$& 
$8.6^{+90.1}_{-7.6}\times10^4$& $50^{+250}_{-0}$& $10^{14}$& $4.3\times10^6$& This paper\\

H$^{35}$Cl& 3&  18.75--150& (1.1-1.3)$\times$10$^{13}$&
&            &            & & This paper\\

H$^{37}$Cl& 3&  18.75--150& (5.3-6.5)$\times$10$^{12}$&
&            &            & & This paper\\

C$^+$&      1& 18.75--150& (5.4-185.2)$\times$10$^{18}$& 
     &       & 1.1$\times$10$^{19}$& & \citet{ossenkopf2013}\\          

SH$^+$&     3& 18.75--150&  (2.8-5.5)$\times$10$^{12}$& 
10$^6$& 200& $10^{13}$& 2.0$\times$10$^8$& \citet{nagy2013a}\\

CH$^+$&     2& 18.75--150&  (2.1-2.3)$\times$10$^{13}$&
10$^5$& 500-1000& $9\times10^{14}$& (5-10)$\times$10$^7$& \citet{nagy2013a}\footnotemark[4]\\

HF&         1&  18.75--150& (1.7-5.5)$\times$10$^{12}$& 
&           100& 10$^{15}$& & \citet{vandertak2012}\\

OH$^+$&     2&  10--160& (5.5-82.1)$\times$10$^{12}$& 
& & $10^{14}$& & \citet{vandertak2013}\footnotemark[5]\\

\hline                         
\end{tabular}
}
\end{minipage}
\end{table*}

\section{Discussion}
\label{sect:discussion}

In the previous sections we summarized observations of the \textit{Herschel}/HIFI spectral line survey carried out toward the CO$^+$ peak of the Orion Bar in the frequency range between 480--1250 GHz and 1410--1910 GHz. 
About 120 lines corresponding to 29 molecules were detected, including isotopologues.
We derived excitation temperatures and column densities for molecules with multiple detected rotational transitions. The derived excitation temperatures are in the range between $\sim$22 K and $\sim$146 K.
For several molecules we also derived physical parameters using a non-LTE analysis, making use of the available collision rates. For species with at least three detected transitions, we derived the best fit kinetic temperatures and H$_2$ volume densities of their emitting region. 
As mentioned above, the  modelling using the RADEX code is a simplification as it assumes a single temperature and density, while the observed line intensities are expected to originate in gas with a range of temperatures and densities.

Even though the observations presented in this paper correspond to a single position, the temperatures and densities given by the RADEX fits for various molecules -- especially those with the largest number of transitions -- give an indication of what kind of gas component these molecules originate in. In Sect. \ref{discussion:models} we discuss a few groups of species based on the best fit kinetic temperatures and H$_2$ volume densities derived from RADEX. In Sect. \ref{discussion:kinematics} we discuss what the kinematical properties of the molecules suggest about their origin in the PDR.

\subsection{Temperature and density differentiation of the Orion Bar}
\label{discussion:models}

\begin{itemize}

\item \textit{Tracers of hot gas at average Orion Bar density} \\
Very hot (500-1000 K) CH$^+$ gas is found to trace the average Orion Bar density of 10$^5$ cm$^{-3}$ \citep{nagy2013a}. Two more species detected in this line survey trace similar densities based on the RADEX models: HCO$^+$ with a best fit kinetic temperature of 210 K and CN with a best fit kinetic temperature of 283 K.

\item \textit{Tracers of hot and dense gas} \\
Two of the observed molecules correspond  (at least partly)  to very high-density hot gas.
As investigated in \citet{nagy2015b}, a small fraction of C$_2$H traces unresolved hot (400 K) and dense (5$\times$10$^6$ cm$^{-3}$) structures, and therefore provides additional evidence for clumpiness of the Orion Bar.
Similar to the hot component of C$_2$H, the o-H$_2$CO transitions observed in this line survey are most consistent with hot (315 K) and dense ($1.4\times10^6$ cm$^{-3}$) gas.
Though the H$_2$CO transitions observed by \citet{leurini2006} toward a different position are also consistent with a high kinetic temperature and density (150~K and 5$\times$10$^5$ cm$^{-3}$), these values are below the best fit parameters which the transitions observed in this line survey suggest.
Some of the transitions covered in this line survey likely trace gas close to the ionization front, which includes the hot and dense structures revealed by C$_2$H \citep{nagy2015b} and OH \citep{goicoechea2011} emission.

\item \textit{Species tracing warm and dense gas} \\
Most molecules covered in this line survey, such as CO, $^{13}$CO, C$^{18}$O, o-H$_2$S, CS, and the bulk of C$_2$H, trace temperatures of 100-150~K and densities of 10$^5$-10$^6$ cm$^{-3}$. These temperatures are below what can be expected from species tracing the PDR surface, but are higher than what can be expected deep in the cloud.
These species represent most of the gas detected toward the CO$^+$ peak and are very likely associated with the pattern of high-density substructures recently revealed by ALMA in the HCO$^+$ $J$=4-3 line \citep{goicoechea2016}.

\item \textit{Species tracing lower temperature ($<$100~K) gas} \\
For C$^{17}$O, SO, and HCN the best fit kinetic temperatures are below the 100-150~K range which most species observed in this line survey trace. 
The lowest kinetic temperature among the species modelled in this paper (56~K) was found for SO.
This may be interpreted as tracing gas which is not as directly exposed to UV irradiation as the species tracing higher temperature gas.
\end{itemize}
Table \ref{species_classification} includes a summary of the species discussed above with their possible origin toward the Orion Bar based on the transitions observed in this line survey.

\subsection{Structure of the Orion Bar as suggested by the observed kinematical properties}
\label{discussion:kinematics}

Apart from the kinetic temperatures estimated with RADEX, the line properties of the species discussed in Sect. \ref{sect_line_properties} can also be used to trace the origin of the species detected in this line survey. As mentioned in Sect. \ref{sect_line_properties}, for some species we expect a contribution of emission from the Orion Ridge region. The LSR velocity of the Orion Ridge is around 9 \kms \citep{vandertak2013}; therefore, the species detected at the lowest velocities are expected to have a contribution from gas toward the Orion Ridge. These species are OH$^+$ (as also seen in maps analysed in \citealt{vandertak2013}), HCl, part of the emission in the lowest $J$ CO, and $^{13}$CO transitions covered by this line survey, and possibly part of the SO line emission.
Apart from the groups related to the dense PDR and the Orion Ridge material, some of the species are likely to trace lower density / diffuse gas, such as CF$^+$, HF, and CN. Based on their physical properties and their expected origin close to the surface of the PDR, part of CH$^+$ and HCO$^+$ may also trace low density gas.

In addition to the velocity of the species, their line width may also give an indication to some detected molecules tracing the same gas component.
The molecules tracing hot gas at average density include CH$^+$; however, as discussed in \citet{nagy2013a}, its width is related to its chemistry rather than the conditions of its emitting region. The other two species tracing hot gas at average density are CN and HCO$^+$ with average FWHM widths (weighted by the signal-to-noise ratio of the lines) of 2.5 \kms and 2.7 \kms, respectively. The other two species tracing hot gas, probably the surfaces of high-density substructures \citep{goicoechea2016}, are C$_2$H (partly) and H$_2$CO with average line widths of 2.5 \kms and 2.3 \kms, respectively.
Most species detected toward the CO$^+$ peak correspond to warm (100-150~K) and dense (10$^5$-10$^6$ cm$^{-3}$) gas, such as CO, $^{13}$CO, C$^{18}$O, CS, and C$_2$H. The large CO average line width of 3.9 \kms is mostly due to the optical depth of the CO transitions. The difference in the line width of the $^{13}$CO and C$^{18}$O lines (2.5 \kms\ and 2.0 \kms, respectively) may be due to the difference in the kinetic temperature of their emitting region, and therefore means that C$^{18}$O is less exposed to UV irradiation than is $^{13}$CO. The $^{13}$CO lines are also more optically thick than the C$^{18}$O lines.
Despite the best fit kinetic temperature of 123~K, the CS lines have a very low line width of 1.7 \kms. 
The species with the lowest kinetic temperatures include C$^{17}$O, with a 2.0 \kms line width as measured for C$^{18}$O. As expected, the SO lines are narrow (FWHM$\sim$1.8 \kms); however, the HCN lines are broader with an average line width of 2.5 \kms. 
Based on the above, there is no correlation between the line width of the species and the best fit kinetic temperatures found using RADEX (Fig. \ref{fwhm_kintemp_species}). 
This may be related to the fact that physical conditions in the different regions of the PDR change rapidly, and the species trace temperature gradients and different density components around the best fit values found with RADEX.

Based on the results of the RADEX models we conclude that the species detected in this line survey allow us to trace multiple temperature and density components of a PDR with a complex morphology.
The large HIFI beam (as shown in Fig \ref{orionbar_map}) captures gas from several components of the PDR, and the transitions covered in this line survey correspond to temperature and density gradients including the values found as best fits in the RADEX models. The presence of dense substructures is also required to explain the existence of some of the observed line intensities; the high-density and hot component of C$_2$H and the high-density and high-temperature gas traced by the H$_2$CO transitions covered in this line survey may trace the same dense, unresolved structures close to the ionization front of the PDR. Such small, dense structures were detected in HCO$^+$ 4-3 at a high angular resolution with ALMA \citep{goicoechea2016}.

\begin{table*}[ht]
\begin{minipage}[!h]{\linewidth}\centering
\caption{Origin of the molecules discussed in Sect. \ref{sect:discussion} based on their transitions detected in this line survey.
The temperatures and densities are based on the RADEX models in Sect \ref{sect_radex}.}
\label{species_classification}
\renewcommand{\footnoterule}{}
\renewcommand{\thefootnote}{\alph{footnote}}
\begin{tabular}{lllll|ll}
\hline
& Hot gas at& Hot and dense&      Warm and dense& Lower (<100 K)&              Diffuse& Orion Ridge\\
& average density&  gas& gas&             temperatures&  gas&     material\\
\hline
         &             CH$^+$&      C$_2$H&            C$_2$H&                    SO& 
                       CF$^+$, HF, CN&            OH$^+$, HCl\\
Molecules&             HCO$^+$&     H$_2$CO&           CS&                        HCN&
                       HCO$^+$, CH$^+$& CO, $^{13}$CO\\
         &             CN&          &                  CO, $^{13}$CO, C$^{18}$O&  C$^{17}$O&
                       & SO\\
$T_{\rm{kin}}$ (K)&    $>$200&      $\gtrsim$300&      100-150&                   50-80\\
$n$ (cm$^{-3}$)&       $\sim$10$^5$&  $\gtrsim 10^6$&  $10^5-10^6$&   $5\times10^5-10^6$\\
\hline
\end{tabular}
\end{minipage}
\end{table*}

\subsection{Comparison with Orion South}
\label{sect:orionbar_orions}

\begin{figure}[h!]
\centering
\includegraphics[width=\textwidth, angle=0]{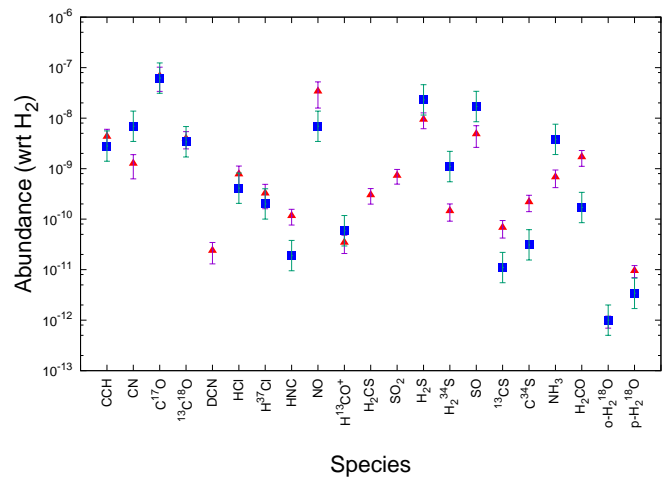}
\caption{Comparison of molecular abundances in the Orion Bar (blue squares) and those in Orion South (red triangles).}
\label{bar_south}
\end{figure}

Figure~\ref{bar_south} shows the abundances of molecular species in the Orion Bar, derived by dividing the column densities in Table \ref{coldens} by the reference $N$(H$_2$) value of 2.9$\times$10$^{22}$\,cm$^{-2}$ as derived from our C$^{17}$O data (Appendix~\ref{appendix_rotdiagram}), which is basically the same as the value from \citet{cuadrado2015}. 
Although the column densities in Table~\ref{coldens} should be accurate to $\sim$20\%, we adopt an uncertainty of a factor of 2 for the abundances to allow for the uncertainty in $N$(H$_2$).
No values are plotted for DCN, H$_2$CS, and SO$_2$, which have been detected in APEX line surveys of the Orion Bar (\citealp{leurini2006}; \citealp{parise2009}), but not in our HIFI survey.
Since the APEX line surveys were carried out at another position in the Bar and at higher angular resolution, including their results in Figure~\ref{bar_south} would add an extra uncertainty.

Overplotted in Figure~\ref{bar_south} are the molecular abundances derived from the HIFI line survey of Orion South \citep{tahani2016}. 
This object is a massive protostellar envelope under strong external irradiation by the Trapezium stars, also affected by internal irradiation of outflow cavity walls.
The Orion Bar and Orion South are compared to determine the role of the different excitation mechanisms since both objects are at the same distance and since both surveys were carried out with the same instrument.
Further comparison of abundances in Orion South with those for the various components of the Orion-KL \citep{crockett2014} massive star-forming region is presented in \citet{tahani2016}.

Many species are seen to have similar abundances in the two sources (Fig. \ref{bar_south}), including classic PDR tracers such as C$_2$H (\citealp{nagy2015b}; \citealp{cuadrado2015}). 
This chemical uniformity of Orion, observed previously on larger scales \citep{ungerechts1997}, is somewhat surprising as PDR chemistry is expected to favour small radicals and ions as a result of UV irradiation, whereas in protostellar envelopes, grain surface chemistry is expected to lead to higher abundances of more highly saturated species.
The higher abundances of H$_2$CO and isotopic CS in Orion South compared with the Bar seem to follow this expectation, but the higher NH$_3$ abundance is in disagreement with this picture.
We conclude that the chemistry of Orion South is actually more PDR-like and not like the chemistry of other protostellar envelopes such as AFGL 2591 \citep{kazmierczakbarthel2015}, as also suggested by the lack of `hot core' signatures in Orion South \citep{tahani2016}.

\section{Summary}

We have carried out a line survey with \textit{Herschel}/HIFI toward the CO$^+$ peak of the Orion Bar in the frequency range between 480-1250 GHz and 1410-1910 GHz. About 120 lines corresponding to 29 molecules were detected, including isotopologues.

The average FWHM line width for each detected species as a function of their average LSR velocity shows a decreasing trend for the line widths with velocity. This trend may be due to a combination of the changing beam size, the effect of opacity, and different species tracing different temperature and density components toward the observed position.

For species with at least three transitions we have created rotational diagrams to get a column density and an excitation temperature estimate. The excitation temperatures from the rotational diagrams are in the range between $\sim$22 and $\sim$146~K and the column densities are in the range between 1.8$\times$10$^{12}$ cm$^{-2}$ and 4.5$\times$10$^{17}$ cm$^{-2}$.

For species with at least three transitions and available collisional excitation rates we also performed a non-LTE analysis with RADEX and derived best fit kinetic temperatures and H$_2$ volume densities. Most detected species have temperatures in the range between 100 and 150~K and densities in the range between 10$^5$ and 10$^6$ cm$^{-3}$. The exceptions include SO, which has the lowest kinetic temperature among the species detected in the line survey, 56~K, probably tracing gas deeper into the PDR. Another exception is H$_2$CO, which -- based on the transitions observed in this line survey -- involves a high-density ($\sim$1.4$\times$10$^6$ cm$^{-3}$) component but also a higher temperature (315 K) than the other studied species.

The physical conditions in the Orion Bar encompass a range of temperatures and densities, indicating that the morphology of the PDR is more complex than the clump/interclump model, which is consistent with the complex morphology revealed by the recent ALMA observations \citep{goicoechea2016}.
Furthermore, we warn against classifying a source and its chemistry based on the mere detection of one or a few species. 
The comparison made in Sect. \ref{sect:orionbar_orions} with Orion S shows that such simple comparisons may be rather misleading.

\begin{acknowledgements}
We thank the referee for the useful comments which helped to improve the paper.
HIFI has been designed and built by a consortium of institutes and university departments from across
Europe, Canada, and the US under the leadership of SRON Netherlands Institute for Space Research, Groningen, The Netherlands, with major contributions from Germany, France, and the US. Consortium members are: Canada: CSA, U.Waterloo; France: CESR, LAB, LERMA, IRAM; Germany: KOSMA,MPIfR,MPS; Ireland, NUIMaynooth; Italy: ASI, IFSI-INAF, Arcetri-INAF; Netherlands: SRON, TUD; Poland: CAMK, CBK; Spain: Observatorio Astronomico Nacional (IGN), Centro de Astrobiolog\'ia (CSIC-INTA); Sweden: Chalmers University of Technology - MC2, RSS \& GARD, Onsala Space Observatory, Swedish National Space Board, Stockholm University - Stockholm Observatory; Switzerland: ETH Z\"urich, FHNW; USA: Caltech, JPL, NHSC. \\
HIPE is a joint development by the Herschel Science Ground Segment Consortium, consisting of ESA, the NASA Herschel Science Center, and the HIFI, PACS, and SPIRE consortia. \\
V. O. was supported by Deutsche Forschungsgemeinschaft (DFG) via the collaborative research grant SFB 956, project C1. C. J. and M. G. acknowledge funding from the French space agency CNES. \\
Y. C. was supported by the Basic Science Research Program through the National Research Foundation of Korea (NRF) (grant No. NRF-2015R1A2A2A01004769) and the Korea Astronomy and Space Science Institute under the R\&D program (Project No. 2015-1-320-18) supervised by the Ministry of Science, ICT and Future Planning.
\end{acknowledgements}

\clearpage
\begin{appendix}    

\onecolumn
\section{Observed lines and their parameters}

\setlength\LTcapwidth{18cm}

\begin{small}
\begin{longtable}{lllrrrrlrr}
\caption[]{Gaussian fit and spectroscopic parameters of the lines identified in the spectral line survey. The Gaussian fit parameters without errors have been fixed in the fit, for example in the case of blended lines.}            
\label{line_ident}      \\

\hline
Molecule& Transition& $A_{\rm{ij}}$& $E_{\rm{up}}$& $\nu$ & $\int T_{\rm MB}{\mathrm d}V$ & $V_{\rm LSR}$ & $\Delta V$    & $T_{\rm peak}$& rms\\
&  & (s$^{-1}$)& (K)& (GHz)& (K km s$^{-1}$)& (km s$^{-1}$)& (km s$^{-1}$)& (K)& (K)\\
\hline

\endhead

\hline \multicolumn{10}{|r|}{{Continued on next page}} \\ \hline
\endfoot

\hline \hline
\endlastfoot

\hline

H$_2$Cl$^+$& $1_{11}-0_{00}$, $F$=3/2-3/2& $1.59\times10^{-3}$& 23.3& 
485413.4& 0.22$\pm$0.03& 10.42$\pm$0.16& 2.50& 0.08& 0.02\\

           & $1_{11}-0_{00}$, $F$=5/2-3/2& $1.59\times10^{-3}$& 23.3&
485417.7& 0.24$\pm$0.03& 10.76$\pm$0.25& 2.50& 0.09& 0.02\\

           & $1_{11}-0_{00}$, $F$=1/2-3/2& $1.59\times10^{-3}$& 23.3&
485420.8& 0.10$\pm$0.05& 11.50& 2.50& 0.04& 0.02\\

CS& 10-9& $2.50\times10^{-3}$& 129.3&          
489750.9& 1.17$\pm$0.03& 10.67$\pm$0.02& 1.65$\pm$0.04& 0.67& 0.03\\

o-H$_2$CO& $7_{1,7}-6_{1,6}$& $3.44\times10^{-3}$& 106.3& 
491968.4& 0.48$\pm$0.03& 10.84$\pm$0.09& 2.62$\pm$0.24& 0.17& 0.02\\

[C{\sc{i}}]& $^3$P$_1$--$^3$P$_0$& $7.99\times10^{-8}$& 23.6& 
492160.7& 38.59$\pm$0.03& 10.41$\pm$0.01& 3.06$\pm$0.01& 11.85& 0.02\\

o-H$_2$S& $2_{21}-2_{12}$& $1.36\times10^{-3}$& 79.4&
505564.8& 1.61$\pm$0.03& 10.29$\pm$0.02& 2.08$\pm$0.05& 0.73& 0.03\\

p-H$_2$CO& $7_{07}-6_{06}$& $3.82\times10^{-3}$& 97.4&
505833.7& 0.26$\pm$0.02& 10.77$\pm$0.09& 1.75$\pm$0.17& 0.14& 0.02\\

o-H$_2$CO& $7_{35}-6_{34}$& $3.20\times10^{-3}$& 203.9&    
510155.7& 0.25$\pm$0.02& 10.81$\pm$0.07& 1.72$\pm$0.17& 0.13& 0.02\\

o-H$_2$CO& $7_{34}-6_{33}$& $3.20\times10^{-3}$& 203.9&    
510237.8& 0.31$\pm$0.02& 10.62$\pm$0.11& 2.00& 0.14& 0.02\\

CF$^+$& 5-4& $8.21\times10^{-4}$& 73.8&          
512846.5& 0.10$\pm$0.02& 10.99$\pm$0.18& 1.55$\pm$0.30& 0.06& 0.02\\

SO& $11_{12}-10_{11}$& $1.78\times10^{-3}$& 167.6& 
514853.8& 0.32$\pm$0.03& 10.44$\pm$0.08& 2.04$\pm$0.18& 0.15& 0.02\\

SO& $12_{12}-11_{11}$& $1.80\times10^{-3}$& 174.2&
516335.8& 0.25$\pm$0.02& 10.29$\pm$0.07& 1.80$\pm$0.15& 0.14& 0.01\\

SO& $13_{12}-12_{11}$& $1.82\times10^{-3}$& 165.8&
517354.5& 0.36$\pm$0.07& 10.35$\pm$0.16& 1.69$\pm$0.40& 0.20& 0.02\\

C$_2$H& $N=6-5$, $J=13/2-11/2$& $6.61\times10^{-4}$& 88.0& 
523971.6& 4.39$\pm$0.03& 10.59$\pm$0.01& 2.62$\pm$0.02& 1.57& 0.02\\        
                              
C$_2$H& $N=6-5$, $J=11/2-9/2$& $6.50\times10^{-4}$& 88.0& 
524033.9& 3.72$\pm$0.03& 10.58$\pm$0.01& 2.66$\pm$0.03& 1.31& 0.02\\

o-H$_2$CO& $7_{16}-6_{15}$& $4.20\times10^{-3}$& 112.8&
525665.8& 0.49$\pm$0.02& 10.72$\pm$0.05& 1.95$\pm$0.11& 0.23& 0.02\\

SH$^+$& $N$=0--1, $J$=1--2, $F$=1/2--3/2& $7.99\times10^{-4}$& 25.3&
526038.7& 0.32$\pm$0.03& 10.76$\pm$0.10& 2.14$\pm$0.22& 0.14& 0.02\\

SH$^+$& $N$=0--1, $J$=1--2, $F$=3/2--5/2& $9.59\times10^{-4}$& 25.2&
526047.9& 0.71$\pm$0.04& 10.66$\pm$0.08& 3.23$\pm$0.19& 0.21& 0.02\\

SH$^+$& $N$=0--1, $J$=1--2, $F$=3/2--3/2& $1.60\times10^{-4}$& 25.3& 
526124.9& 0.15$\pm$0.03& 10.12$\pm$0.26& 2.53$\pm$0.53& 0.06& 0.02\\

HCN& 6-5& $7.20\times10^{-3}$& 89.3&
531716.4& 3.53$\pm$0.04& 10.59$\pm$0.01& 2.55$\pm$0.04& 1.30& 0.02\\

CH& $^2\Pi_{3/2}$ $J=3/2-1/2$& $2.07\times10^{-4}$& 25.7&
532723.9& 16.22$\pm$0.03& 10.60$\pm$0.01& 4.20$\pm$0.01& 3.63& 0.02\\

CH&     $^2\Pi_{3/2}$ $J=3/2-1/2$& $4.14\times10^{-4}$& 25.7&
532793.3& 7.05$\pm$0.04& 10.49$\pm$0.01& 3.62$\pm$0.02& 1.83& 0.03\\

HCO$^+$& $6-5$& $1.25\times10^{-2}$& 89.9& 
535061.6& 11.08$\pm$0.04& 10.63$\pm$0.01& 2.82$\pm$0.01& 3.70& 0.02\\

CH& $^2\Pi_{3/2}$ $J=3/2-1/2$& $6.38\times10^{-4}$& 25.8&  
536761.2& 14.04$\pm$0.03& 10.56$\pm$0.01& 3.92$\pm$0.01& 3.37& 0.02\\

CH& $^2\Pi_{3/2}$ $J=3/2-1/2$& $2.13\times10^{-4}$& 25.8& 
536782.0& 4.02$\pm$0.03& 10.59$\pm$0.01& 3.70$\pm$0.04& 1.02& 0.02\\

CH& $^2\Pi_{3/2}$ $J=3/2-1/2$& $4.25\times10^{-4}$& 25.8& 
536795.7& 6.42$\pm$0.03& 10.54$\pm$0.01& 3.65$\pm$0.02& 1.65& 0.02\\

CS& $11-10$& $3.34\times10^{-3}$& 155.2&        
538689.0& 0.81$\pm$0.02& 10.69$\pm$0.02& 1.73$\pm$0.06& 0.44& 0.02\\

HNC& $6-5$& $8.04\times10^{-3}$& 91.4&
543897.6& 0.25$\pm$0.02& 10.92$\pm$0.14& 3.01$\pm$0.34& 0.08& 0.02\\

o-H$_2^{18}$O& $1_{1,0}-1_{0,1}$& $3.29\times10^{-3}$& 60.5& 
547676.4& 0.31$\pm$0.03& 10.26$\pm$0.16& 2.92$\pm$0.34& 0.10& 0.02\\

C$^{18}$O& $5-4$& $1.06\times10^{-5}$& 79.0&  
548831.0&  14.60$\pm$0.07& 10.51$\pm$0.01& 2.07$\pm$0.01& 6.62& 0.03\\

$^{13}$CO& $5-4$& $5.37\times10^{-6}$& 79.3&
550926.3& 120.71$\pm$0.03& 10.39$\pm$0.01& 2.63$\pm$0.01& 43.15& 0.03\\

NO& $J=11/2-9/2$& $2.16\times10^{-5}$& 84.2&
551187.5& 0.58$\pm$0.03& 10.22$\pm$0.05& 2.11$\pm$0.10& 0.26& 0.02\\

NO& $J=11/2-9/2$& $2.23\times10^{-5}$& 84.3&
551534.0& 0.31$\pm$0.02& 10.92$\pm$0.08& 2.16$\pm$0.16& 0.13& 0.02\\

o-H$_2$O& $1_{10}-1_{01}$& $3.43\times10^{-3}$& 61.0&
556936.0& 22.81$\pm$0.10& 10.18$\pm$0.01& 4.42$\pm$0.02& 4.85& 0.04\\

C$^{17}$O& 5-4& $1.14\times10^{-5}$& 80.9&          
561712.8& 4.51$\pm$0.03& 10.62$\pm$0.01& 2.03$\pm$0.02& 2.09& 0.02\\

o-H$_2$CO& $8_{18}-7_{17}$& $5.20\times10^{-3}$&  133.3&
561899.3& 0.89$\pm$0.05& 10.82$\pm$0.08& 2.93$\pm$0.23& 0.29& 0.03\\

CN& $N$=5-4, $J$=9/2-7/2, $F$=11/2-9/2& $1.98\times10^{-3}$& 81.6&                  
566730.0& 3.14$\pm$0.03& 10.72$\pm$0.01& 2.35$\pm$0.03& 1.25& 0.02\\

CN& $N$=5-4, $J$=11/2-9/2, $F$=9/2-7/2& $1.95\times10^{-3}$& 81.6&
566947.3& 3.92$\pm$0.11& 11.10$\pm$0.03& 2.38$\pm$0.08& 1.55& 0.03\\

CN& $N$=5-4, $J$=11/2-9/2, $F$=9/2-11/2& $6.66\times10^{-7}$& 81.6&     
566978.9& 1.29$\pm$0.04& 10.88$\pm$0.06& 3.89$\pm$0.14& 0.31& 0.02\\

NH$_3$& 1-0& $3.43\times10^{-3}$& 27.5& 
572498.2& 3.23$\pm$0.04& 10.43$\pm$0.02& 3.22$\pm$0.04& 0.94& 0.03\\

CO& 5-4& $1.23\times10^{-5}$& 83.0&
576267.9& 502.93$\pm$0.01& 10.08$\pm$0.01& 4.07$\pm$0.01& 116.15& 0.03\\

o-H$_2$CO& $8_{36}-7_{35}$& $5.08\times10^{-3}$& 231.9&
583144.6& 0.20$\pm$0.03& 11.14$\pm$0.10& 1.46$\pm$0.20& 0.13& 0.03\\

o-H$_2$CO& $8_{35}-7_{34}$& $5.08\times10^{-3}$& 231.9&
583308.6& 0.25$\pm$0.03& 9.74$\pm$0.20& 3.04$\pm$0.35& 0.08& 0.02\\

CS& 12-11& $4.33\times10^{-3}$& 183.4&
587616.5& 0.42$\pm$0.03& 10.74$\pm$0.06& 1.47$\pm$0.13& 0.27& 0.03\\

o-H$_2$CO& $8_{17}-7_{16}$& $6.34\times10^{-3}$& 141.6&
600330.6& 0.35$\pm$0.03& 9.84$\pm$0.18& 3.31$\pm$0.34& 0.10& 0.02\\

C$_2$H& $N=7-6$, $J=15/2-13/2$& $1.05\times10^{-3}$& 117.4& 
611267.2& 2.91$\pm$0.02& 10.77$\pm$0.01& 2.44$\pm$0.02& 1.12& 0.02\\

C$_2$H& $N=7-6$, $J=13/2-11/2$& $1.04\times10^{-3}$& 117.4& 
611329.7& 2.52$\pm$0.03& 10.70$\pm$0.01& 2.49$\pm$0.03& 0.95& 0.02\\

HCN& 7-6& $1.16\times10^{-2}$& 119.1&
620304.0& 1.59$\pm$0.04& 10.58$\pm$0.02& 2.27$\pm$0.06& 0.66& 0.03\\

HCO$^+$& 7-6& $2.01\times10^{-2}$& 119.8& 
624208.5& 7.40$\pm$0.03& 10.69$\pm$0.01& 2.59$\pm$0.01& 2.69& 0.02\\

H$^{37}$Cl& 1-0, 1.5--1.5& $1.16\times10^{-3}$& 30.0&          
624964.4& 0.84$\pm$0.12& 10.32$\pm$0.13& 1.79$\pm$0.29& 0.44& 0.02\\

&           1-0, 2.5--1.5& $1.16\times10^{-3}$& 30.0&
624977.8& 1.13$\pm$0.10& 10.23$\pm$0.08& 1.87$\pm$0.19& 0.57& 0.02\\

&           1-0, 0.5--1.5& $1.16\times10^{-3}$& 30.0&
624988.3& 0.46$\pm$0.03& 10.34$\pm$0.05& 1.77$\pm$0.14& 0.25& 0.02\\

H$^{35}$Cl& 1-0, 1.5--1.5& $1.17\times10^{-3}$& 30.0&
625901.6& 1.82$\pm$0.24& 10.21$\pm$0.13& 1.97$\pm$0.30& 0.87& 0.03\\

&    1-0, 2.5--1.5& $1.17\times10^{-3}$& 30.0&
625918.8& 2.34$\pm$0.20& 10.29$\pm$0.08& 1.95$\pm$0.19& 1.13& 0.03\\

&    1-0, 0.5--1.5& $1.17\times10^{-3}$& 30.0&
625932.0& 1.21$\pm$0.27& 10.30$\pm$0.21& 1.92$\pm$0.52& 0.59& 0.03\\

o-H$_2$CO& $9_{1,9}-8_{1,8}$& $7.46\times10^{-3}$& 163.6&
631702.8& 0.48$\pm$0.04& 10.78$\pm$0.09& 2.39$\pm$0.24& 0.19& 0.03\\

CS& 13-12& $5.52\times10^{-3}$& 213.9&
636531.8& 0.23$\pm$0.04& 10.95$\pm$0.16& 1.77$\pm$0.29& 0.12& 0.04\\

NO& $J$=13/2-11/2& $3.74\times10^{-5}$& 115.4&
651433.1& 0.51$\pm$0.04& 10.43$\pm$0.08& 1.92$\pm$0.19& 0.25& 0.04\\ 

NO& $J$=13/2-11/2& $3.25\times10^{-4}$& 115.5&
651771.4& 0.45$\pm$0.04& 9.89$\pm$0.10& 2.40$\pm$0.22& 0.18& 0.03\\

C$^{18}$O& 6-5& $1.86\times10^{-5}$& 110.6& 
658553.3& 14.50$\pm$0.05& 10.48$\pm$0.01& 2.01$\pm$0.01& 6.78& 0.05\\

$^{13}$CO& 6-5& $9.40\times10^{-6}$& 111.1&
661067.3& 112.85$\pm$0.06& 10.50$\pm$0.01& 2.53$\pm$0.01& 41.85& 0.04\\

C$^{17}$O& 6-5& $1.99\times10^{-5}$& 113.2&
674009.3& 3.76$\pm$0.05& 10.56$\pm$0.01& 1.99$\pm$0.03& 1.77& 0.05\\

CN& 6-5 11/2-9/2& $3.36\times10^{-3}$& 114.2&
680047.4& 1.32$\pm$0.05& 10.73$\pm$0.04& 2.12$\pm$0.10& 0.59& 0.05\\

CN& 6-5 13/2-11/2& $3.47\times10^{-3}$& 114.3&        
680263.9& 1.58$\pm$0.05& 10.83$\pm$0.03& 2.15$\pm$0.09& 0.69& 0.05\\

p-H$_2$S& $2_{02}-1_{11}$& $9.33\times10^{-4}$& 54.7& 
687303.5& 2.73$\pm$0.04& 10.47$\pm$0.02& 2.06$\pm$0.04& 1.25& 0.04\\

CO& 6-5& $2.14\times10^{-5}$& 116.2&
691473.1& 481.35$\pm$0.21& 10.16$\pm$0.01& 3.96$\pm$0.01& 114.33& 0.04\\

C$_2$H& $N$=8-7, J=19/2-17/2& $1.59\times10^{-3}$& 150.9&
698544.8& 1.55$\pm$0.05& 10.67$\pm$0.03& 2.14$\pm$0.07& 0.68& 0.04\\

C$_2$H& $N$=8-7, J=17/2-15/2& $1.57\times10^{-3}$& 150.9&
698607.5& 1.56$\pm$0.04& 10.66$\pm$0.03& 2.28$\pm$0.06& 0.64& 0.04\\

HCN& 8-7& $1.74\times10^{-2}$& 153.1&
708877.0& 0.43$\pm$0.05& 10.82$\pm$0.12& 2.06$\pm$0.22& 0.20& 0.05\\

HCO$^+$& 8-7& $3.02\times10^{-2}$& 154.1&
713341.2& 4.23$\pm$0.08& 10.60$\pm$0.02& 2.68$\pm$0.07& 1.48& 0.05\\

o-H$_2$S& $2_{12}-1_{01}$& $1.33\times10^{-3}$& 55.1&
736034.1& 7.34$\pm$0.01& 10.42$\pm$0.01& 2.48$\pm$0.02& 2.78& 0.04\\

p-H$_2$O& $2_{11}-2_{02}$& $6.98\times10^{-3}$& 136.9&
752033.1& 8.50$\pm$0.12& 10.39$\pm$0.02& 2.46$\pm$0.04& 3.25& 0.08\\        

C$^{18}$O& 7-6& $7.98\times10^{-5}$& 147.5&
768251.6& 12.90$\pm$0.11& 10.53$\pm$0.01& 1.89$\pm$0.02& 6.42& 0.12\\
                     
$^{13}$CO& 7-6& $1.50\times10^{-5}$& 148.1&
771184.1& 112.69$\pm$0.12& 10.58$\pm$0.01& 2.39$\pm$0.01& 44.34& 0.09\\

C$_2$H& N=9-8, J=19/2-17/2& $1.58\times10^{-3}$& 188.6& 
785802.1& 1.38$\pm$0.07& 10.36$\pm$0.06& 2.48$\pm$0.15& 0.52& 0.06\\

C$_2$H& N=9-8, J=17/2-15/2&     $1.56\times10^{-3}$& 188.6&
785865.0& 1.09$\pm$0.07& 10.57$\pm$0.07& 2.41$\pm$0.17& 0.42& 0.06\\

C$^{17}$O& 7-6& $3.18\times10^{-5}$& 151.0&
786280.8& 3.25$\pm$0.08& 10.56$\pm$0.02& 1.95$\pm$0.06& 1.57& 0.08\\

CN& $N$=7-6, $J$=13/2-11/2, $F$=15/2-13/2& $5.64\times10^{-3}$& 152.3&
793336.1& 0.95$\pm$0.15& 10.67$\pm$0.27& 4.12$\pm$0.98& 0.22& 0.09\\

CN& $N$=7-6, $J$=15/2-13/2, $F$=17/2-15/2& $5.71\times10^{-3}$& 152.4&
793553.3& 1.07$\pm$0.10& 10.58$\pm$0.10& 2.24$\pm$0.24& 0.45& 0.10\\        
                       
HCO$^+$& 9-8& $4.33\times10^{-2}$& 192.6&
802458.2& 2.17$\pm$0.11& 10.71$\pm$0.05& 2.12$\pm$0.12& 0.96& 0.11\\

CO& 7-6& $3.42\times10^{-5}$& 154.9& 
806651.8& 451.54$\pm$0.21& 10.31$\pm$0.01& 3.86$\pm$0.01& 109.99& 0.13\\

[C{\sc{i}}]&    $^3$P$_2$--$^3$P$_1$& $2.67\times10^{-7}$& 62.5&  
809342.0& 51.22$\pm$0.08& 10.64$\pm$0.01& 2.57$\pm$0.01& 18.72& 0.08\\

CH$^+$& 1-0& $6.36\times10^{-3}$& 40.1& 
835137.5& 31.37$\pm$0.15& 10.51$\pm$0.01& 5.29$\pm$0.03& 5.57& 0.10\\

C$_2$H& $N=10-9$, $J=21/2-19/2$& $2.19\times10^{-3}$& 230.5& 
873036.4& 0.98$\pm$0.11& 11.13$\pm$0.13& 2.22$\pm$0.27& 0.42& 0.13\\

C$_2$H& $N=10-9$, $J=19/2-17/2$& $2.17\times10^{-3}$& 230.5& 
873099.5& 1.58$\pm$0.15& 10.12$\pm$0.19& 4.08$\pm$0.41& 0.37& 0.12\\
                 
C$^{18}$O& 8-7& $4.47\times10^{-5}$& 189.6&
877922.0& 9.67$\pm$0.09& 10.61$\pm$0.01& 1.81$\pm$0.02& 5.02& 0.10\\

$^{13}$CO& 8-7& $2.26\times10^{-5}$& 190.4&         
881272.8& 93.63$\pm$0.16& 10.59$\pm$0.01& 2.24$\pm$0.01& 39.26& 0.15\\

U-line& & & &
890034.0& 0.86$\pm$0.07& 10.03$\pm$0.07& 1.50$\pm$0.13& 0.54& 0.09\\

HCO$^+$& 10-9& $5.97\times10^{-2}$& 235.4&
891557.3& 2.00$\pm$0.10& 10.50$\pm$0.06& 2.60$\pm$0.17& 0.72& 0.10\\

C$^{17}$O& 8-7& $4.77\times10^{-5}$& 194.1&
898523.0& 2.42$\pm$0.07& 10.72$\pm$0.02& 1.72$\pm$0.06& 1.32& 0.09\\

CO& 8-7& $5.13\times10^{-5}$& 199.1& 
921799.7& 453.35$\pm$0.18& 10.43$\pm$0.01& 3.70$\pm$0.01& 115.17& 0.10\\
                                        
OH$^+$& 1,2,3/2-0,1,1/2& $1.52\times10^{-2}$& 46.7& 
971805.3& 3.55$\pm$0.21& 11.70$\pm$0.19& 6.04$\pm$0.39& 0.55& 0.16\\

HCO$^+$& 11-10& $7.98\times10^{-2}$& 282.4&
980636.5& 0.57$\pm$0.07& 10.88$\pm$0.09& 1.53$\pm$0.24& 0.35& 0.09\\

C$^{18}$O& $9-8$& $6.38\times10^{-5}$& 237.0&
987560.4& 5.33$\pm$0.07& 10.74$\pm$0.01& 1.75$\pm$0.03& 2.86& 0.09\\

p-H$_2$O& $2_{0,2}-1_{1,1}$& $5.79\times10^{-3}$& 100.8& 
987926.8& 8.12$\pm$0.09& 10.39$\pm$0.02& 2.90$\pm$0.03& 2.63& 0.09\\

$^{13}$CO& $9-8$& $3.22\times10^{-5}$& 237.9&
991329.3& 63.52$\pm$0.02& 10.68$\pm$0.01& 2.16$\pm$0.01& 27.69& 0.08\\

o-H$_2$S& $3_{0,3}-2_{1,2}$& $3.73\times10^{-3}$& 102.8& 
993108.3& 2.03$\pm$0.07& 10.69$\pm$0.04& 2.28$\pm$0.09& 0.84& 0.09\\

C$^{17}$O& 9-8& $6.82\times10^{-5}$& 242.6&
1010731.8& 1.21$\pm$0.07& 11.00$\pm$0.04& 1.51$\pm$0.12& 0.75& 0.10\\

OH$^+$& 1,1,1/2-0,1,3/2& $1.76\times10^{-2}$& 49.6&
1033118.6& 1.12$\pm$0.10& 10.03$\pm$0.19& 4.06$\pm$0.33& 0.26& 0.10\\

CO& 9-8& $7.33\times10^{-5}$& 248.9&
1036912.4& 391.05$\pm$0.40& 10.48$\pm$0.01& 3.60$\pm$0.01& 102.05& 0.27\\                     
                        
C$^{18}$O& 10-9& $8.76\times10^{-5}$& 289.7& 
1097162.9& 3.39$\pm$0.08& 10.71$\pm$0.02& 1.46$\pm$0.04& 2.18& 0.11\\

o--H$_2$O& $3_{12}-3_{03}$& $1.63\times10^{-2}$& 249.4& 
1097364.8& 2.27$\pm$0.10& 10.17$\pm$0.04& 2.04$\pm$0.10& 1.04& 0.13\\

$^{13}$CO& 10-9& $4.43\times10^{-5}$& 290.8&
1101349.6& 55.40$\pm$0.01& 10.59$\pm$0.01& 1.98$\pm$0.01& 26.29& 0.13\\

p--H$_2$O& $1_{11}-0_{00}$& $1.83\times10^{-2}$& 53.4& 
1113343.0& 11.46$\pm$0.15& 9.96$\pm$0.03& 5.02$\pm$0.08& 2.15& 0.13\\

CO& $10-9$& $1.01\times10^{-4}$& 304.2&
1151985.5& 336.82$\pm$0.33& 10.47$\pm$0.01& 3.45$\pm$0.01& 91.70& 0.30\\

$^{13}$CO& 11-10& $5.89\times10^{-5}$& 348.9& 
1211329.7& 37.32$\pm$0.29& 10.68$\pm$0.01& 1.81$\pm$0.02& 19.33& 0.42\\

HF& 1-0& $2.42\times10^{-2}$& 59.1&
1232476.3& 5.89$\pm$0.35& 11.23$\pm$0.11& 3.39$\pm$0.23& 1.63& 0.37\\

CO& 11--10& $1.34\times10^{-4}$& 365.0& 
1267014.5& 318.48$\pm$0.34& 10.41$\pm$0.01& 3.29$\pm$0.01& 90.96& 0.31\\

CO& 13--12& $2.20\times10^{-4}$& 503.1&
1496922.9& 187.71$\pm$0.29& 10.38$\pm$0.01& 2.91$\pm$0.01& 60.59& 0.36\\         
                             
CO& 14--13& $2.74\times10^{-4}$& 580.5&
1611793.5& 161.08$\pm$0.34& 10.34$\pm$0.01& 2.89$\pm$0.01& 52.39& 0.46\\
                         
CH$^+$& 2-1& $6.10\times10^{-2}$& 120.2&            
1669281.3& 8.96$\pm$0.32& 10.56$\pm$0.07& 4.10$\pm$0.17& 2.05& 0.35\\

o-H$_2$O& $2_{12}-1_{01}$& $5.54\times10^{-2}$& 114.4& 
1669904.8& 7.28$\pm$0.42& 10.40$\pm$0.11& 4.01$\pm$0.30& 1.70& 0.44\\         

CO& 15--14& $3.35\times10^{-4}$& 663.4&
1726602.5& 123.18$\pm$0.29& 10.39$\pm$0.01& 2.62$\pm$0.01& 44.15& 0.48\\

CO& 16--15& $4.05\times10^{-4}$& 751.7&              
1841345.5& 66.37$\pm$0.25& 10.42$\pm$0.01& 2.39$\pm$0.01& 26.14& 0.37\\

U-line& & & &
1871194.0&  4.31$\pm$0.29& 9.92$\pm$0.09& 2.63$\pm$0.21& 1.54& 0.37\\

$^{13}$C$^+$& $^2$P$_{3/2}$--$^2$P$_{1/2}$, $F$=1--1& $7.73\times10^{-7}$& 91.2&
1900136.0& 5.72$\pm$0.30& 9.92$\pm$0.07& 2.94$\pm$0.20& 1.83& 0.35\\         
         
$^{13}$C$^+$& $^2$P$_{3/2}$--$^2$P$_{1/2}$, $F$=2--1& $2.32\times10^{-6}$& 91.2&         
1900466.1& 24.26$\pm$0.19& 10.72$\pm$0.01& 2.38$\pm$0.02& 9.57& 0.30\\          
                            
C$^+$& $^2$P$_{3/2}$--$^2$P$_{1/2}$& $2.32\times10^{-6}$& 91.2&             
1900536.9& 1109.3$\pm$2.45& 10.54$\pm$0.01& 3.64$\pm$0.01& 286.03& 0.41\\

$^{13}$C$^+$& ${^2}$P$_{3/2}$--$^2$P$_{1/2}$, $F$=1--0& $1.55\times10^{-6}$& 91.2&      
1900950.0& 9.20$\pm$0.26& 10.17$\pm$0.03& 2.29$\pm$0.08& 3.77& 0.38\\

\hline

\end{longtable}
\end{small}

\newpage
\section{The observed data}
\label{sect:obs_data}

        \begin{center}
         \begin{figure*}[!h]
           \centering
           \includegraphics[width=\textwidth]{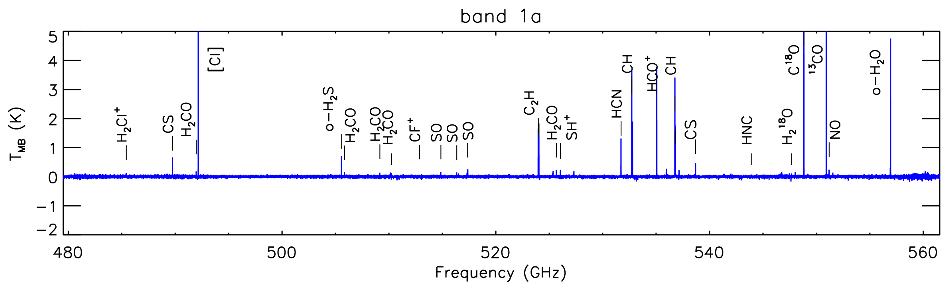}
           \includegraphics[width=\textwidth]{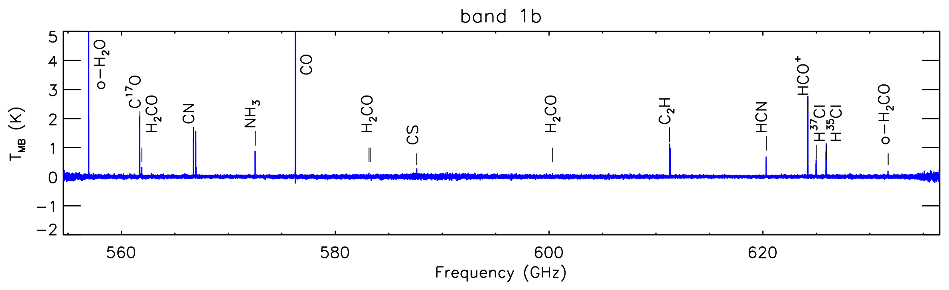}     
           \includegraphics[width=\textwidth]{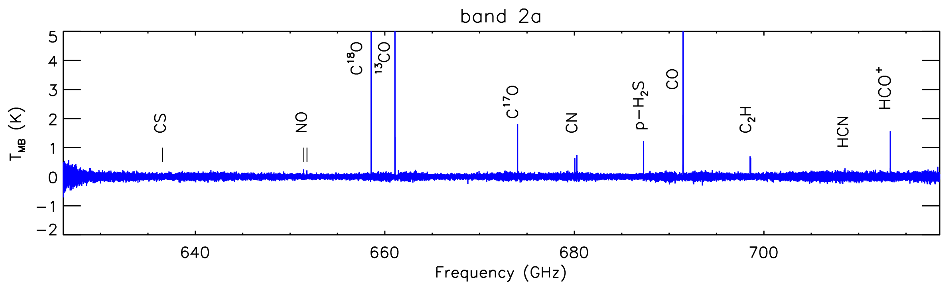}     
           \includegraphics[width=\textwidth]{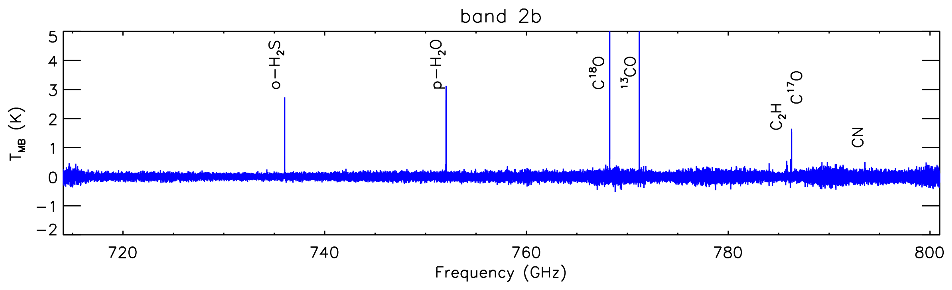}     
      \caption{From top to bottom: Band 1a, 1b, 2a, and 2b.}
         \label{1a}
     \end{figure*}
    \end{center}

        \begin{center}
         \begin{figure*}[!h]
           \centering
           \includegraphics[width=\textwidth]{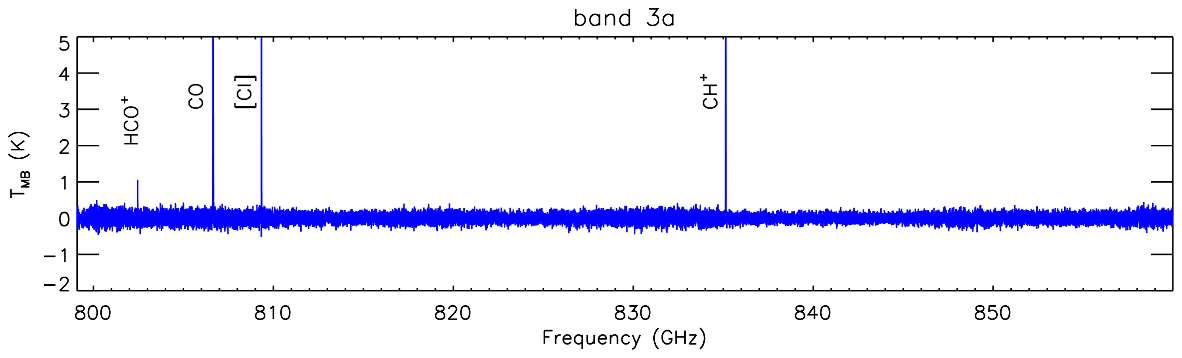}           
           \includegraphics[width=\textwidth]{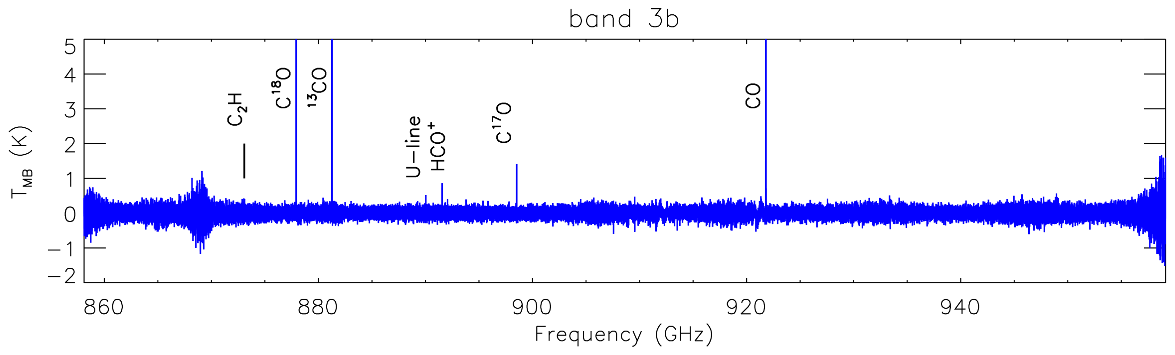} 
           \includegraphics[width=\textwidth]{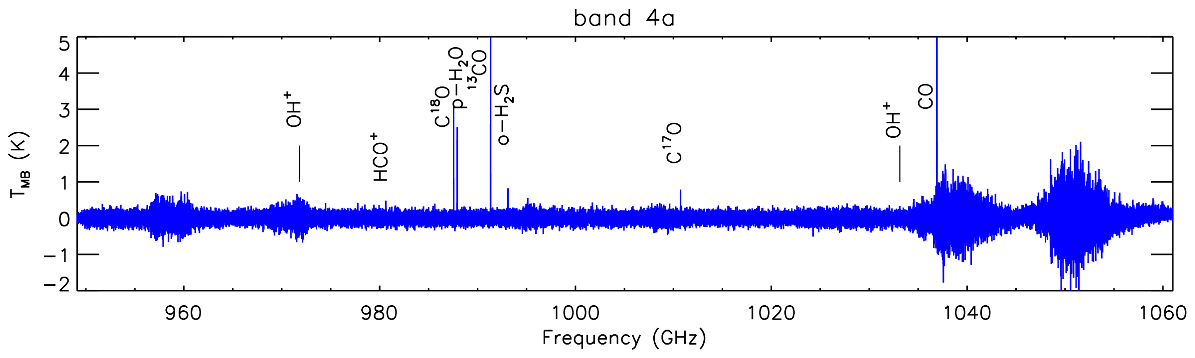} 
           \includegraphics[width=\textwidth]{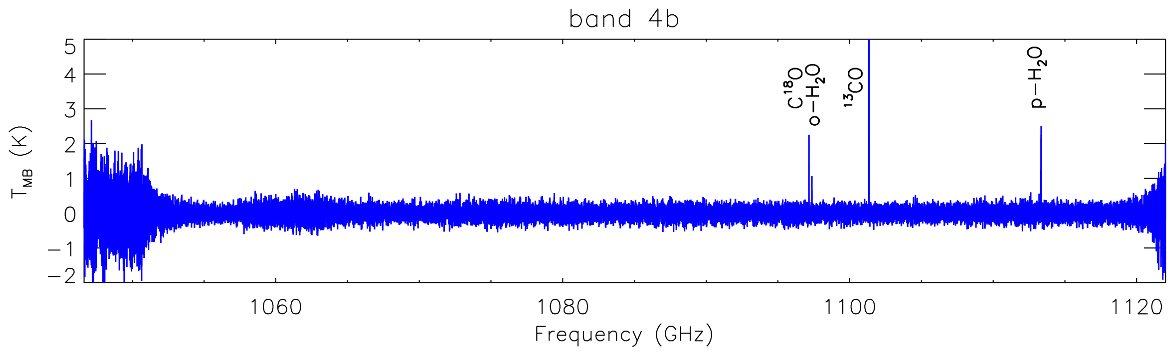}                         
      \caption{From top to bottom: Band 3a, 3b, 4a, and 4b.}
         \label{3a}
     \end{figure*}
    \end{center}
    
        \begin{center}
         \begin{figure*}[!h]
           \centering
           \includegraphics[width=\textwidth]{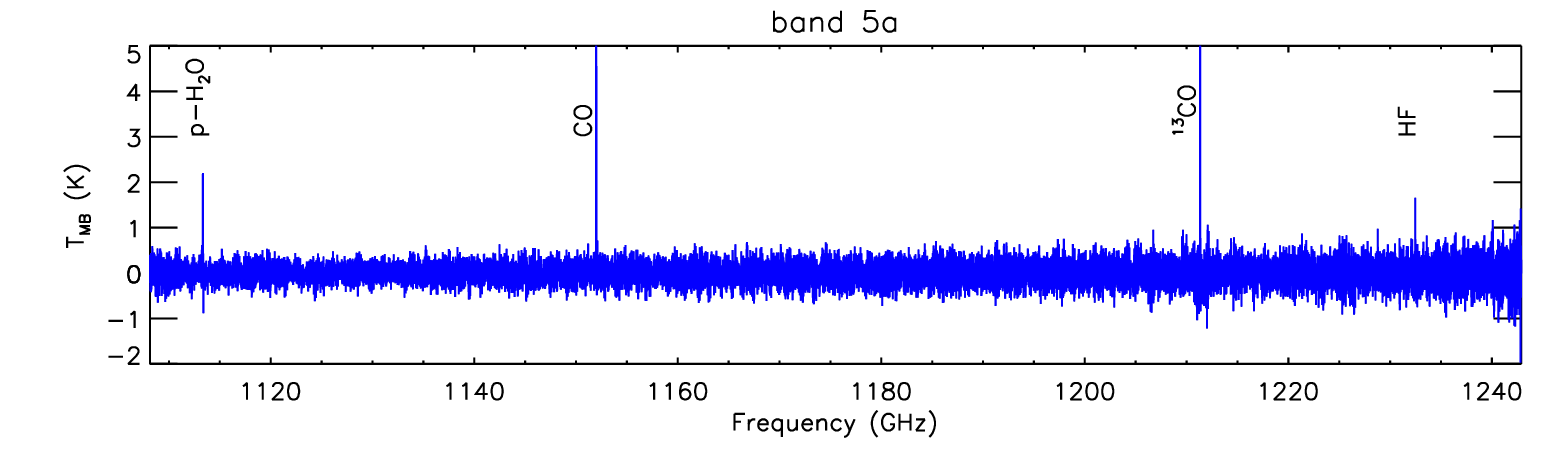} 
           \includegraphics[width=\textwidth]{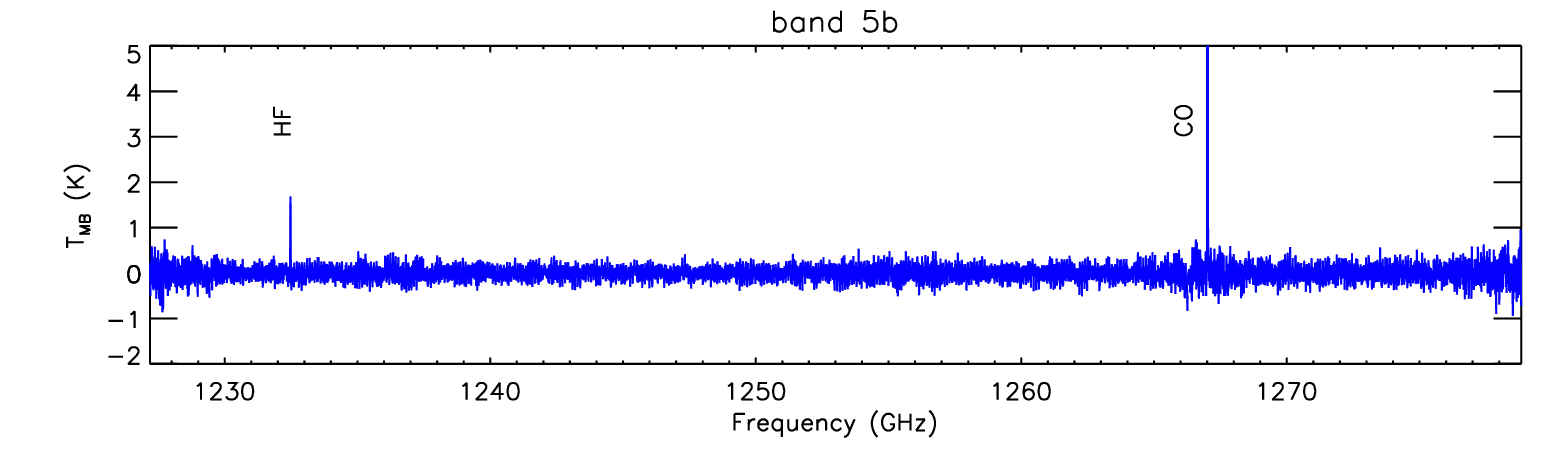} 
           \includegraphics[width=\textwidth]{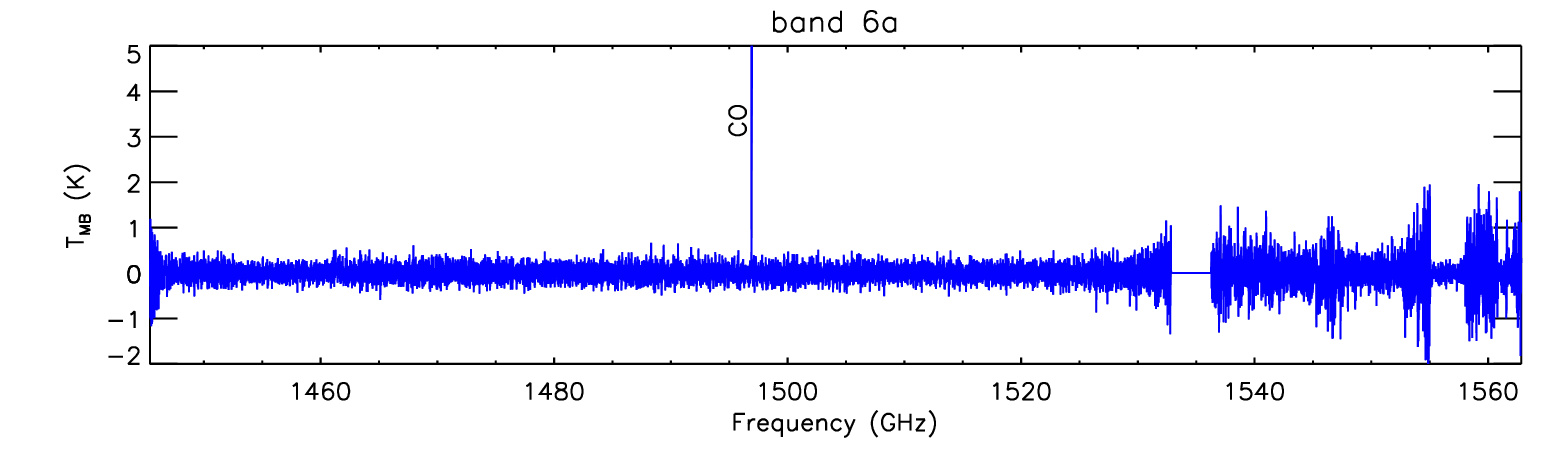}     
           \includegraphics[width=\textwidth]{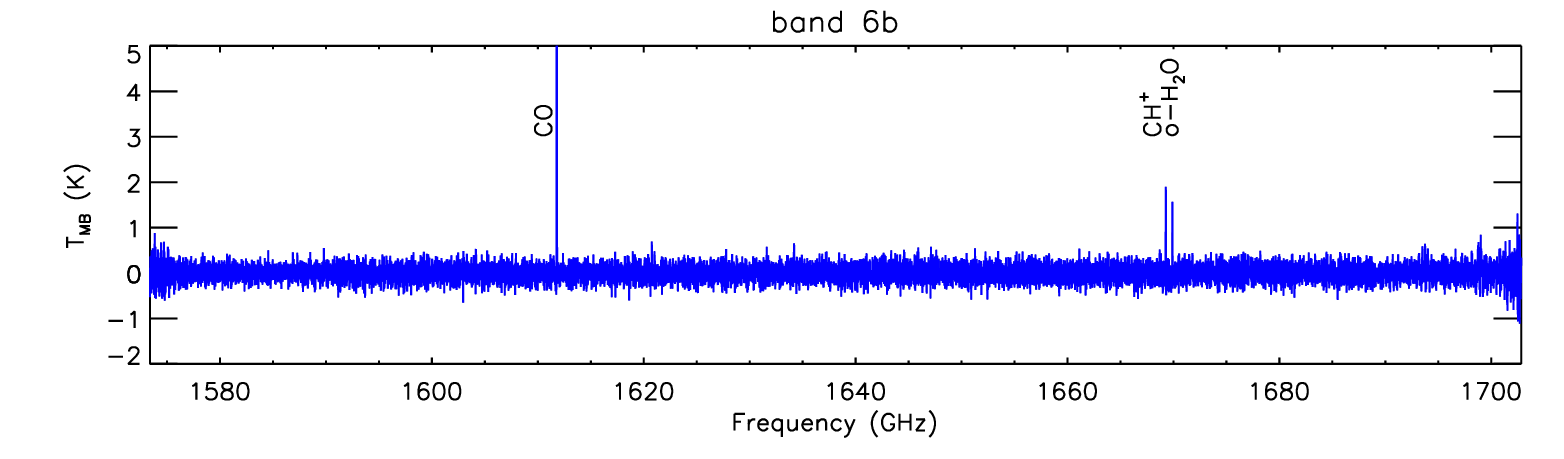}                                          
      \caption{From top to bottom: Band 5a, 5b, 6a, and 6b.  
      Bands 5 and 6 were smoothed from the original velocity resolutions of  $\sim$0.12 \kms\ and $\sim$0.09 \kms\ to the $\sim$1 \kms\ and $\sim$1.5 \kms\ channels, respectively.}
         \label{5a}
     \end{figure*}
    \end{center} 
    
        \begin{center}
         \begin{figure*}[!h]
           \centering
           \includegraphics[width=\textwidth]{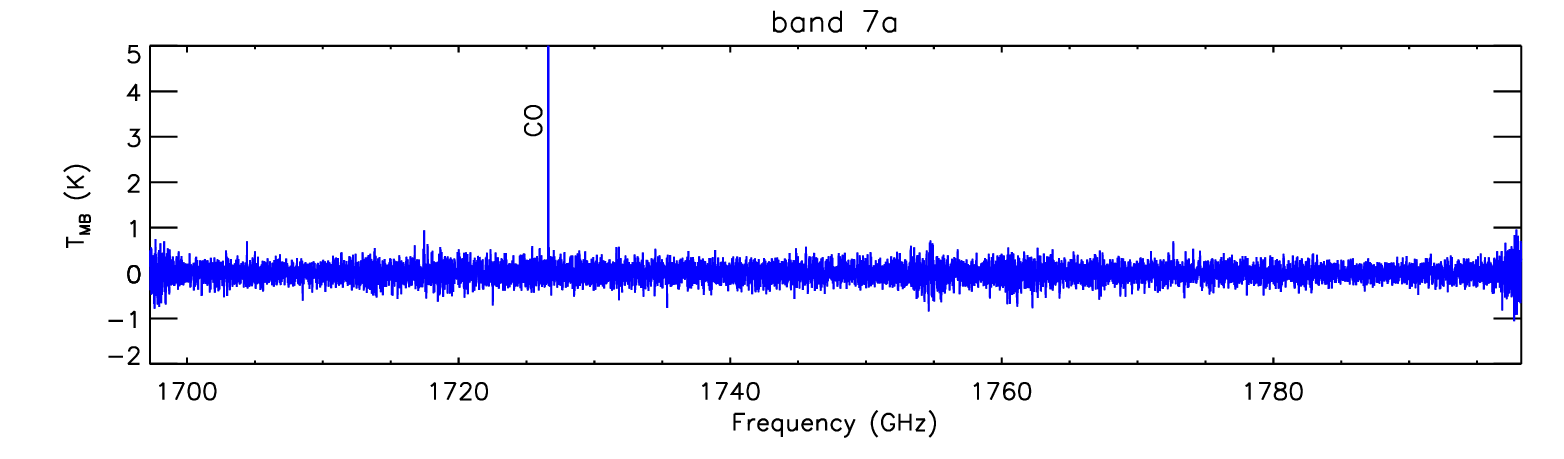} 
           \includegraphics[width=\textwidth]{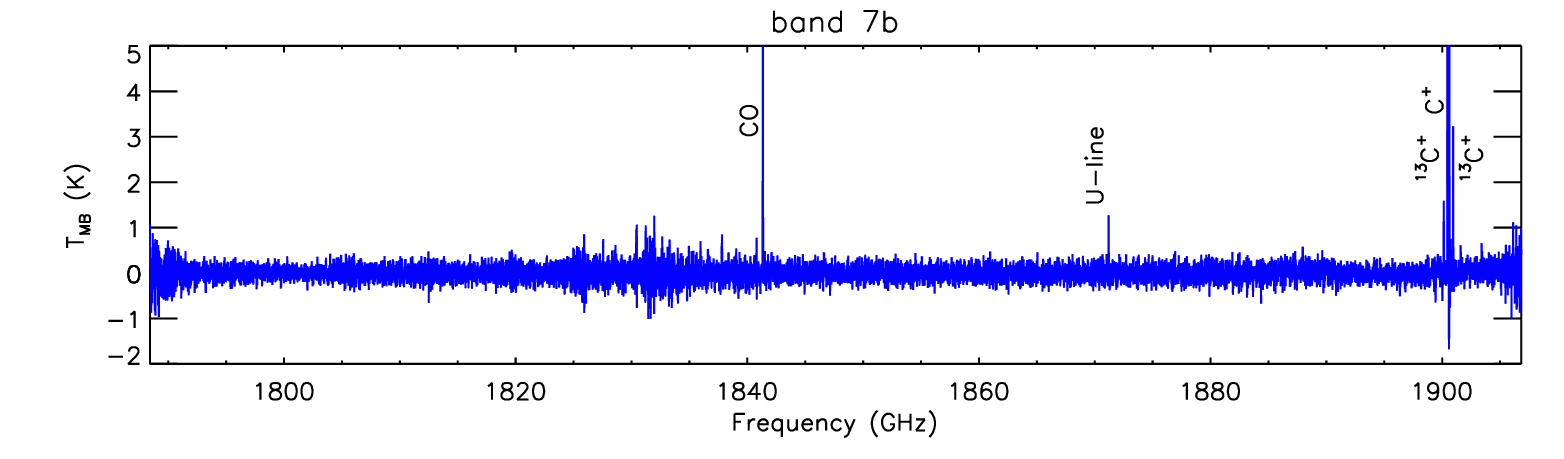}       
      \caption{From top to bottom: Band 7a, 7b.  
      The spectra were smoothed from the original velocity resolution of  $\sim$0.08 \kms\ to the $\sim$1.4 \kms\ channel.}
         \label{7a}
     \end{figure*}
    \end{center} 
    
\clearpage    
\newpage    

\section{Line properties}
\label{appendix:lineparam}

\begin{figure*}[ht]
\centering
\includegraphics[width=17 cm, trim=-0.3cm 0cm 0cm 0cm,clip=true]{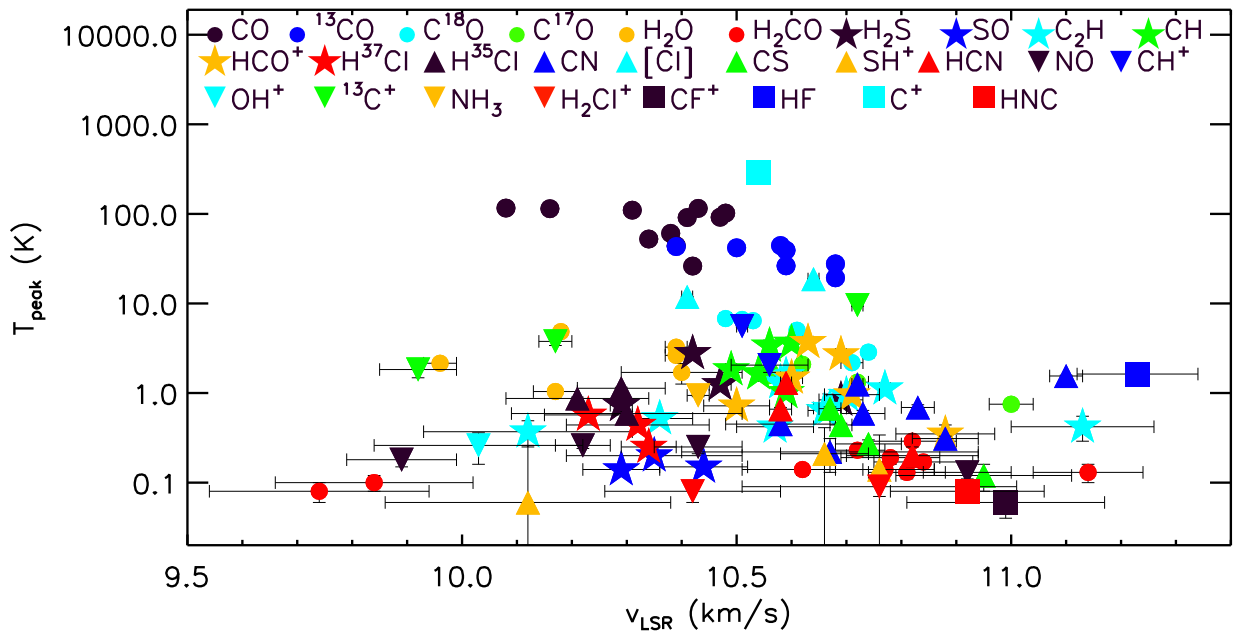}
\caption{Peak intensities of the species shown in Fig. \ref{vlsr_fwhm} as a function $V_{\rm{LSR}}$. For some of the transitions in this figure the error bars are smaller than the symbol sizes.}
\label{vlsr_tpeak}
\end{figure*}

To test whether some of the fitted line parameters are the result of low signal-to-noise, we plotted the fitted velocities as a function of the peak intensities in Fig. \ref{vlsr_tpeak}. Some of the parameters of lower intensity lines may indeed be affected by the low signal-to-noise as there is no trend showing that the lowest intensity lines peak at the lowest or highest velocities of the observed range, the effect of line parameters due to low signal-to-noise on the $V_{\rm{LSR}}$-line width relation is small.\footnote{The velocity difference between the three hyperfine components of the $^{13}$C$^{+}$ line may indicate some problem with the accuracy of the published rest frequencies.}

\begin{figure*}[ht]
\centering
\includegraphics[width=17 cm, trim=0cm 0cm 0cm 0cm,clip=true]{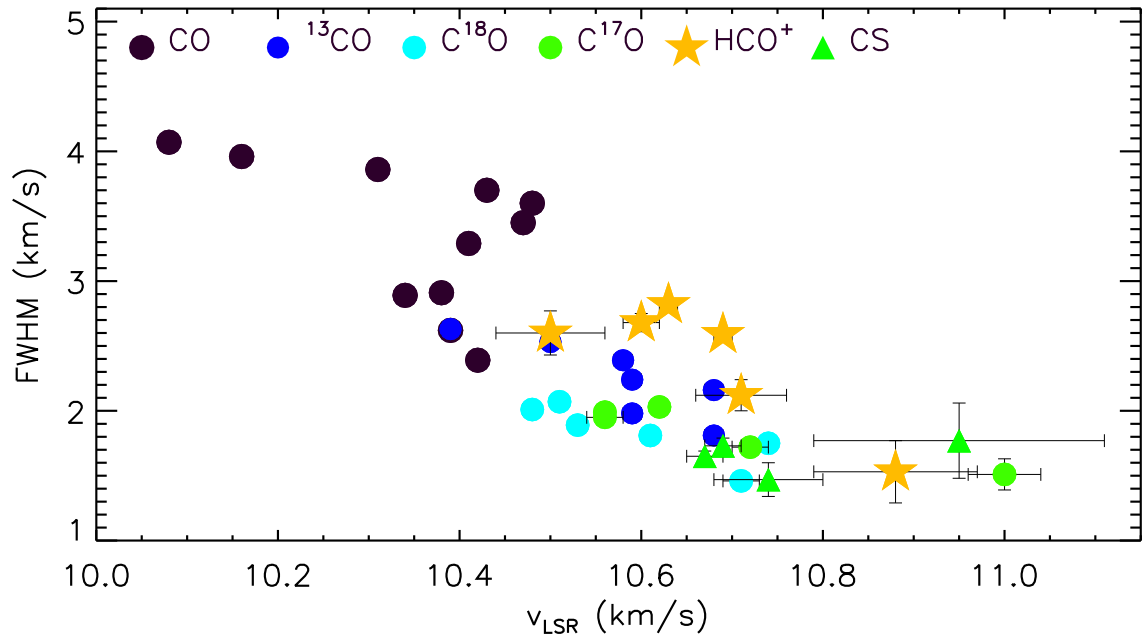}
\caption{Line width versus the LSR velocity of the species with most detected transitions. For most transitions in this figure the error bars are smaller than the symbol sizes.}
\label{vlsr_fwhm_species}
\end{figure*}

\begin{figure*}[ht]
\centering
\includegraphics[width=17 cm, trim=0cm 0cm 0cm 0cm,clip=true]{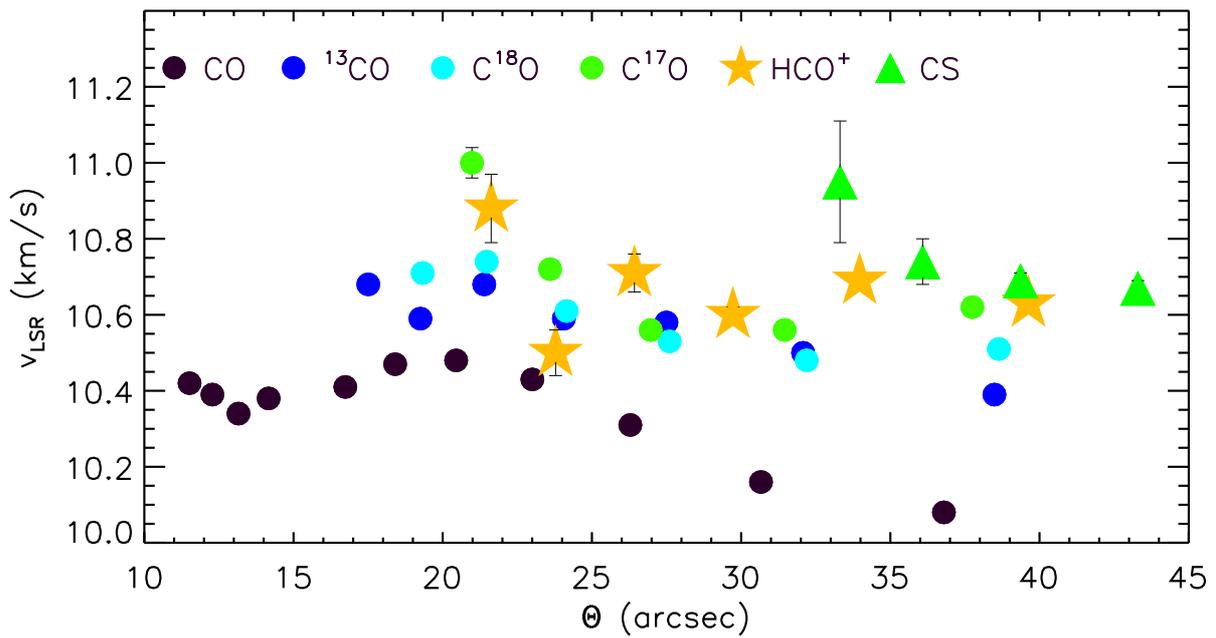}
\caption{ LSR velocity versus the beam size of the species with most detected transitions.
For most transitions in this figure the error bars are smaller than the symbol sizes.}
\label{vlsr_beam}
\end{figure*}

Fig. \ref{vlsr_fwhm_species} shows the line width--velocity plot for the species with the highest number of  detected transitions in the line survey. The general trend that the line widths decrease with the velocity as seen in Fig. \ref{vlsr_fwhm} is also seen in the case of individual transitions.
The CS line at a velocity of $\sim$10.95 \kms\ is a marginal (3-$\sigma$) detection, and is also close to the edge of bands 1b and 2a.
Some of the variations in the observed peak velocities may be due to the changing beam size.  
The lower frequency transitions capture more emission of the Orion Ridge region, which is dominated by lower velocity gas with $V_{\rm{LSR}}$ $<$ 9 \kms\ \citep{vandertak2013} compared to that toward the Orion Bar. 
To test whether the changing beam size has a dominant effect on the trend seen in the $V_{\rm{LSR}}$--line width relation, we plot the observed LSR velocities of the molecules with the most detected transitions as a function of the beam size (Fig. \ref{vlsr_beam}). For some of the plotted molecules there indeed seems to be a decrease in the velocity with increasing beam size as a result of capturing more emission from the Orion Ridge region. These molecules are CO, $^{13}$CO, C$^{18}$O, and CS. The low $J$ CO transitions are dominated by emission from the Orion Ridge. The low $J$ transitions of $^{13}$CO also show a significant contribution from the Orion Ridge.
For C$^{17}$O and HCO$^+$ the velocity does not depend on the beam-size based on the observed transitions as these molecules are not excited in the Orion Ridge.

\begin{figure*}[ht]
\centering
\includegraphics[width=17cm, angle=0]{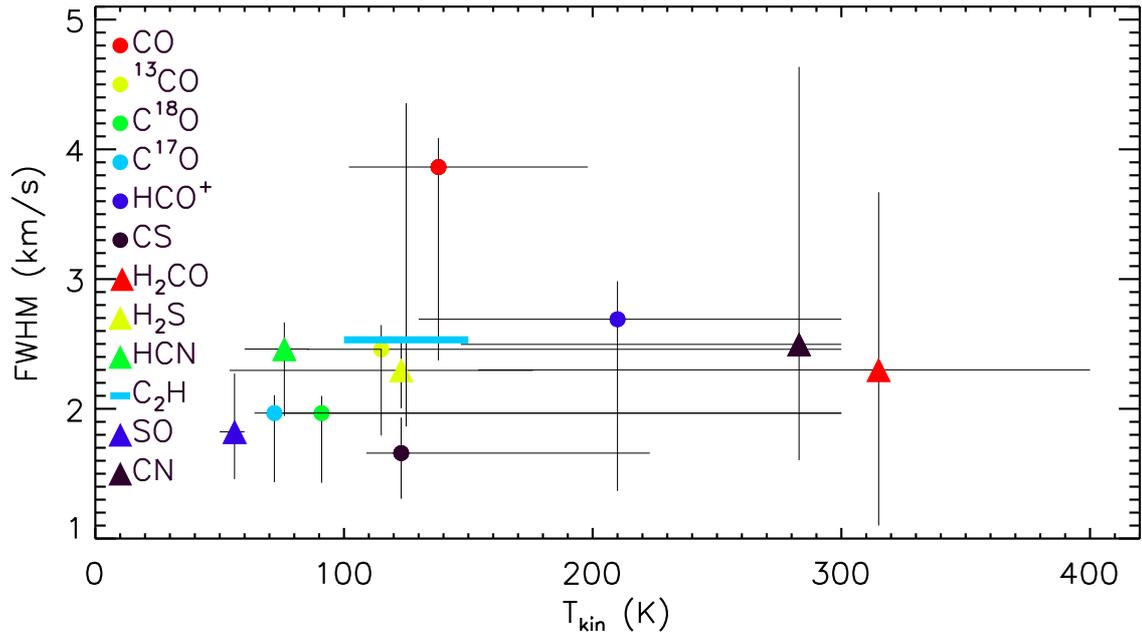}
\caption{Average line widths (weighted by the signal-to-noise ratio of the transitions) of species with kinetic temperature estimates. For C$_2$H the lower kinetic temperature is plotted as it represents most C$_2$H emission for the observed C$_2$H transitions.}
\label{fwhm_kintemp_species}
\end{figure*}

\newpage 
\twocolumn 
 
\section{Rotational diagrams}
\label{appendix_rotdiagram}

For the five C$^{17}$O transitions the rotational diagram gives an excitation temperature estimate of 65$\pm$3 K. The six detected  C$^{18}$O transitions suggest a slightly higher rotational temperature: 71$\pm$2 K. The C$^{18}$O column density of (6.2$\pm$0.2)$\times$10$^{15}$ cm$^{-2}$ is consistent with the value derived from the C$^{17}$O column density ((1.8$\pm$0.1)$\times$10$^{15}$ cm$^{-2}$) and an abundance ratio of $^{18}$O/$^{17}$O=3.2 \citep{wilsonrood1994}. 

The seven $^{13}$CO transitions correspond to a rotational temperature higher than that of the C$^{17}$O and C$^{18}$O lines: 92$\pm$3~K. Its column density is (4.7$\pm$0.2)$\times$10$^{16}$ cm$^{-2}$.
It is expected that at least some of the 11 CO transitions are optically thick; therefore, we include a correction for the opacity for CO, as explained in \citet{nagy2015a}, and evaluated Eqn. 1 in \citet{nagy2015a} for a set of CO column densities, excitation temperatures, and opacities. We assumed a uniform beam filling, as for the rotational diagrams without opacity correction. This method results in an excitation temperature of 137$^{+16}_{-14}$ K, and a best fit CO column density of 4.5$\times$10$^{17}$ cm$^{-2}$. This fit suggests that the CO 5-4,...,11-10 transitions are optically thick and have opacities in the range between 1.2 and 2.4. 
The CO column density of 4.5$\times$10$^{17}$ cm$^{-2}$ is about a factor of 10 below the $3.2\times10^{18}$ cm$^{-2}$ value suggested by the C$^{17}$O column density and the abundance ratios of $^{16}$O/$^{18}$O=560 and $^{18}$O/$^{17}$O=3.2 \citep{wilsonrood1994}. 
The difference between the CO column density derived from the C$^{17}$O column density and the value given by the CO rotation diagram is consistent with the expected high optical depth of $^{12}$CO emission. Non-uniform beam filling with different emitting region sizes for CO and C$^{17}$O could also contribute to the differences.

A CO column density of $3.2\times10^{18}$ cm$^{-2}$ (based on the C$^{17}$O column density and the abundance ratios quoted above) and a CO abundance of 1.1$\times$10$^{-4}$ \citep{johnstone2003} suggest an H$_2$ column density of 2.9$\times$10$^{22}$ cm$^{-2}$. The difference between this value and the 2.1$\times$10$^{22}$ cm$^{-2}$ derived in \citet{nagy2015b} is due to re-calculating the C$^{17}$O column density using the line intensities scaled by the updated HIFI main beam efficiencies (see Sect. \ref{sect:obs}). 
An H$_2$ column density of 2.9$\times$10$^{22}$ cm$^{-2}$ is consistent with the 3$\times$10$^{22}$ cm$^{-2}$ value found by \citet{cuadrado2015} toward a position close to the CO$^+$ peak. \citet{vanderwiel2009} measured an H$_2$ column density of 10$^{23}$ cm$^{-2}$ toward a position further away from the CO$^+$ peak.

The nine detected o-H$_2$CO transitions are consistent with a rotational temperature of 146$\pm$48 K which is consistent with the rotational temperature derived from the CO transitions. The corresponding column density is $(1.5\pm0.5)\times10^{12}$ cm$^{-2}$.
 
For HCO$^+$ the rotational diagram suggests an excitation temperature of 46$\pm$3 K, and a column density of $(5.4 \pm 0.3) \times 10^{12}$ cm$^{-2}$. This column density is consistent with that derived by \citet{jansen1995} toward some of their observed positions.
 
The rotational diagram of the four detected CS transitions results in an excitation temperature of 39$\pm$2 K and a column density of (1.0$\pm$0.1)$\times$10$^{13}$ cm$^{-2}$. 

An even lower value, 27$\pm$10 K is estimated for o-H$_2$S. A similarly low excitation temperature, 22$\pm$4, is suggested for CF$^+$, when creating a rotational diagram using the 5-4 transition observed in this line survey and the 3-2, 2-1, and 1-0 transitions previously observed by \citet{neufeld2006}. The corresponding column density is (1.8$\pm$0.3)$\times$10$^{12}$ cm$^{-2}$.
The difference between the value quoted in \citet{nagy2013a} and in this paper is due to the re-reduction of the data with a more recent HIPE version, which affects the intensity of the low signal-to-noise CF$^+$ 5-4 transition. The rotation temperature value quoted in \citet{nagy2013a} was 32~K, and the column density 2.1$\times$10$^{12}$ cm$^{-2}$.

The three detected HCN transitions are also consistent with a low excitation temperature: 24$\pm$2~K, and a column density of (1.1$\pm$0.1)$\times$10$^{13}$ cm$^{-2}$. The low column density suggested by the rotational diagram compared to that estimated using RADEX (see Sect. \ref{sect_radex}) suggests non-LTE conditions, which is expected as a result of the high critical densities of the HCN transitions, for example 10$^8$ cm$^{-3}$ for the 6-5 transition.

The seven detected CN transitions correspond to three rotational transitions. These transitions suggest a rotational temperature of 27$\pm$8 K and a CN column density of $(4.7\pm1.3)\times10^{13}$ cm$^{-2}$. The [CN]/[HCN] column density ratio based on the rotational diagrams is $\sim$4 which is similar to previous measurements (e.g. \citealp{jansen1995}).

Six H$_2$O transitions were detected in this line survey, with three ortho and three para transitions. The excitation temperatures have large errors in both cases, especially for p-H$_2$O, but the o/p ratio based on the two column densities is $\sim$0.14, which is consistent with the non-LTE estimate of \citet{choi2014}.
The large scatter seen in the H$_2$O rotational diagrams confirms that non-LTE effects are important in the excitation of H$_2$O.   

\begin{figure*}[ht]
\begin{center}
\includegraphics[width=4.7cm, angle=-90]{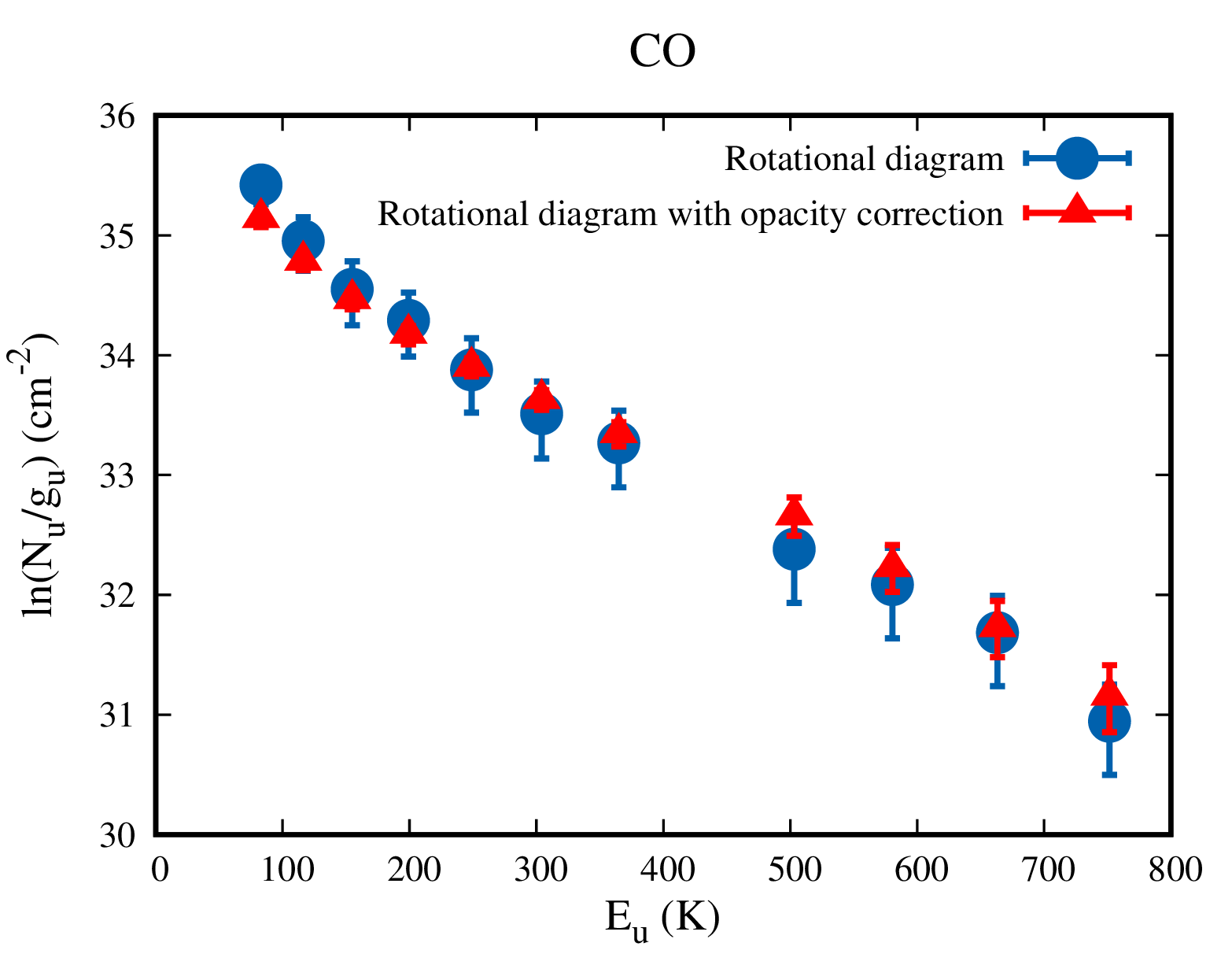}
\includegraphics[width=4.7cm, angle=-90]{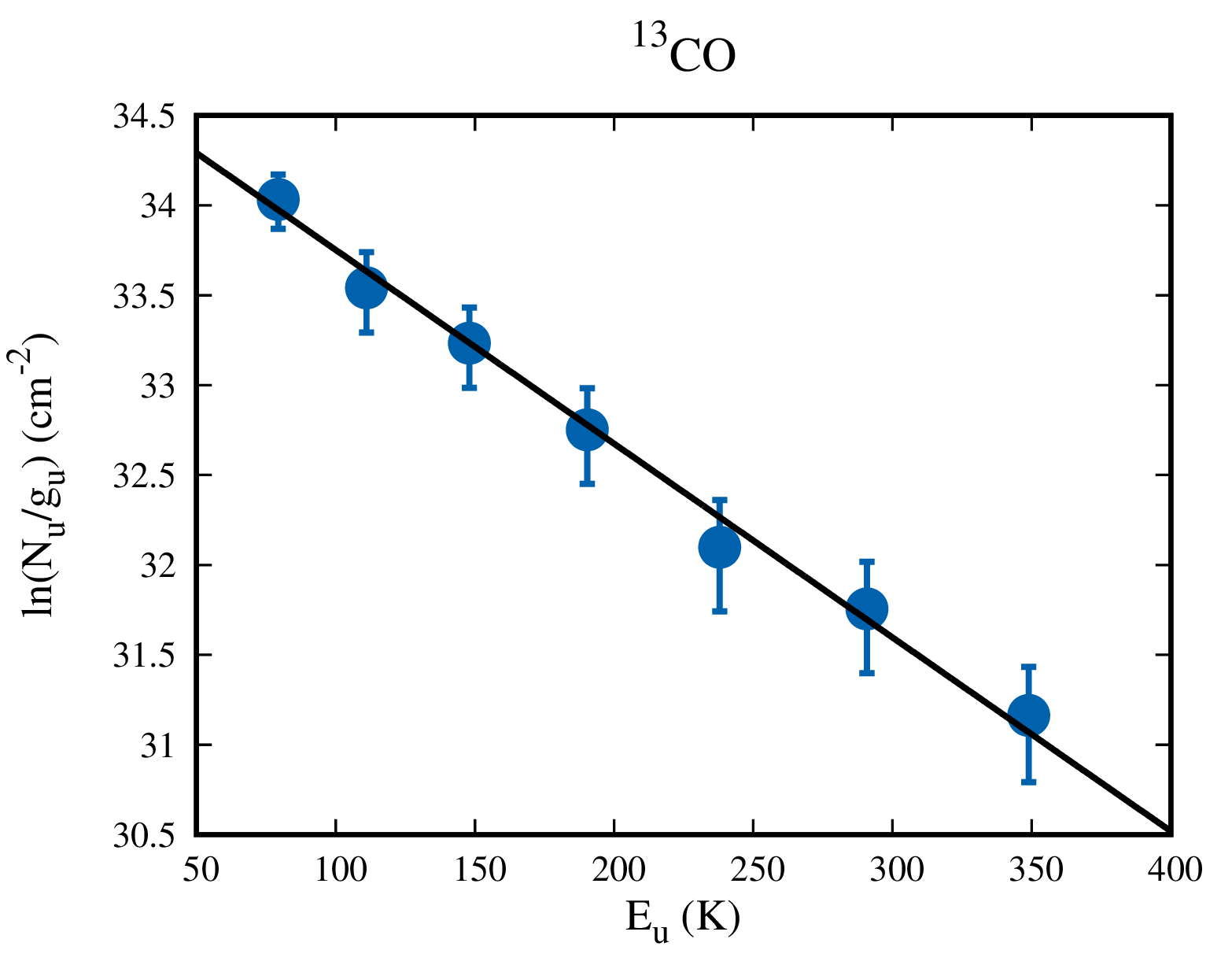} 
\includegraphics[width=4.7cm, angle=-90]{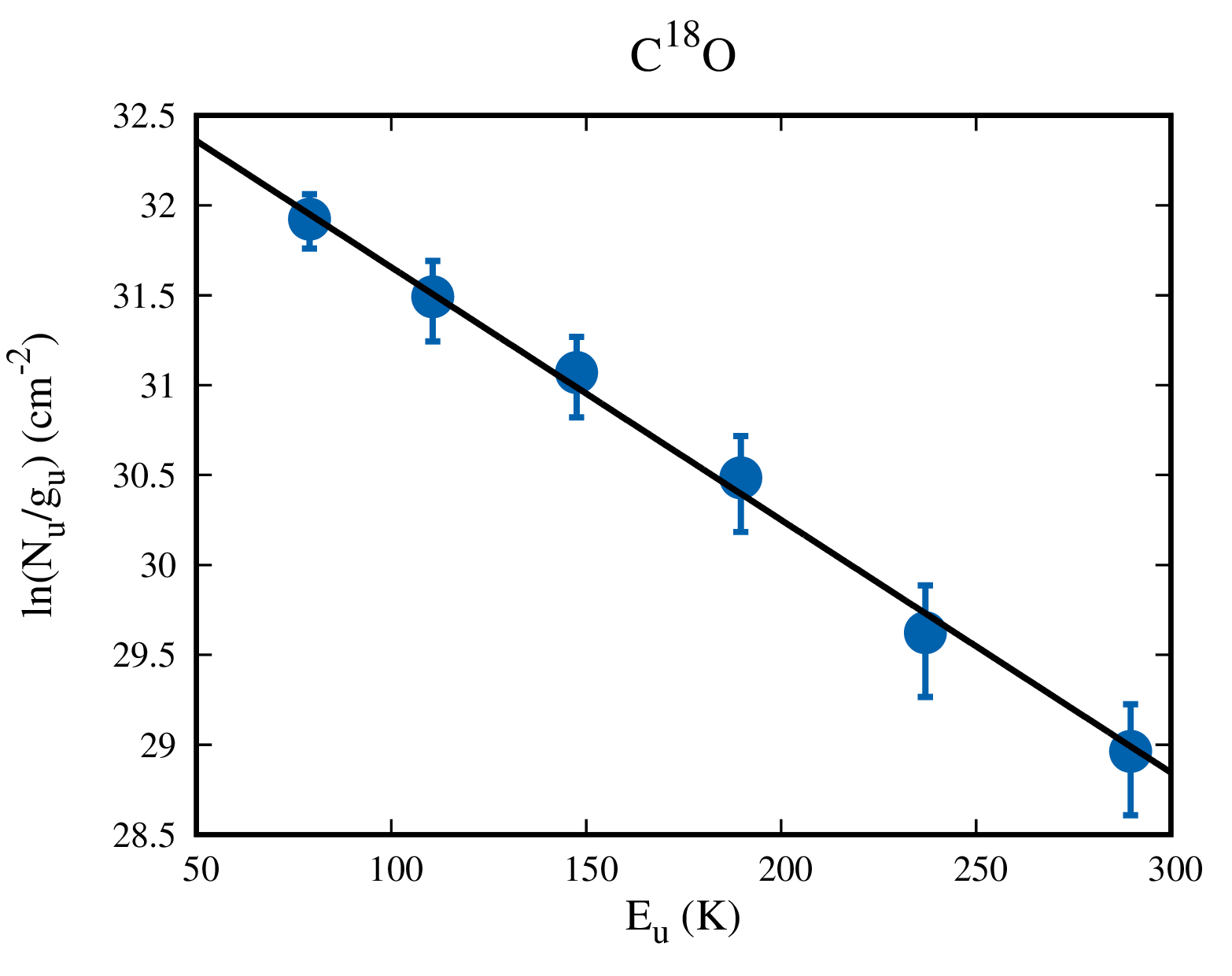}
\includegraphics[width=4.7cm, angle=-90]{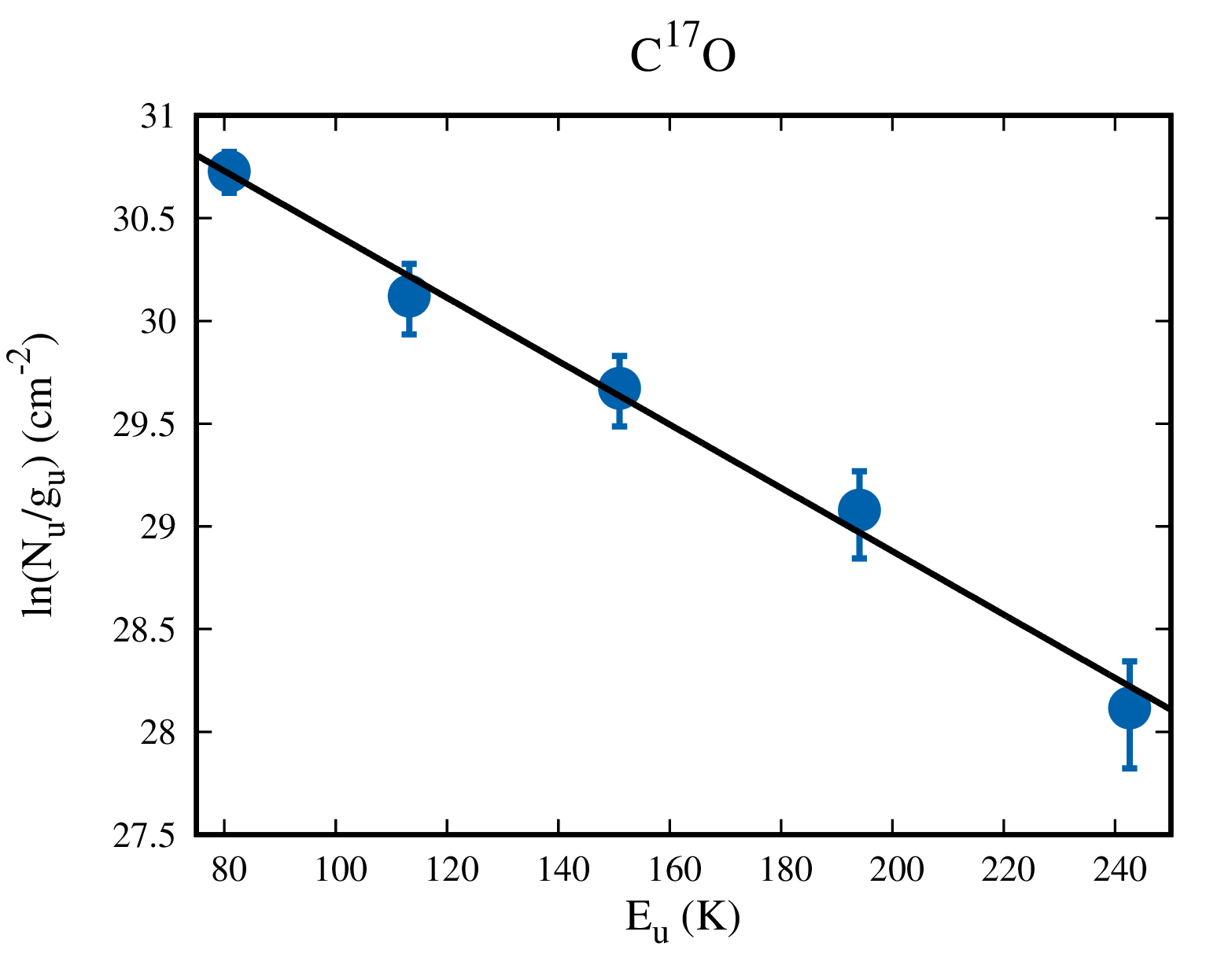}
\includegraphics[width=4.7cm, angle=-90]{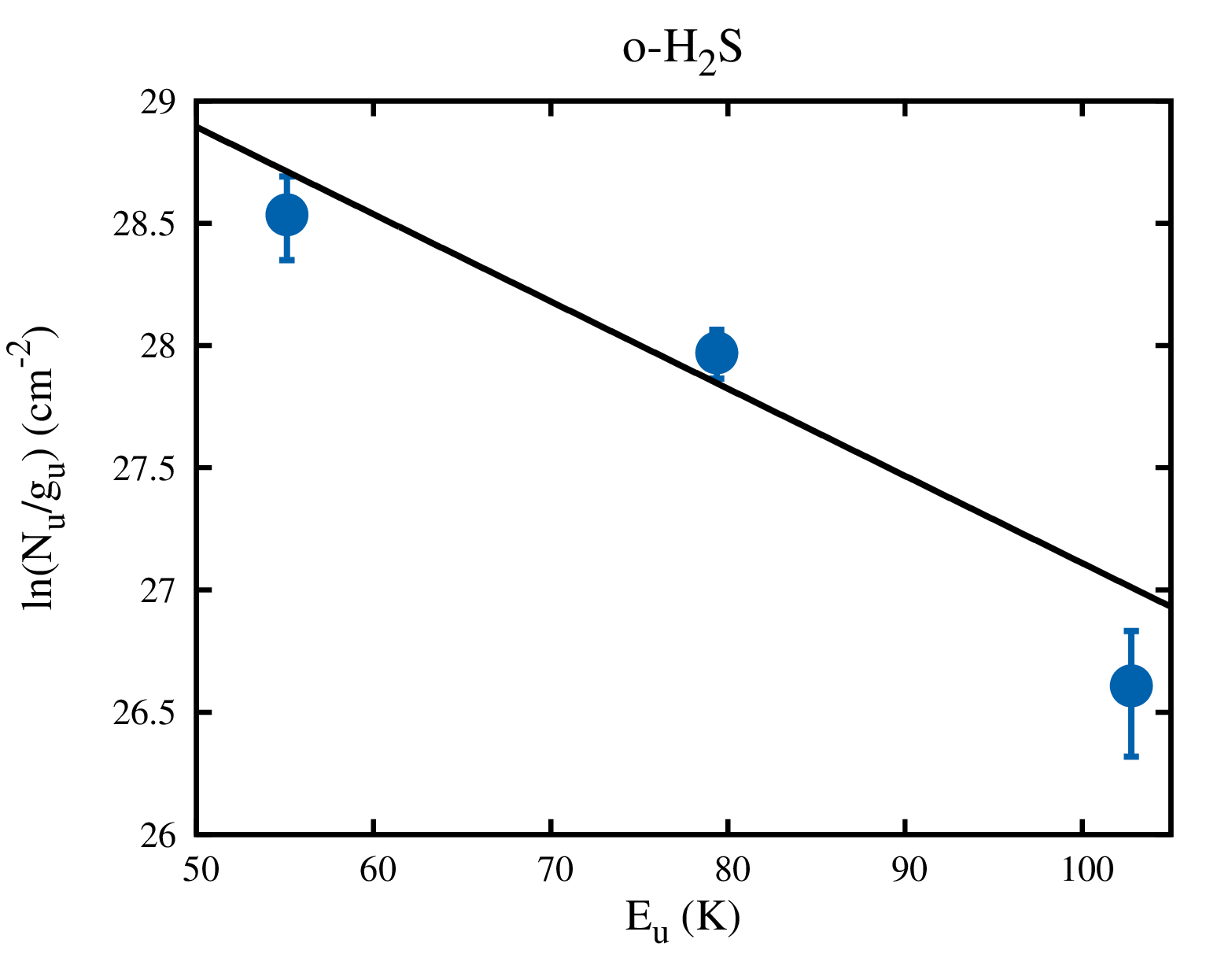}
\includegraphics[width=4.7cm, angle=-90]{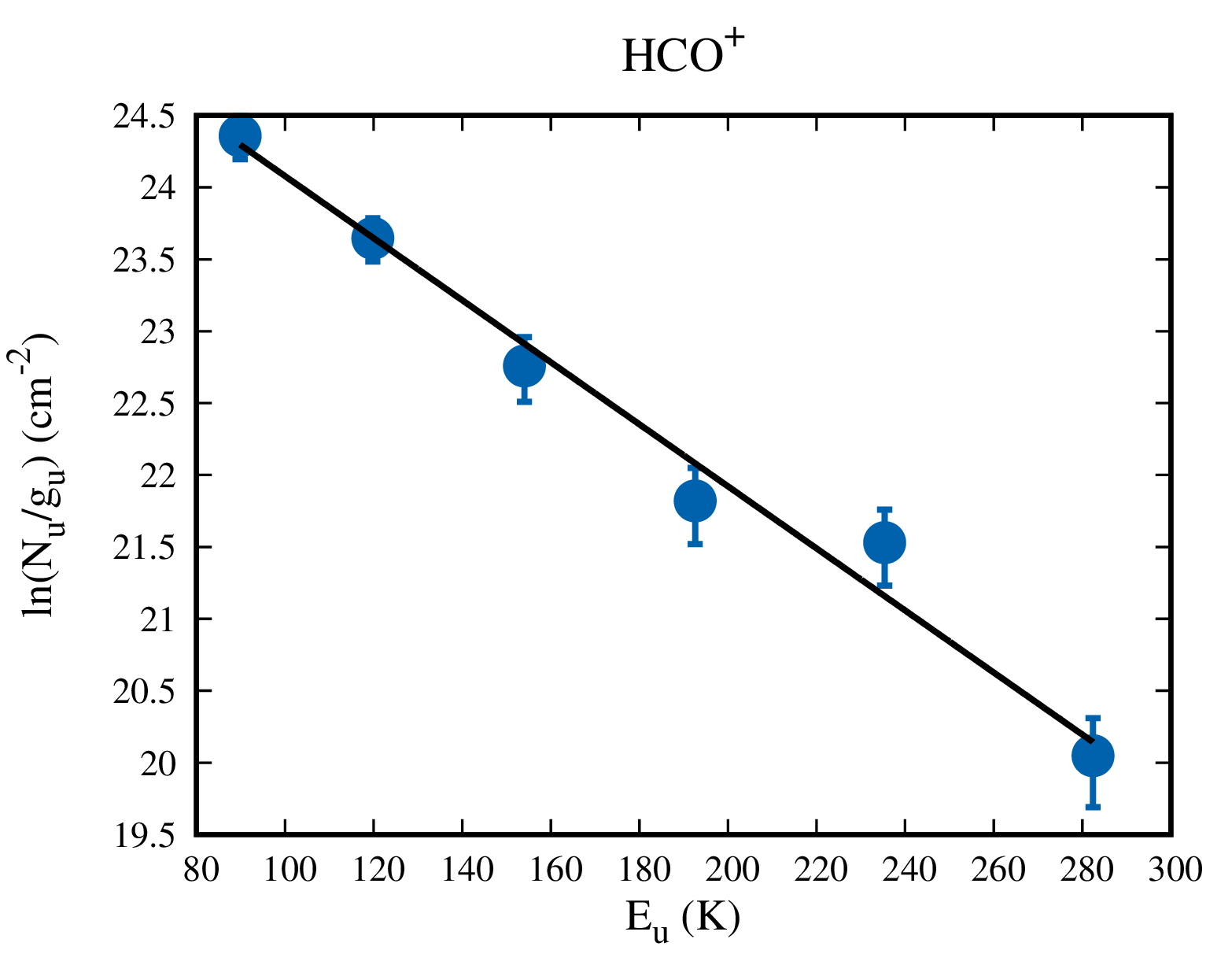}
\includegraphics[width=4.7cm, angle=-90]{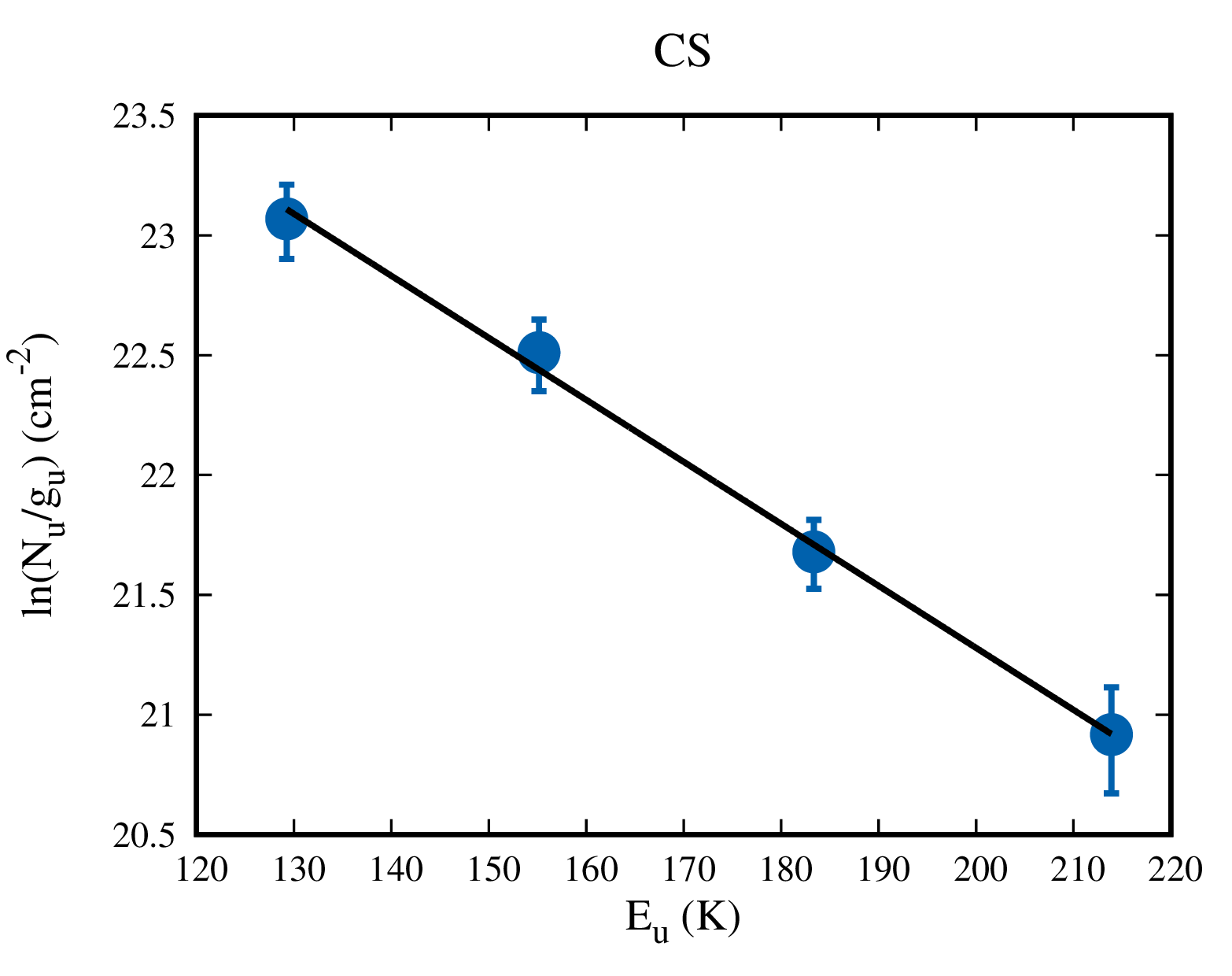}  
\includegraphics[width=4.7cm, angle=-90]{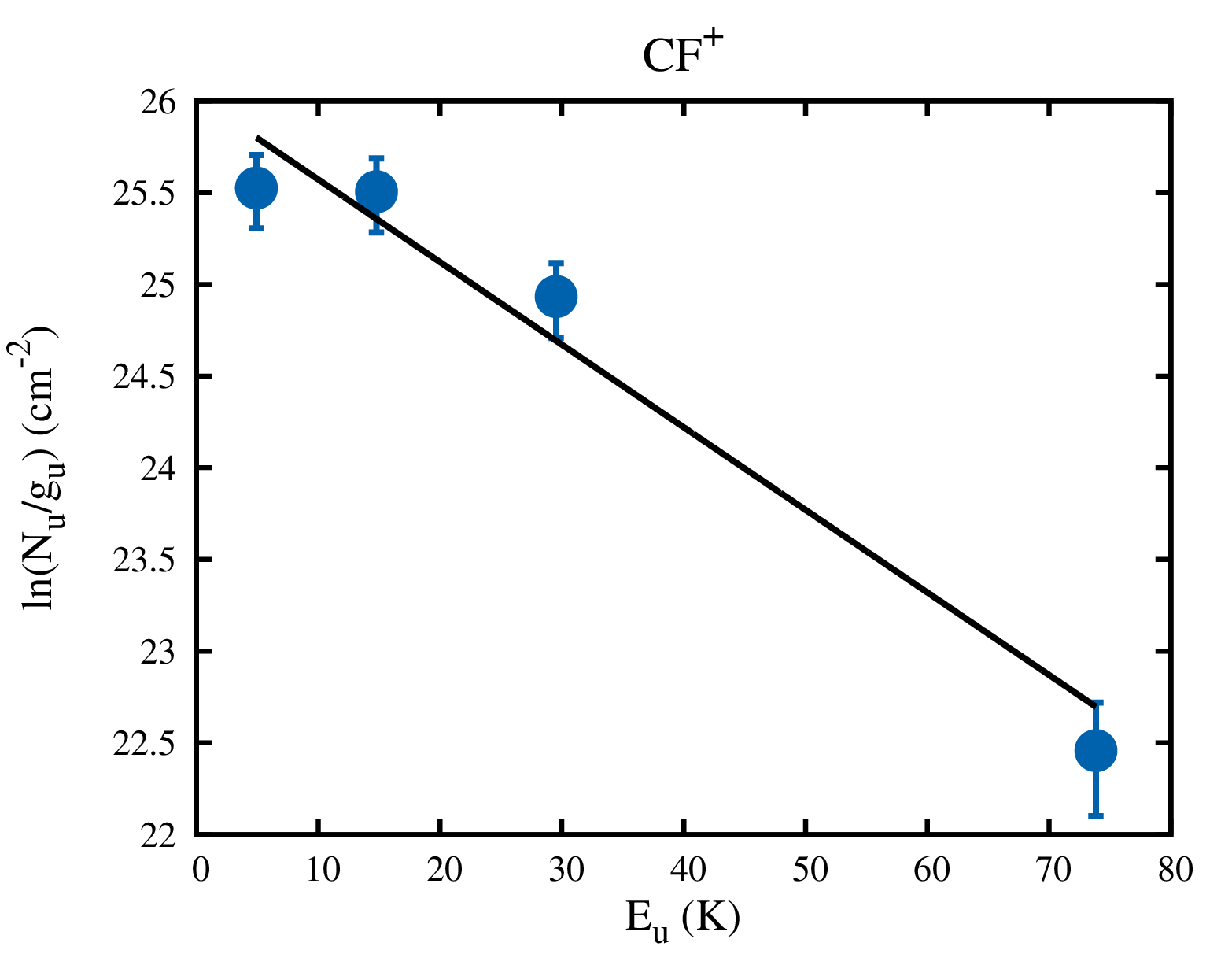} 
\includegraphics[width=4.7cm, angle=-90]{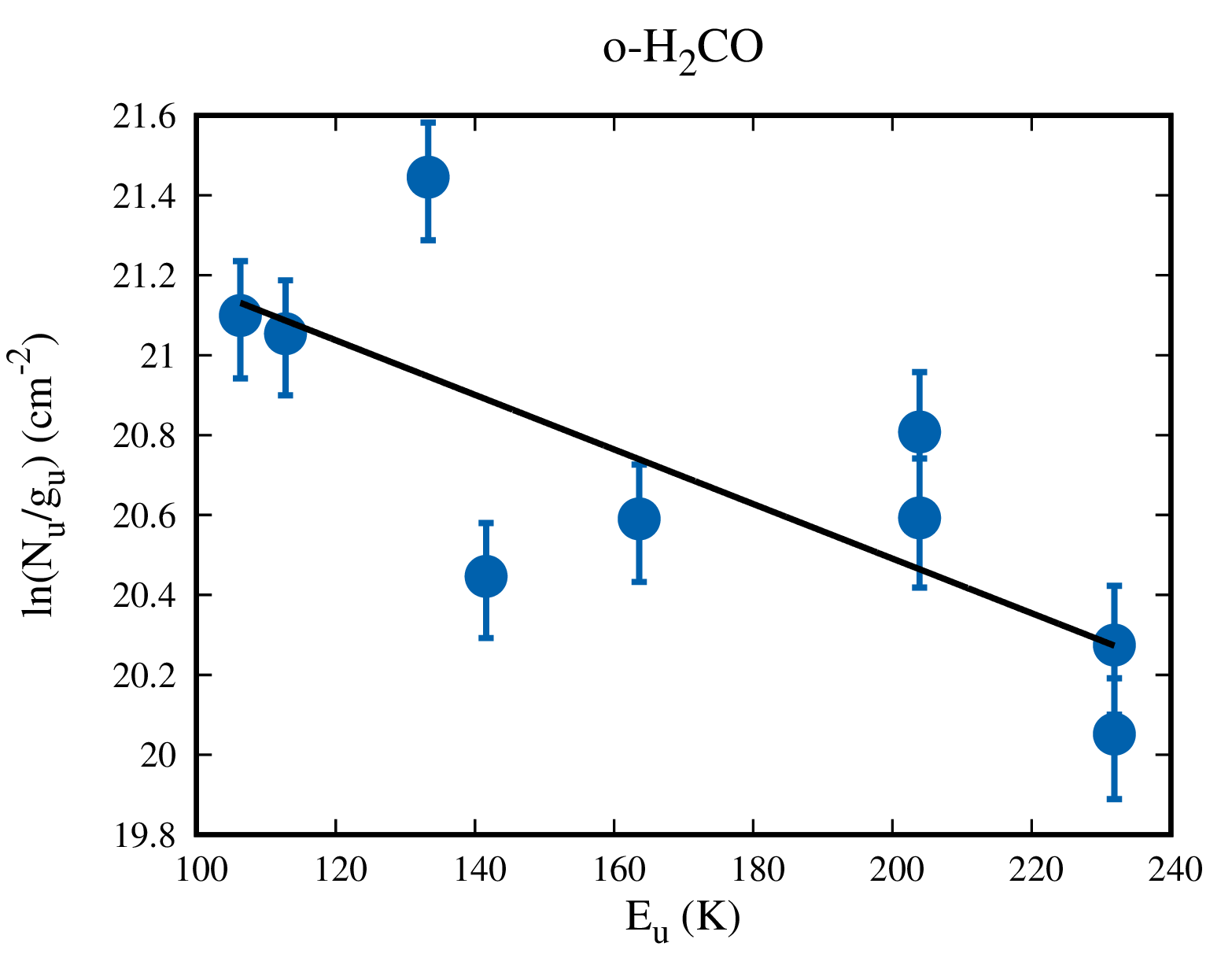}         
\includegraphics[width=4.7cm, angle=-90]{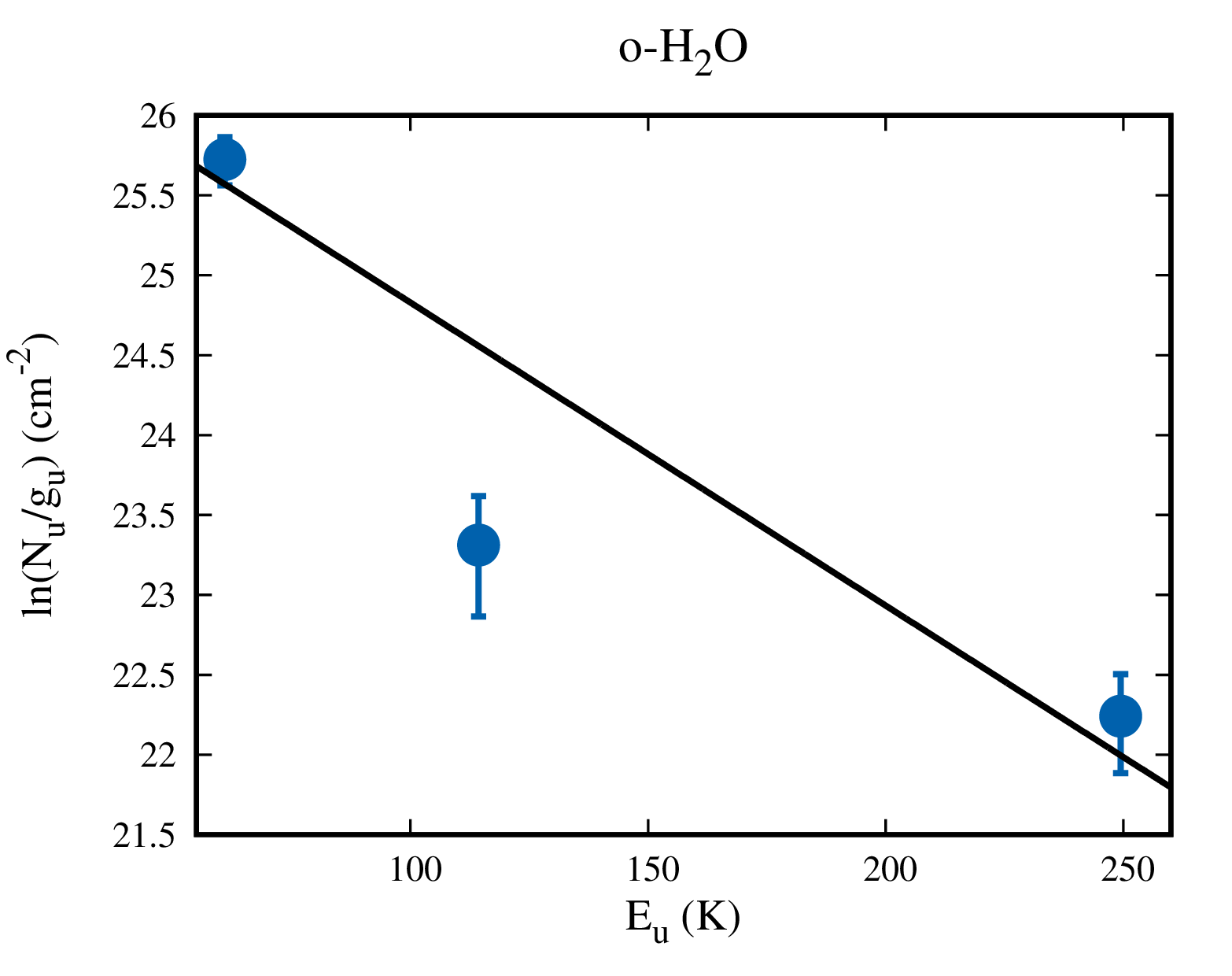}  
\includegraphics[width=4.7cm, angle=-90]{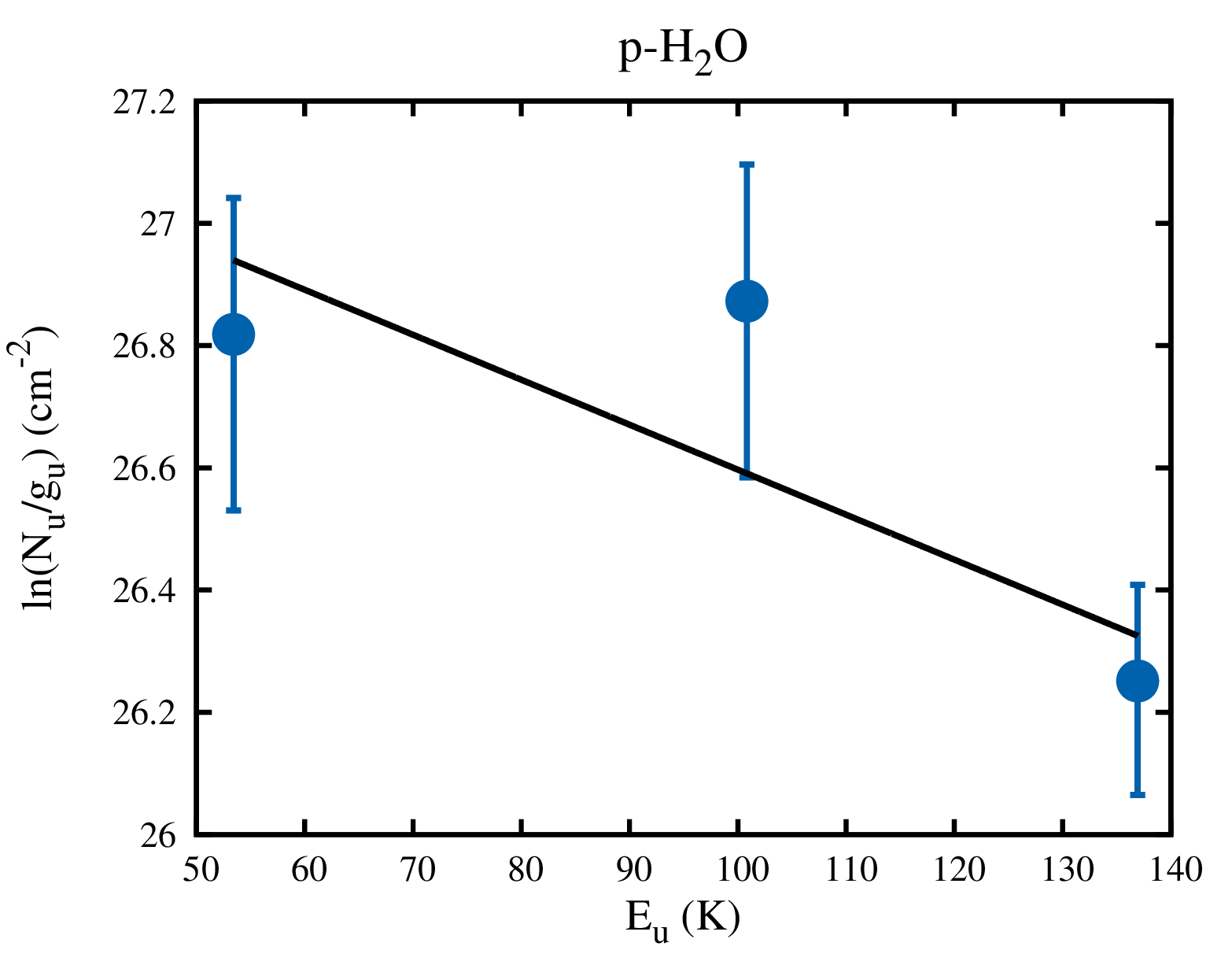} 
\includegraphics[width=4.7cm, angle=-90]{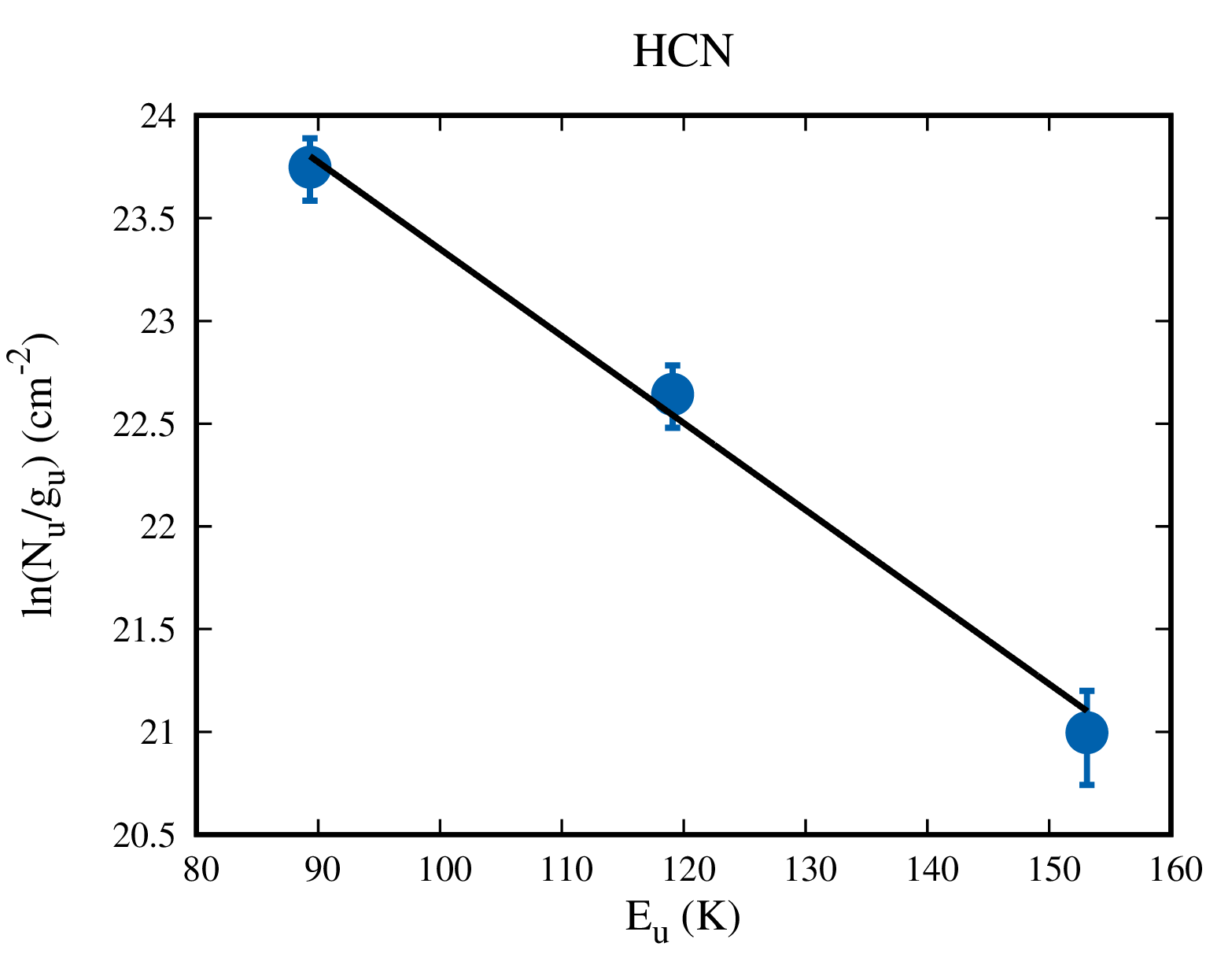}  
\includegraphics[width=4.7cm, angle=-90]{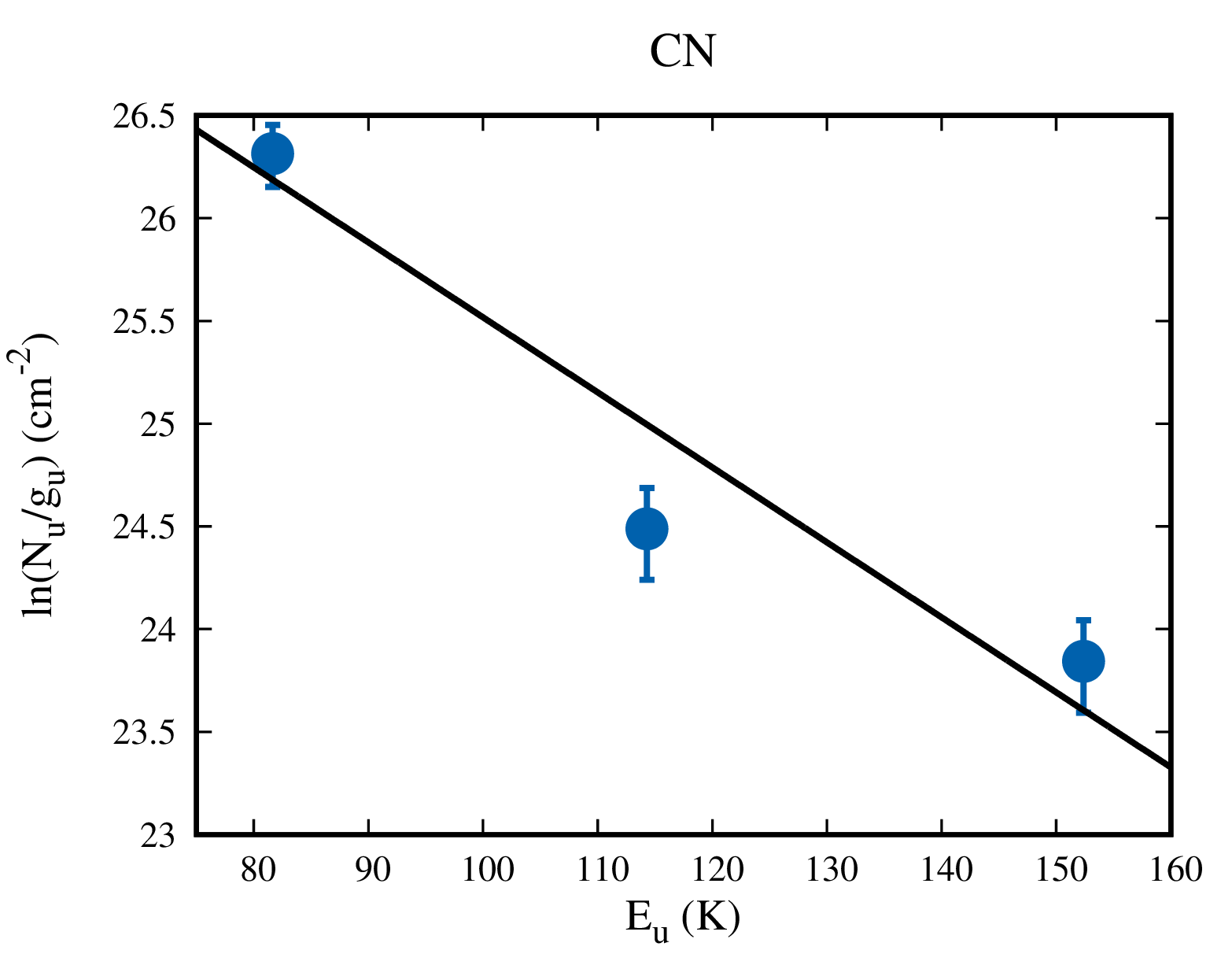}  
\caption{Rotational diagrams of molecules with more than three detected transitions. The red triangles for CO have been corrected for the effect of  opacity.}
\label{rot_diagram}
\end{center}
\end{figure*} 

\section{RADEX results}
\label{appendix_radex}

To model the CO line intensities, we adopt a CO column density based on the C$^{17}$O rotational diagram result and isotopic ratios of C$^{18}$O/C$^{17}$O of 3.2 and CO/C$^{18}$O of 560 \citep{wilsonrood1994}.
With the adopted column density, the CO line intensities are consistent with an H$_2$ volume density of $1.6\times$10$^5$ cm$^{-3}$ and a kinetic temperature of 138~K. This temperature is similar to the excitation temperature derived by the rotational diagram method within the given error bars. This temperature is also close to the value of $\sim$150~K found by \citet{batrlawilson2003} near the ionization front of the Orion Bar.

The H$_2$ volume density of $3.4\times$10$^5$ cm$^{-3}$ corresponding to the $^{13}$CO transitions is similar to the density derived from the CO transitions. The best fit kinetic temperature of 115~K is lower than the kinetic temperature implied by the CO transitions, but higher than the $^{13}$CO excitation temperature of 93$\pm$3~K resulted from the rotational diagram. The $^{13}$CO column density in the fit was slightly higher than the value given by the rotational diagram, in order to better fit the intensity of the lowest-$J$ $^{13}$CO transition covered by this line survey.
 
The observed C$^{18}$O transitions suggest a similar density to that of  CO and $^{13}$CO, with a best fit value of 3.4$\times$10$^5$ cm$^{-3}$. The best fit kinetic temperature of 91~K is below the value given for the $^{13}$CO fit, but above the excitation temperature of 71$\pm$2 given by the rotational diagram fit.

The five detected C$^{17}$O transitions are consistent with a kinetic temperature of 72~K which is close to the average kinetic temperature of 85~K \citep{hogerheijde1995}. 
The corresponding H$_2$ volume density is $8.6\times10^5$ cm$^{-3}$. The difference between the best fit H$_2$ volume density of C$^{17}$O and those given by the other CO isotopologues is likely due to the lower number of C$^{17}$O transitions. The H$_2$ volume densities derived from the other CO isotopologues fall in the 1-$\sigma$ limit range derived for the C$^{17}$O fit.

The six detected HCO$^+$ transitions are most consistent with a kinetic temperature of 210~K, which is much higher than the excitation temperature of 46$\pm$3 K given by the rotational diagram method. The best fit H$_2$ volume density given by the HCO$^+$ transitions is the average H$_2$ volume density toward the Orion Bar, $1.0\times$10$^5$ cm$^{-3}$.

The four detected CS transitions suggest a kinetic temperature of 123~K, much above the 39$\pm$2 K excitation temperature given by the rotational diagram method. The best fit H$_2$ volume density is $8.6\times10^5$ cm$^{-3}$.

To get a better fit for the observed o-H$_2$CO transitions we extended the search range of temperatures and densities to 50-400 K and 10$^4$-10$^7$ cm$^{-3}$, respectively. The nine detected o-H$_2$CO transitions are most consistent with a very high-temperature and high-density component, 315 K and $1.4 \times 10^6$ cm$^{-3}$. The relative intensities of the transitions are not well fitted by the RADEX model.

We use the three detected o--H$_2$S transitions to constrain the physical parameters of the H$_2$S emitting region. The best fit H$_2$ volume density of $1.4\times10^5$ cm$^{-3}$ and kinetic temperature of 123 K are similar to those derived from the RADEX models for CO and $^{13}$CO.

The seven detected CN lines which correspond to three rotational transitions suggest a best fit kinetic temperature of 283 K, which is similar to the value obtained for o-H$_2$CO. 
The best fit H$_2$ volume density of $1.8\times10^5$ cm$^{-3}$ is close to the values derived from the CO isotopologues and HCO$^+$.
The chemically related molecule HCN requires a component with a high density ($1.4^{+0.7}_{-0.2}\times10^6$ cm$^{-3}$) but a moderate temperature ($76^{+10}_{-16}$ K). The volume density above 10$^6$ cm$^{-3}$ H$_2$  is due to the fact that we extended the upper limits of the search range in the case of HCN to get an idea of the error bars of its best fit density.
The [CN]/[HCN] column density ratio in these models is $\sim$3, which is close to what was measured earlier toward the Orion Bar (e.g. \citealp{jansen1995}).

Among the molecules discussed in this paper, SO is consistent with the lowest kinetic temperature, $56^{+4}_{-6}$ K. This is similar to the 60~K used by \citet{choi2014} when deriving parameters for the observed H$_2^{18}$O transitions. The best fit H$_2$ volume density is $4.6^{+0.8}_{-0.7}\times10^5$ cm$^{-3}$, similar to the values derived from the $^{13}$CO and C$^{18}$O transitions.
As no rotational diagram result is available as a first guess for the SO column density, we ran models with several different column densities in the range given by the excitation temperatures used in the LTE calculation. The temperature and density shown above correspond to the best fit.

The four observed NO transitions which correspond to two rotational transitions suggest an H$_2$ volume density of 2.9$\times$10$^5$ cm$^{-3}$, close to that derived for the CO isotopologues.
The kinetic temperature is not well constrained by the fit, and is in the range between 150 and 300 K.

The o-NH$_3$ (J, K)=(1, 0)$\rightarrow$(0, 0) transition was observed with a 2$'$ beam using Odin by \citet{larsson2003}, who conclude an NH$_3$ abundance of NH$_3$/H$_2$=$5 \times 10^{-9}$. This abundance and an $N$(H$_2$)=$2.2\times10^{22}$ cm$^{-2}$ is equivalent to $N$(NH$_3$)=$1.1\times10^{14}$ cm$^{-2}$. Assuming a kinetic temperature of $\sim$145 K \citep{wilson2001}, the NH$_3$ line emission observed with HIFI originates in a $\sim2\times10^{5}$ cm$^{-3}$ density gas component.

Assuming typical Orion Bar conditions ($T_{\rm{kin}}=$100~K and $n$(H$_2$)=10$^5$ cm$^{-3}$) for HNC, the intensity of the observed ($J$=6-5) HNC transition can be reproduced by a column density of $5\times10^{12}$ cm$^{-2}$, which is slightly above the LTE estimate.

Two transitions of [C{\sc{i}}] are covered by the HIFI line survey. Based on the comparison of the observed ratio of the transitions to those calculated with RADEX the best fit kinetic temperature is 86~K and the best fit H$_2$ volume density is $8.6\times10^4$ cm$^{-3}$ when assuming a [C{\sc{i}}] column density of 10$^{18}$ cm$^{-2}$. These parameters are not well constrained, as is shown by the large error bars in Table \ref{coldens}.

Five transitions of CH, which correspond to two rotational transitions, were detected in this line survey. The available collision rates \citep{marinakis2015} did not take into account the hyperfine structure of CH; therefore, the temperature and density of the gas traced by CH can only be constrained by two rotational transitions. We searched for the temperatures and densities which fit the observed line intensity ratio best. The obtained temperature and density cover a broad range of parameters with temperatures of $50_{-0}^{+250}$ K and densities of $8.6^{+90.1}_{-7.6} \times 10^4$ cm$^{-3}$. Given that these parameters are not well constrained by the two observed transitions, we do not consider this molecule in Sect. \ref{sect:discussion}.

We used the CF$^+$ 5-4 line intensity measured as a part of this line survey and the CF$^+$ 1-0, 2-1, and 3-2 line intensities observed by \citet{neufeld2006} toward a position ($05^{\rm{h}}35^{\rm{m}}22.8^{\rm{s}}$, $-5^\circ25'01''$) close to the CO$^+$ peak. Using the column density obtained from the rotational diagram the H$_2$ volume density given by the fit is consistent with $3.4^{+0.6}_{-2.2}\times10^5$ cm$^{-3}$, like the values traced by the CO isotopologues and HCO$^+$. The kinetic temperature of $50_{-0}^{+250}$ K is not well constrained by the fit, but provides a lower limit on the kinetic temperature of the gas traced by CF$^+$.

\begin{figure*}[ht]
\begin{center}
\includegraphics[width=4.7cm, angle=-90]{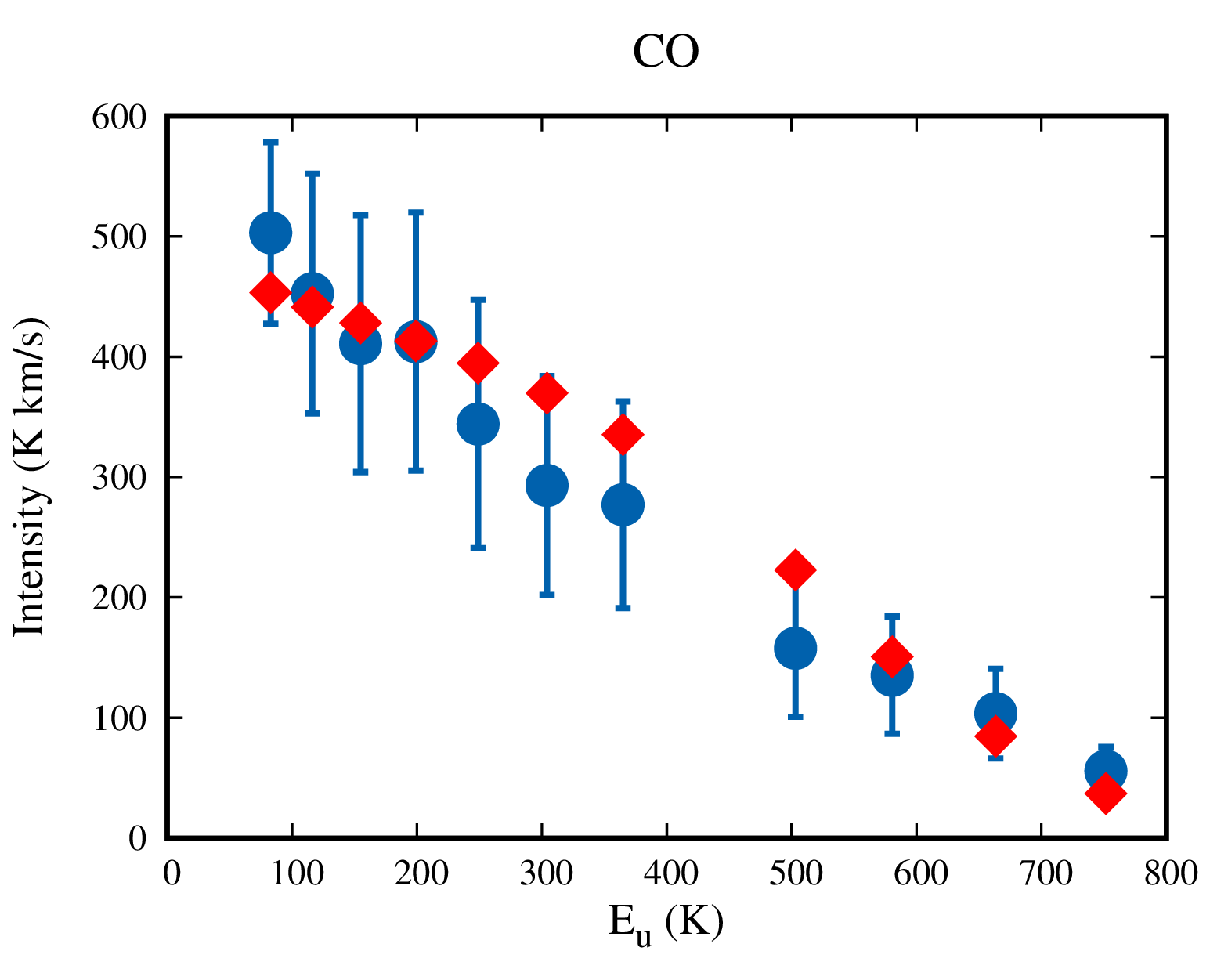}
\includegraphics[width=4.7cm, angle=-90]{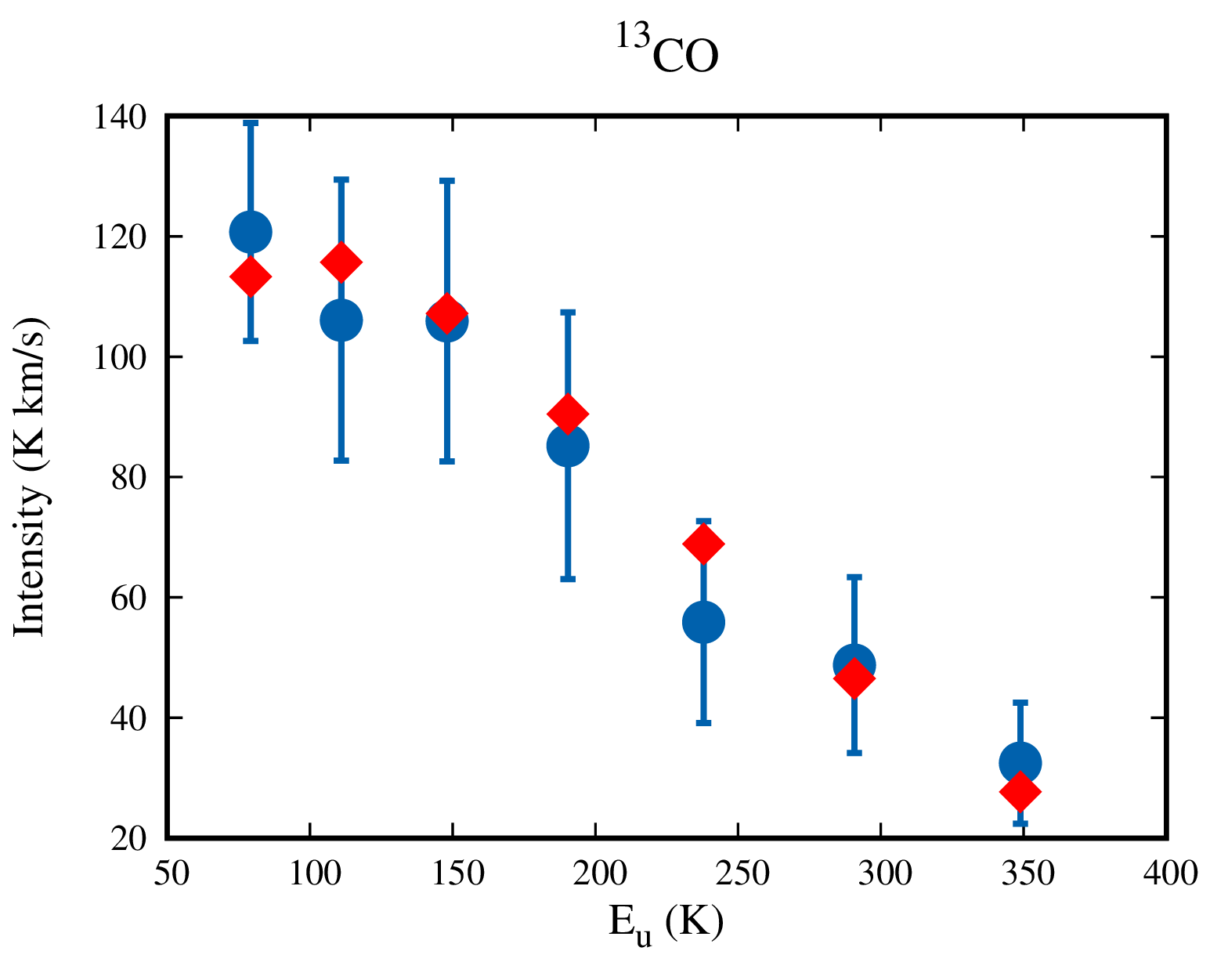}
\includegraphics[width=4.7cm, angle=-90]{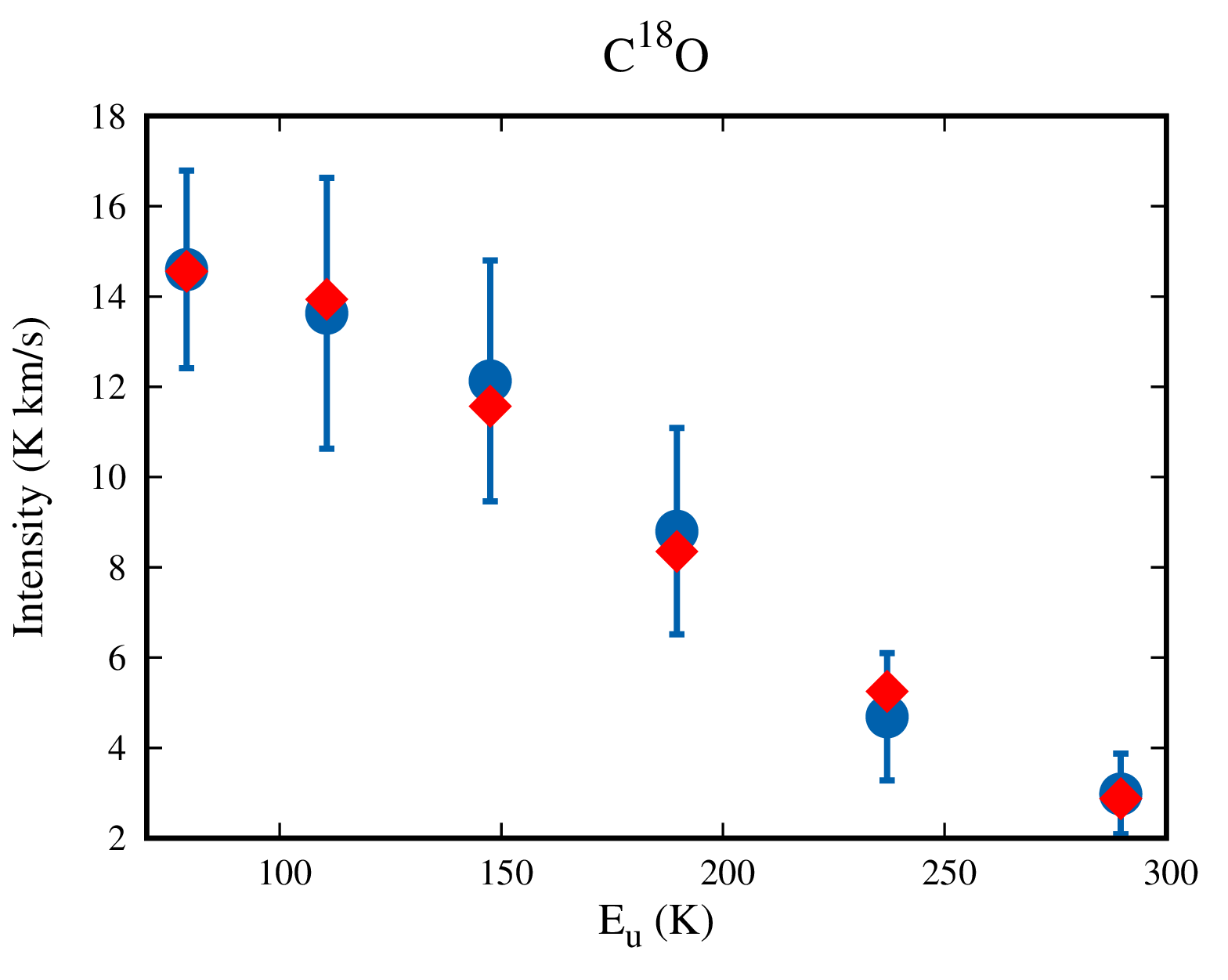}
\includegraphics[width=4.7cm, angle=-90]{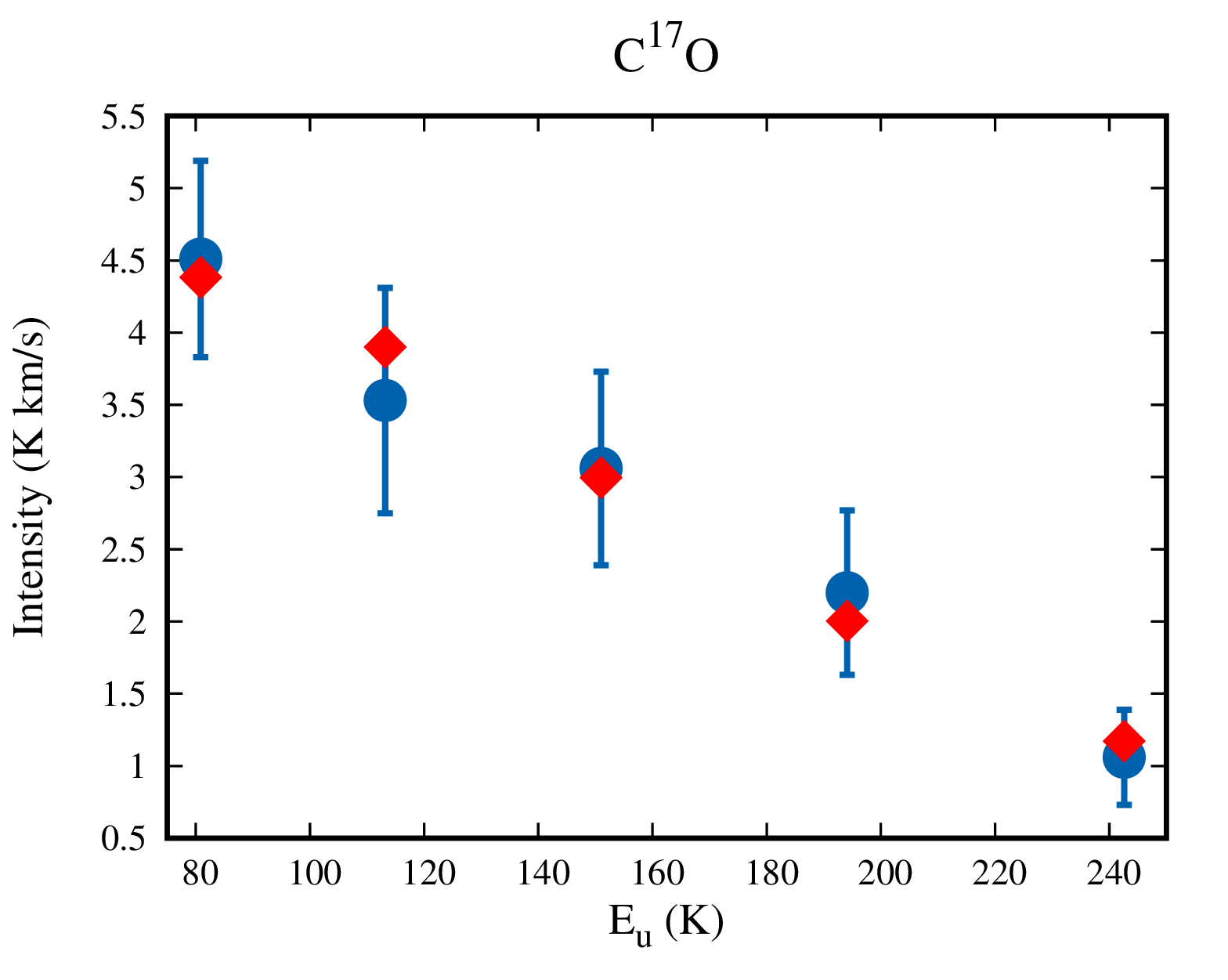}
\includegraphics[width=4.7cm, angle=-90]{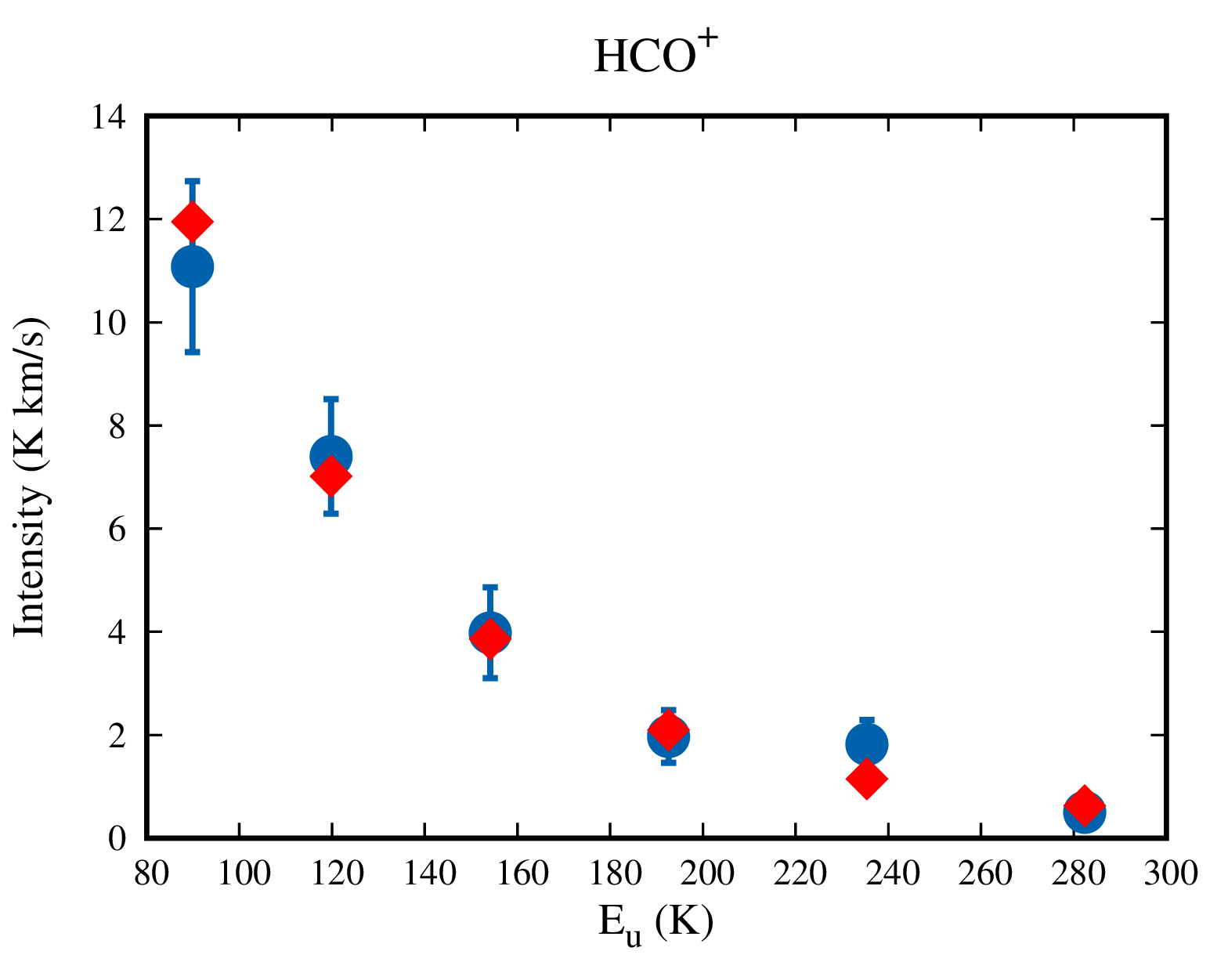}        
\includegraphics[width=4.7cm, angle=-90]{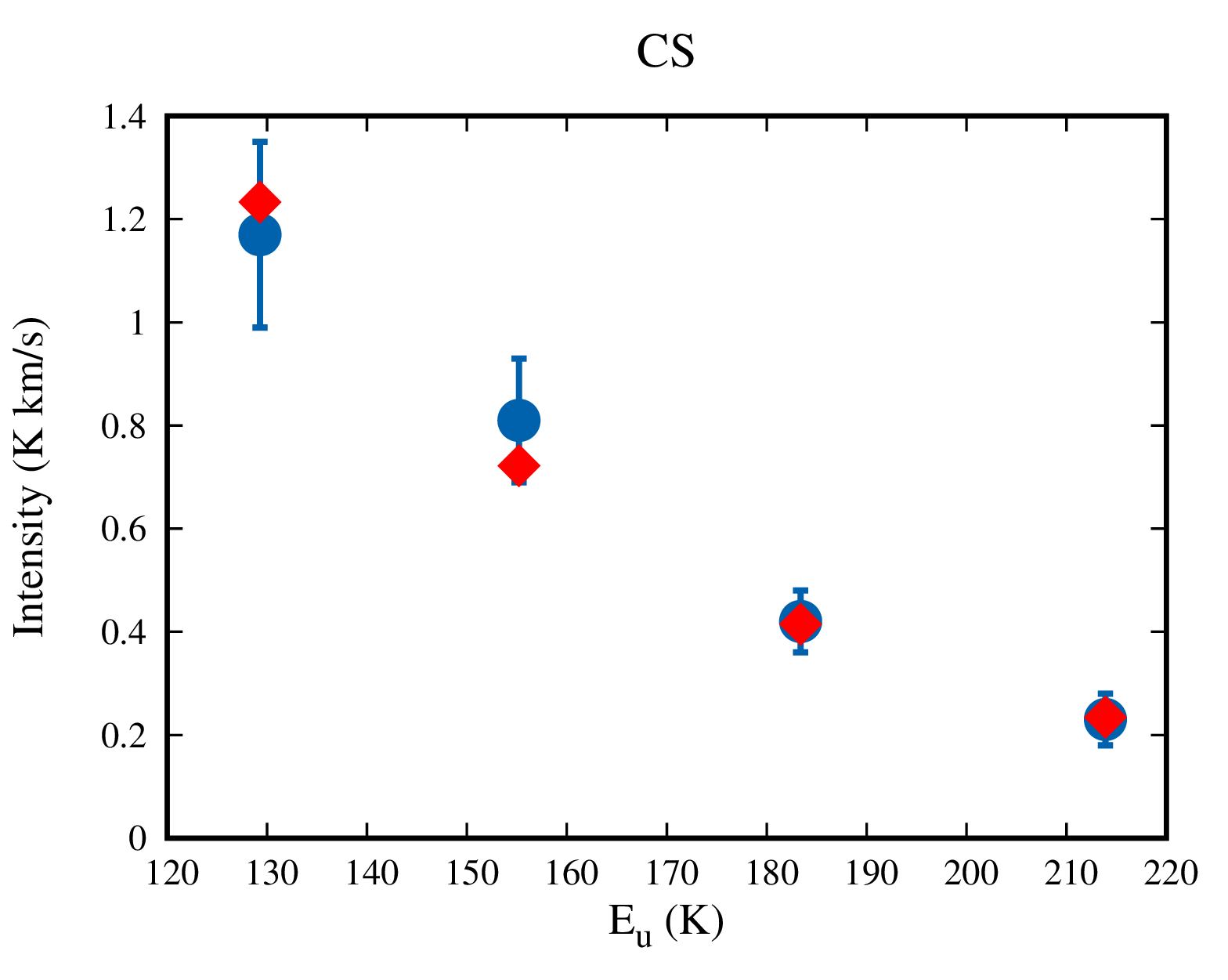}
\includegraphics[width=4.7cm, angle=-90]{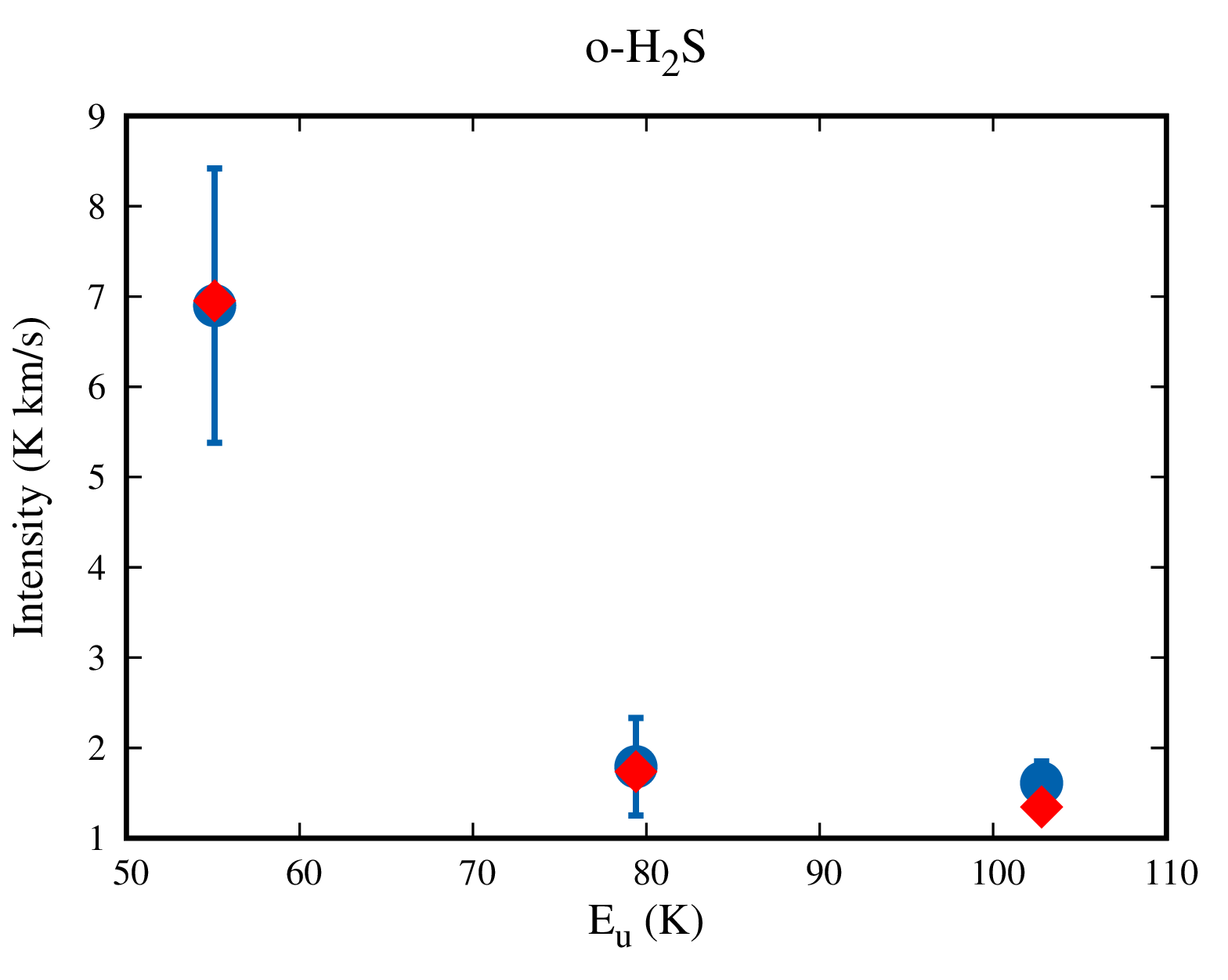}
\includegraphics[width=4.7cm, angle=-90]{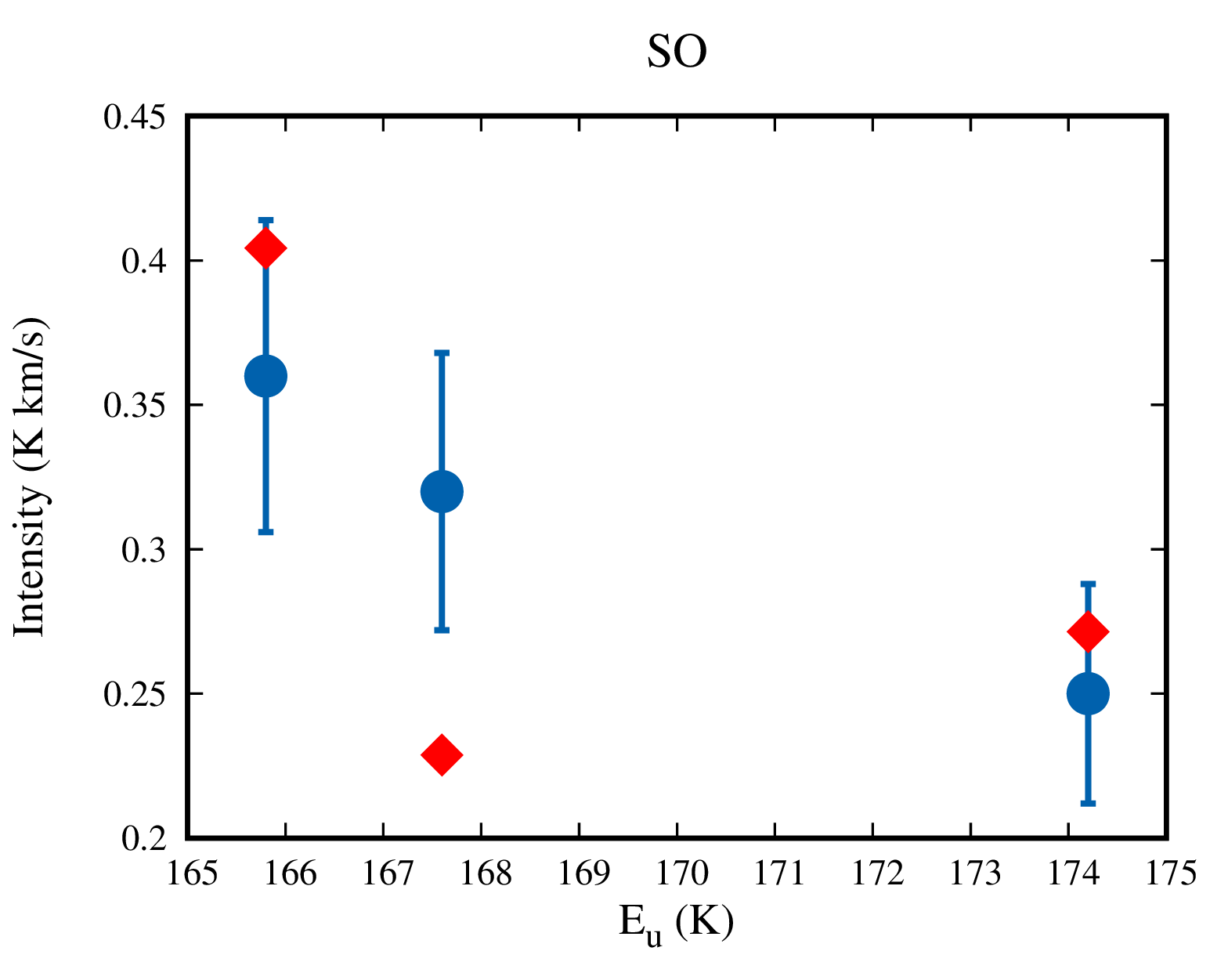}
\includegraphics[width=4.7cm, angle=-90]{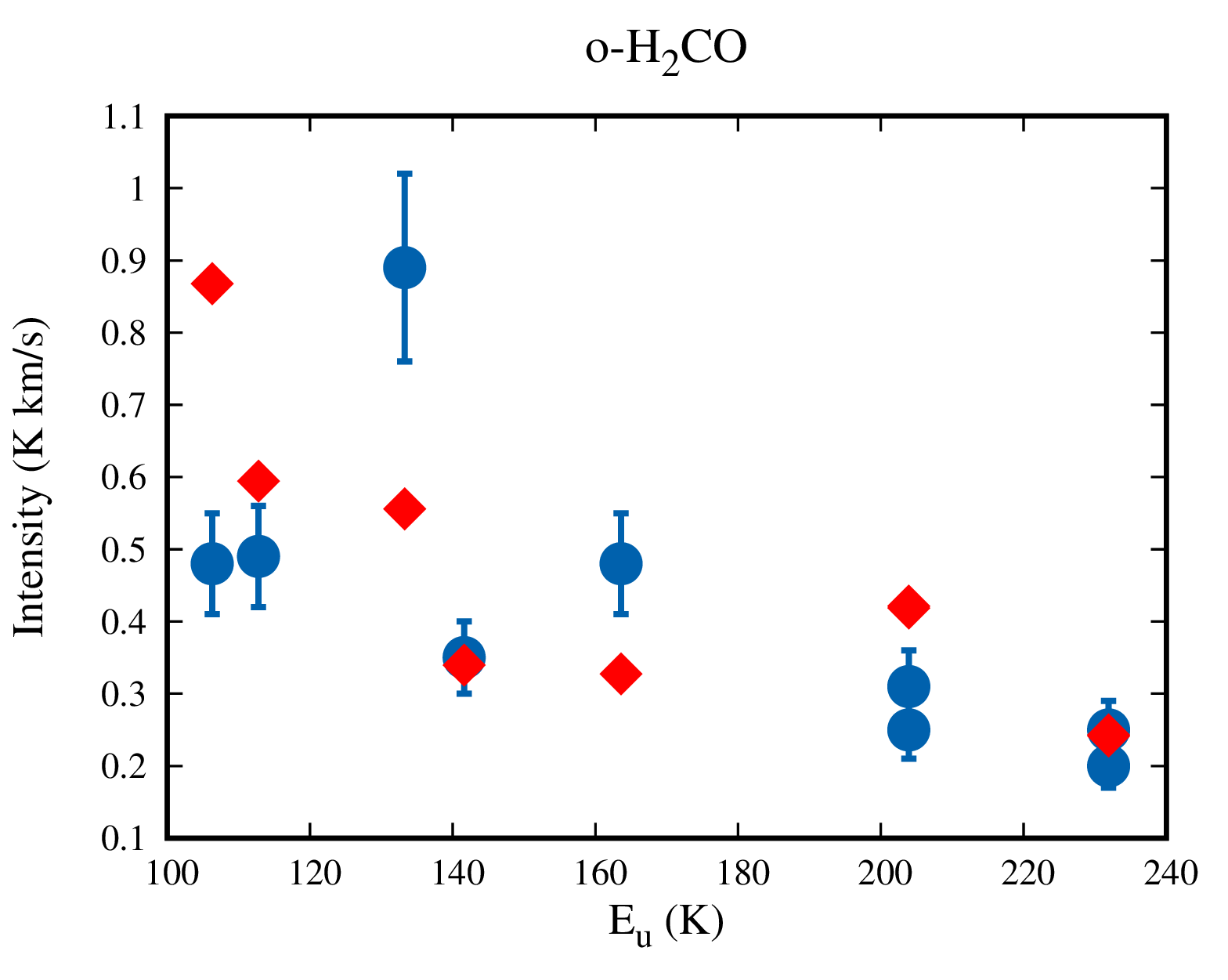}
\includegraphics[width=4.7cm, angle=-90]{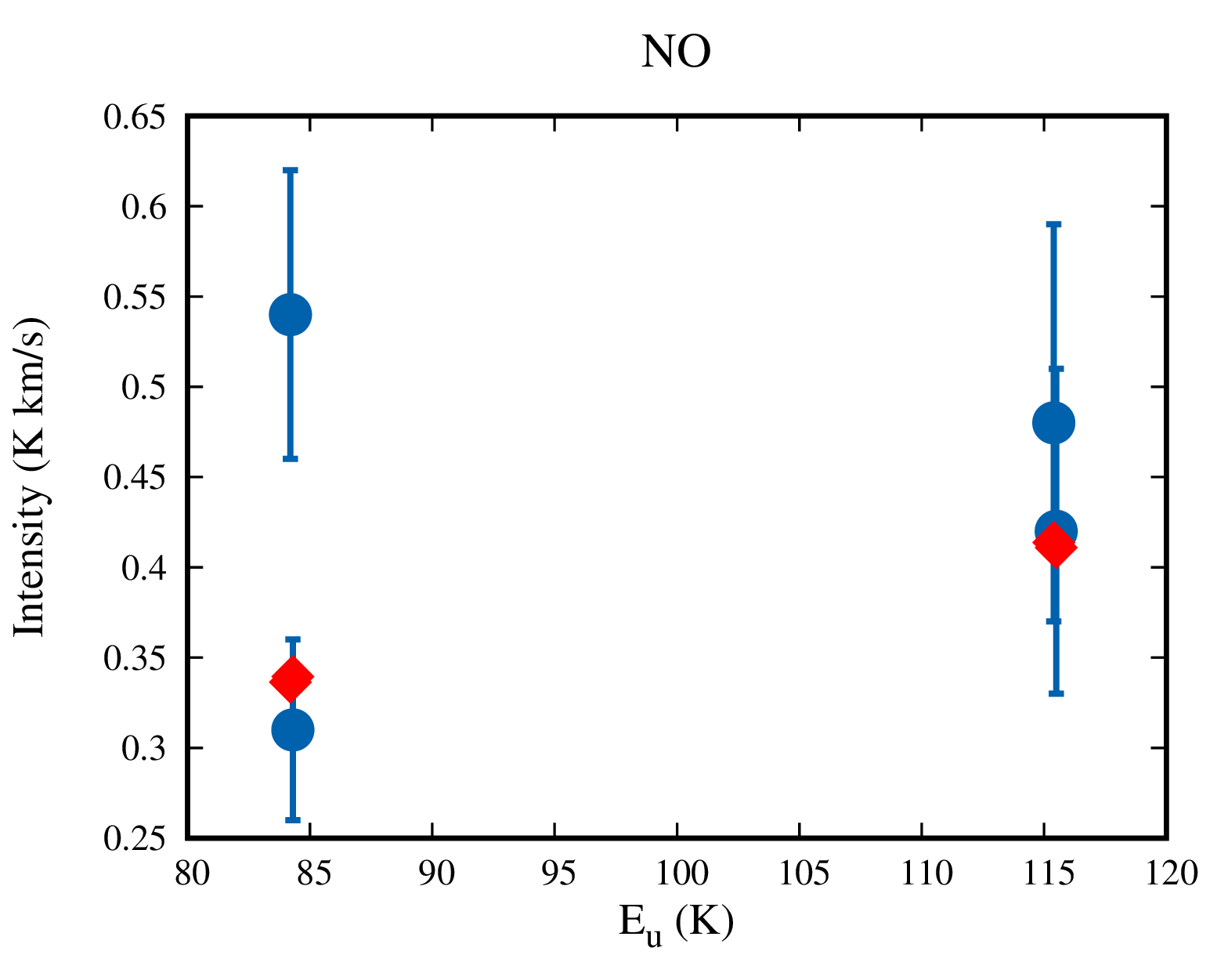}
\includegraphics[width=4.7cm, angle=-90]{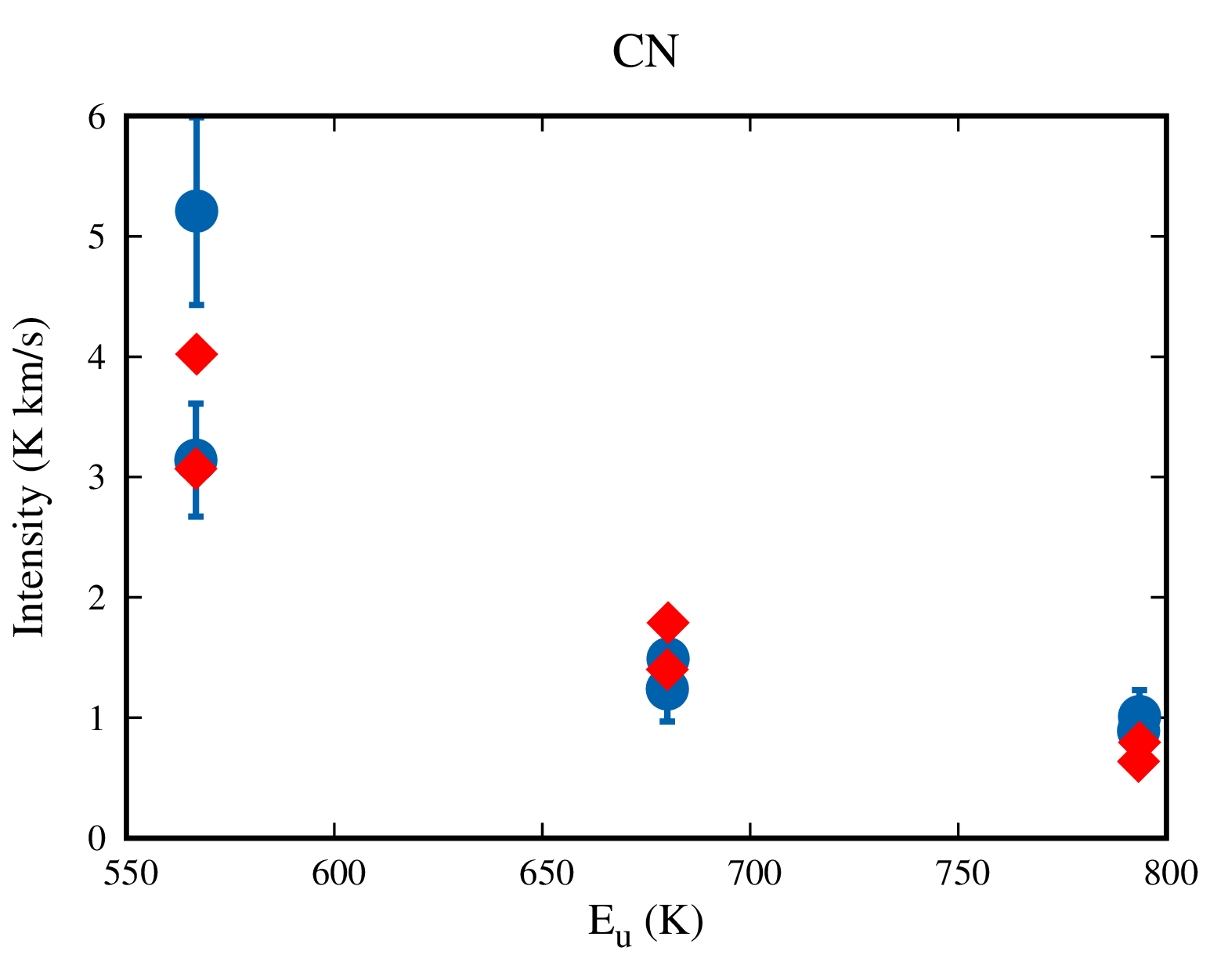}
\includegraphics[width=4.7cm, angle=-90]{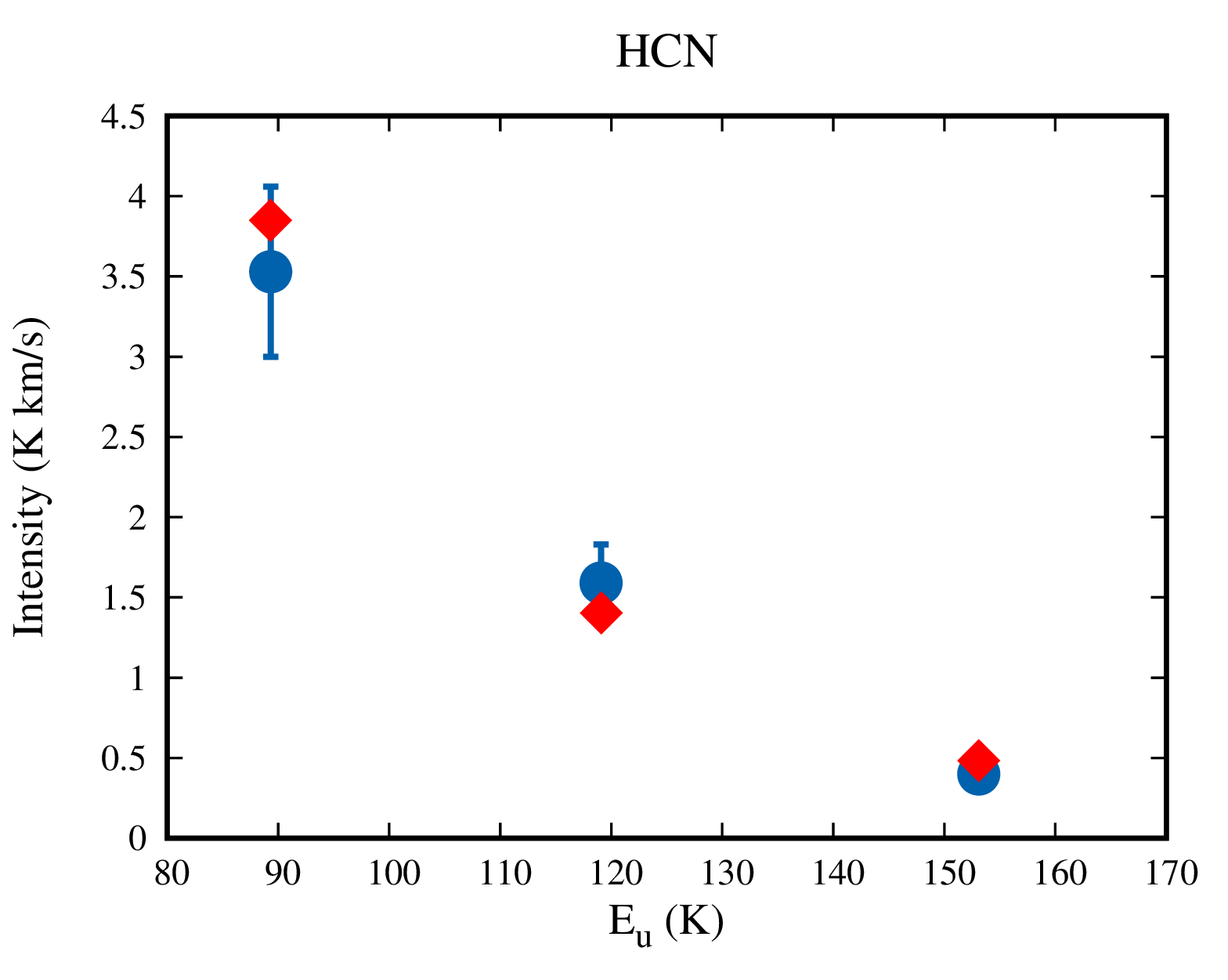}
\includegraphics[width=4.7cm, angle=-90]{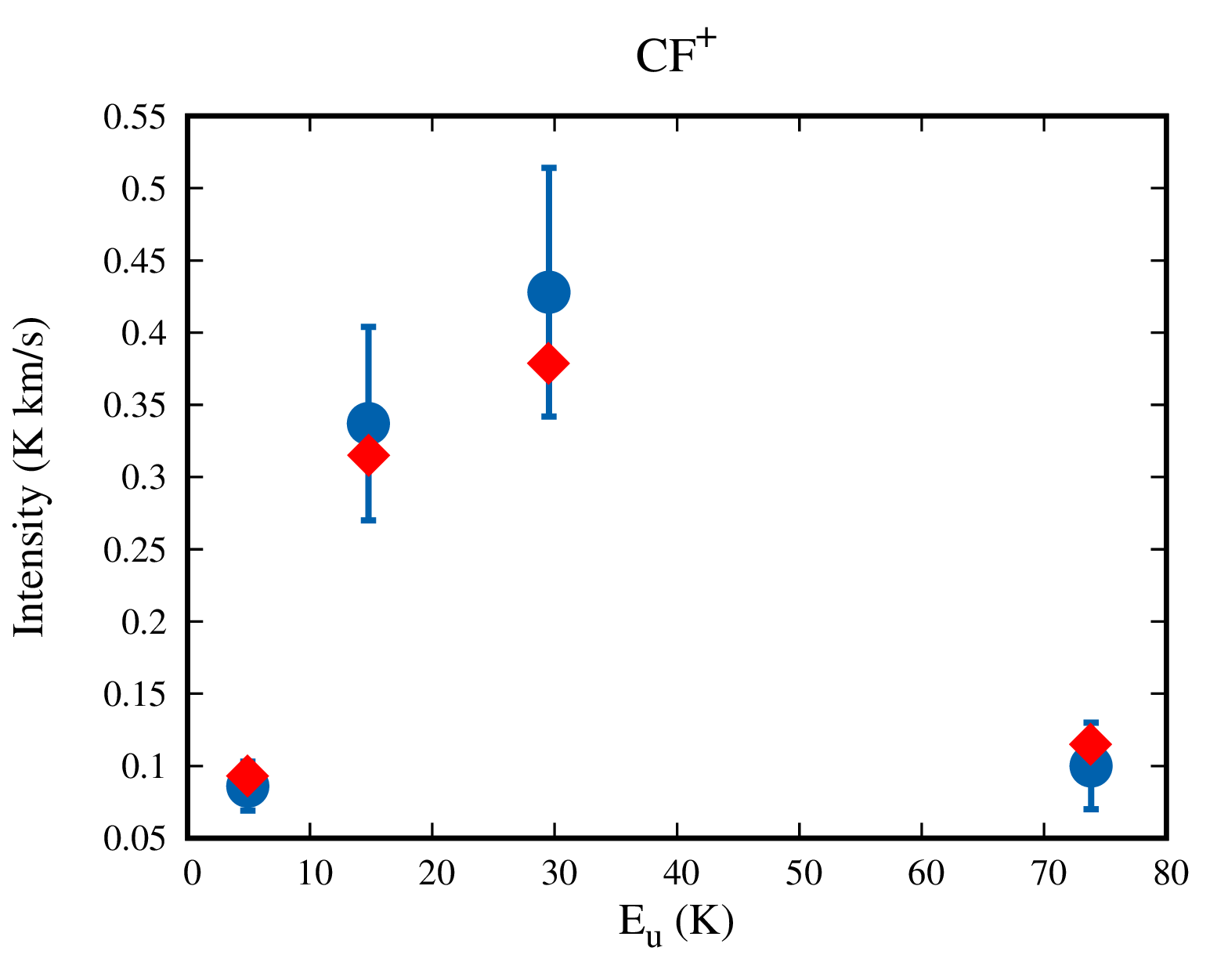}
\caption{RADEX fits of molecules with multiple detected transitions. The blue circles are the observed line intensities and the red diamonds are the line intensities predicted using RADEX.}
\label{radex_fits}
\end{center}
\end{figure*}

\end{appendix}

\end{document}